\documentclass[12pt, a4paper, oneside]{Thesis} 
\usepackage[greek,english]{babel}
\usepackage[square, numbers, comma, sort&compress]{natbib} 
\usepackage{longtable}
\usepackage{lipsum}
\usepackage[strict]{changepage}
\usepackage{graphicx}
\usepackage{array}
\usepackage[export]{adjustbox}[2011/08/13]
\usepackage{amsmath}
\usepackage{amsfonts}
\usepackage{amssymb}
\usepackage{lscape}
\usepackage{pdflscape}
\graphicspath{{Pictures/}} 

\hypersetup{urlcolor=blue, colorlinks=true} 
\title{\ttitle} 

\begin{document}

\frontmatter 

\setstretch{1.3} 
\setlength{\parindent}{1cm}
\setlength{\parskip}{0cm}

\fancyhead{}  
\rhead{\thepage}  
\lhead{}  

\pagestyle{fancy}  

\newcommand{\HRule}{\rule{\linewidth}{0.5mm}}  
\def\be{\begin{equation}}
\def\ee{\end{equation}}
\def\bq{\begin{eqnarray}}
\def\eq{\end{eqnarray}}
\def\beq{\begin{eqnarray*}}
\def\eeq{\end{eqnarray*}}
\def\f{\phi}
\def\r{\rho}
\def\a{\alpha}
\def\b{\beta}
\def\g{\gamma}
\def\G{\Gamma}
\def\d{\delta}
\def\l{\lambda}
\def\L{\Lambda}
\def\k{\kappa}
\def\m{\mu}
\def\n{\nu}
\def\na{\nabla}
\def\pa{\partial}
\def\z{\zeta}
\def\e{\eta}
\def\t{\theta}
\def\s{\sigma}
\def\ep{\epsilon}

\hypersetup{pdftitle={\ttitle}}
\hypersetup{pdfsubject=\subjectname}
\hypersetup{pdfauthor=\authornames}
\hypersetup{pdfkeywords=\keywordnames}


\begin{titlepage}
\begin{center}

\begin{figure}[htbp]
  \centering
    \includegraphics[width=35mm]{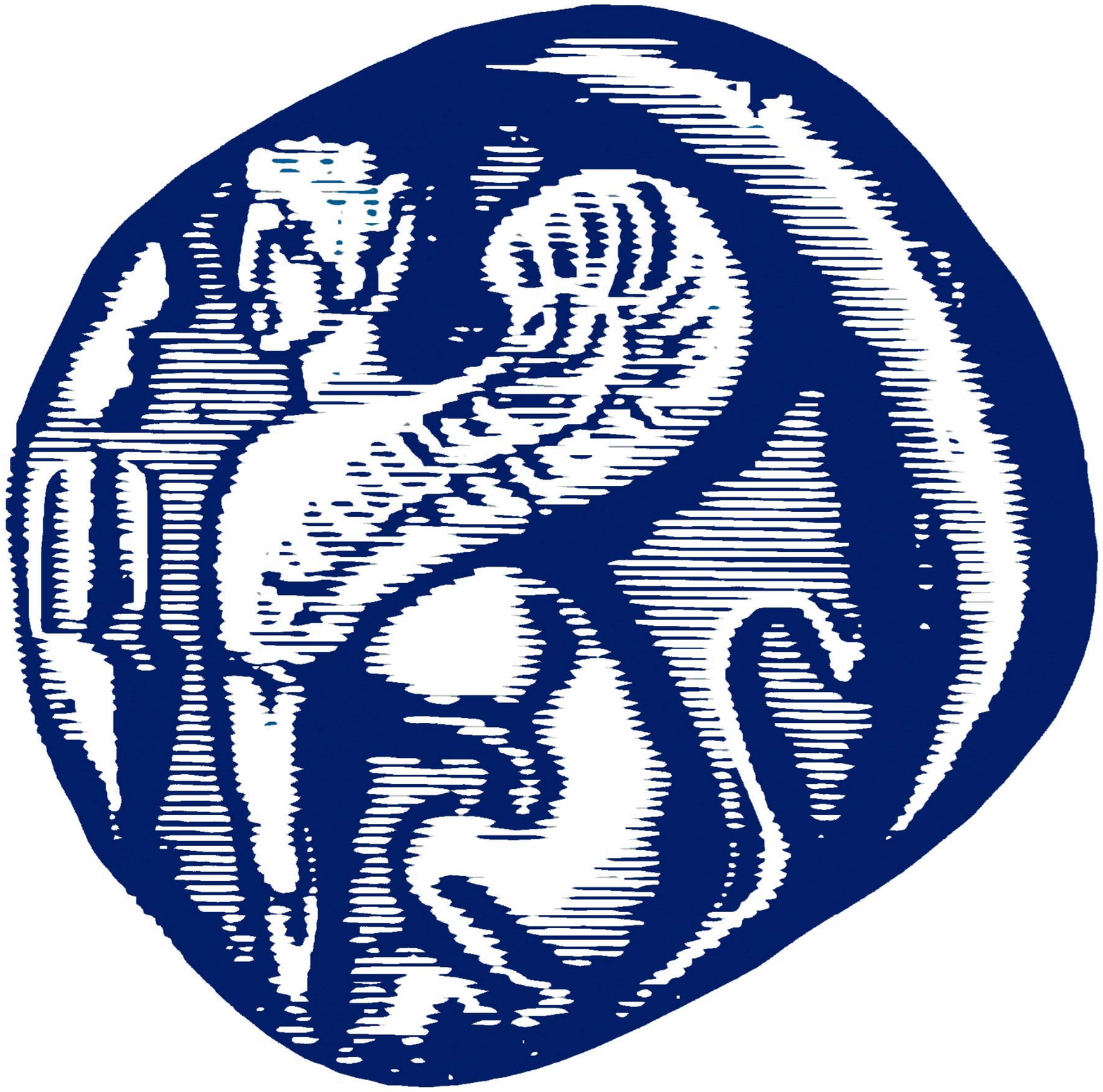}    
\end{figure}

\textsc{\LARGE \univname}\\[1.2cm] 
\textsc{\Large Doctoral Thesis}\\[0.5cm] 

\HRule \\[0.4cm] 
{\huge \bfseries \ttitle}\\[0.4cm] 
\HRule \\[1.5cm] 
 
\Large
\begin{center}
\href{http://www.icsd.aegean.gr/group/members-data.php?group=L5&member=1178}{\authornames}\\[2cm] 
\end{center}
 
 
\large \textit{A thesis submitted in fulfilment of the requirements\\ for the degree of \degreename}\\[0.3cm] 
\textit{in the}\\[0.4cm]
\groupname\\\deptname\\[2cm] 
 
{\large Samos, May 2016}\\[4cm] 

\vfill
\end{center}

\end{titlepage}


\begin{center}

\thispagestyle{empty}   
\begin{figure}[htbp]
  \centering
    \includegraphics[width=35mm]{Figures/Logo.pdf}    
\end{figure}

\textsc{\LARGE \textgreek{PANEPISTHMIO AIGAIOU}}\\[1.1cm] 
\textsc{\Large \textgreek{DIDAKTORIKH DIATRIBH}}\\[0.5cm] 

\HRule \\[0.4cm] 
{\huge \bfseries \textgreek{To tupik'o an'aptugma thc analutik'hc tropopoihm'enhc bar'uthtac}}\\[0.4cm] 
\HRule \\[1.5cm] 

\large

\begin{center}
\href{http://www.icsd.aegean.gr/group/members-data.php?group=L5&member=1178}{\Large \bfseries \textgreek{Dhm'htrioc Traq'ilhc}}\\[2.5cm] 
\end{center}
 
\large \textgreek{Didaktorik'h diatrib'h upoblhje'isa sto}\\[0.3cm] 
\textgreek{Tm'hma Mhqanik'wn Plhroforiak'wn kai Epikoinwniak'wn Susthm'atwn}\\
\textgreek{tou Panepisthm'iou Aiga'iou}\\[0.3cm]
\textgreek{gia thn ap'okthsh tou t'itlou tou}\\
\textgreek{Did'aktora tou Panepisthm'iou Aiga'iou}\\[2.5cm]

\large \textgreek{S'amoc}, \textgreek{M'aioc 2016}\\ 

\vfill
\end{center}

\clearpage


%

\addtotoc{Advisory Committee} 
\thispagestyle{empty}  

\vspace*{\fill} 
\begin{center} 
\begin{minipage}{\textwidth} 
\centering{
\textbf{The advisory committee}
\vspace{0.5cm}\\
Georgios Kofinas, Assistant Professor, University of the Aegean\\Supervisor\\
John Miritzis, Associate Professor, University of the Aegean\\Member\\
Nicolas Hadjisavvas, Professor Emeritus, University of the Aegean\\Member\\}
\vspace{3cm}
\end{minipage}
\end{center} 
\vfill

\clearpage


\thispagestyle{empty}  
\vspace*{\fill} 
\begin{center} 
\begin{minipage}{\textwidth} 
\centering{
\textbf{\textgreek{H sumbouleutik'h epitrop'h}}
\vspace{0.5cm}\\
\textgreek{Ge'wrgioc Kofin'ac}, \textgreek{Ep'ikouroc Kajhght'hc}, \textgreek{Panepist'hmio Aiga'iou}\\ \textgreek{Epibl'epwn}\\
\textgreek{Gi'annhc Muritz'hc}, \textgreek{Anaplhrwt'hc Kajhght'hc}, \textgreek{Panepist'hmio Aiga'iou}\\ \textgreek{M'elos}\\
\textgreek{Nik'olac Qatzhs'abbac}, \textgreek{Om'otimoc Kajhght'hc}, \textgreek{Panepist'hmio Aiga'iou}\\ \textgreek{M'elos}\\}
\vspace{3cm}
\end{minipage} 
\end{center} 
\vfill

\clearpage


\Declaration{

I, \authornames, declare that this thesis entitled, '\ttitle' and the work presented in it are my own. I confirm that:

\begin{itemize} 
\item[\tiny{$\blacksquare$}] This work was done wholly or mainly while in candidature for a research degree at this University.
\item[\tiny{$\blacksquare$}] Where any part of this thesis has previously been submitted for a degree or any other qualification at this University or any other institution, this has been clearly stated.
\item[\tiny{$\blacksquare$}] Where I have consulted the published work of others, this is always clearly attributed.
\item[\tiny{$\blacksquare$}] Where I have quoted from the work of others, the source is always given. With the exception of such quotations, this thesis is entirely my own work.
\item[\tiny{$\blacksquare$}] I have acknowledged all main sources of help.
\end{itemize}

\vspace{10mm}

Signed:\\
\rule[1em]{25em}{0.5pt} 

Date:\\
\rule[1em]{25em}{0.5pt} 
}

\clearpage 


\pagestyle{empty} 

\null\vfill 

\begin{flushright}
\textit{$E\nu \hspace{1mm} \alpha\rho\chi\grave{\eta} \hspace{1mm} \eta\nu \hspace{1mm} o \hspace{1mm} \Lambda \grave{o}\gamma o\varsigma$...}
\end{flushright}

\begin{flushright}
$K\alpha\tau\grave{\alpha} \hspace{1mm} I\omega\grave{\alpha}\nu\nu\eta\nu \hspace{1mm} \alpha' \hspace{1mm} 1.$
\end{flushright}


\vfill\vfill\vfill\vfill\vfill\vfill\null 

\clearpage 

\addtotoc{Abstract} 

\abstract{ In this Thesis, we treat the problem of the existence of generic perturbations of the regular and singular state in higher-order gravity in cases of vacuum and radiation models that derives from the lagrangian $R+\ep R^2$. We show that there is a regular state of the theory in vacuum in the form of a formal series expansion having the same number of free functions as those required for a general solution of the theory, while this is not true for the case of radiation. This means that there exists an open set in the space of analytic initial data of the theory in vacuum that leads to a regular solution having the correct number of free functions to qualify as a general solution. Further, we show that a singular state of the theory in vacuum cannot be admitted, while in the case of radiation we obtain a particular solution. 

To achieve this, we need to develop a number of results in various directions. We prove that there exists a first-order formulation of the theory with the Cauchy-Kovalevskaya property. This formulation of the quadratic theory $R+\ep R^2$  evolves an initial data set $(\mathcal{M},\g_{\a\b},K_{\a\b},D_{\a\b},W_{\a\b})$ (plus energy density $\rho$ and velocities $u_\a$ for the radiation case), through a set of four evolution equations and two constraint equations to build the time development $(\mathcal{V},g)$. We then prove that if we start with an initial data set in which the metric has the asymptotic form (regular or singular) and evolve, then we can build an asymptotic development in the form of a formal series expansion which satisfies the evolution and constraint equations and has the same number of free functions as those of a general solution of the theory in vacuum, with less free functions for the case of radiation. In other words, we show that regularity is a generic feature of the $R+\ep R^2$ theory in vacuum under the assumption of analyticity, but this cannot happen for generic radiation models. 
}

\clearpage 


\vspace*{\fill} 
\begin{center} 

\begin{minipage}{\textwidth} 
\centering{ 
\normalsize
\textsc{\textgreek{PANEPISTHMIO AIGAIOU}}\\ 
\textgreek{Tm'hma Mhqanik'wn Plhroforiak'wn kai Epikoinwniak'wn Susthm'atwn }\\

\bfseries {\large \textit{\textgreek{Per'ilhyh}}}\\ 
\normalfont\textgreek{Didaktorik'hc diatrib'hc}\\
\large\bfseries {\textgreek{To tupik'o an'aptugma thc analutik'hc tropopoihm'enhc bar'uthtac}}\\[0.1cm] 
\normalfont\textsc{\textgreek{DHMHTRIOU TRAQILH}}\\
}
\end{minipage} 
\end{center} 

\normalsize
\textgreek{Se aut'h thn ergas'ia pragmateu'omaste to pr'oblhma thc 'uparxhc twn genik'wn diataraq'wn thc omal'hc kai idi'omorfhc kat'astashc se uyvhl'oterhc t'axhc jewr'iac bar'uthtac, sthn per'iptwsh tou keno'u kai tou mont'elou aktinobol'iac, to opo'io par'agetai ap'o thn lagkratzian'h} $R+\ep R^2$. \textgreek{De'iqnoume 'oti up'arqei m'ia omal'h kat'asta-\\sh thc jewr'iac sto ken'o up'o th morf'h seir'wn, 'eqontac ton 'idio arijm'o ele'ujerwn sunart'hsewn me eke'inec pou apaito'untai ap'o th genik'h l'ush thc jewr'iac kai to opo'io den e'inai alhj'ec gia thn per'iptwsh thc aktinobol'iac. Aut'o shma'inei 'oti up'arqei 'ena anoiqt'o s'unolo analutik'wn arqik'wn dedom'enwn thc jewr'iac sto ken'o to opo'io odhge'i se omal'h l'ush 'eqontac ton swst'o arijm'o ele'ujerwn sunart'hsewn 'etsi 'wste aut'h na perigrafe'i wc genik'h l'ush. Epipl'eon, de'iqnoume 'oti m'ia idi'omorfh kat'astash thc jewr'iac sto ken'o de mpore'i na g'inei apodekt'h, kaj'wc sthn per'iptwsh thc aktinobol'iac odhge'i se eidik'h l'ush.

Gia na to pet'uqoume aut'o, qreiaz'omaste na parousi'asoume 'enan arijm'o apotelesm'atwn se poik'ilec kateuj'unseic. Apodeikn'uoume 'oti up'arqei m'ia pr'wthc t'axhc montelopo'ihsh thc jewr'iac me thn idi'othta} Cauchy-Kovalevskaya. \textgreek{Aut'h h montelopo'ihsh thc tetragwnik'hc jewr'iac} $R+\ep R^2$ \textgreek{exel'issei 'ena s'unolo arqik'wn dedom'enwn} $(\mathcal{M},\g_{\a\b}, K_{\a\b},D_{\a\b},W_{\a\b})$ \textgreek{(kaj'wc kai thn pukn'othta en'ergeiac} $\rho$ \textgreek{kai tic taq'uthtec} $u_\a$ \textgreek{sthn per'iptwsh thc aktinobol'iac), dia m'esou en'oc sun'olou tess'arwn exis'wsewn ex'elixhc kai d'uo exis'wsewn desm'wn gia na kataskeu'asei th qronik'h ex'elixh} $(\mathcal{V},g)$. \textgreek{'Epeita apodeikn'uoume 'oti an xekin'hsoume me 'ena s'unolo arqik'wn dedom'enwn, sto opo'io h metrik'h 'eqei asumptwtik'h morf'h (omal'h 'h idi'omorfh) kai anapt'uxoume, t'ote mporo'ume na kataskeu'asoume m'ia asumptwtik'h ex'elixh up'o th morf'h seir'wn h opo'ia ikanopoie'i tic exis'wseic ex'elixhc kai tic exis'wseic desm'wn kai peri'eqei ton 'idio arijm'o ele'ujerwn sunart'hsewn 'opwc eke'inwn thc genik'hc l'ushc thc jewr'iac sto ken'o, kai lig'oterec ele'ujerec sunart'hseic gia thn per'iptwsh thc aktinobol'iac. Me 'alla l'ogia, de'iqnoume 'oti h omal'othta e'inai 'ena tupik'o qarakthristik'o thc jewr'iac} $R+\ep R^2$ \textgreek{sto ken'o me thn up'ojesh thc analutik'othtac, k'ati t'etoio en to'utic parabi'azetai se tupik'a mont'ela aktinobol'iac.}

\vfill     
\clearpage 


\setstretch{1.3} 

\acknowledgements{\addtocontents{toc}{\vspace{1em}} 

First and foremost I want to thank my supervisor Prof. Spiros Cotsakis for admitting me as his PhD student and supporting me at every step of this effort. I would have never achieved to finish this challenging work without his guidance and ideas. He is an inspiring teacher and I am truly grateful. 

I would also like to thank the members of my advisory committee, Assistant Professor Georgios Kofinas, Associate Professor Johh Miritzis and Professor Emeritus Nicolas Hadjisavvas for their kind support.

I am deeply grateful to Dr. Antonios Tsokaros for his determined help and stimulating contribution in all these years of research. 

I shall also give many thanks to Assistant Professor Ifigeneia Klaoudatou for our useful conversations and guidance in the first years of my research.

Many thanks should also be offered to the Department of Information and Communication Systems Engineering of the University of the Aegean for offering suitable conditions for the completion of this work.

I am also grateful to my wife Eleni Mountricha for insisting on working on this PhD thesis and for her constant encouragement and patience all these years.

My deep gratitude to my friends Giorgos Kolionis and Georgia Kittou for all those years of working together, the fun we shared with and their support altogether.

I would also want to express my gratitude to my parents whose encouragement served as an incentive to open my horizons.

Special thanks to my friend Ioannis Stamatiou for his motivation and constructive ideas throughout our endless hours of discussion.

My final thanks to Paraskevi Alikari, Kalia Anastasopoulou, and all the company in the Papanikolaou and Lymperis building for the humour, the fun and the ideas. 
}  

\clearpage 


\thispagestyle{empty}  
\vspace*{\fill} 
\begin{center} 
\begin{minipage}{\textwidth} 
\centering{
This thesis is partly based on the following publications:
\begin{itemize}

\item S. Cotsakis, D. Trachilis and A. Tsokaros. \textit{Generic regular cosmological solutions in higher order gravity theories}. In Proceedings of the Thirteenth Marcel Grossman Meeting on General Relativity, K.  Rosquist, R. T Jantzen, R. Ruffini (eds.), World Scientific,  2015, p. 1865-1867.\\ 
\href{http://arxiv.org/abs/1302.6674}{arXiv:1302.6674}.
\item Preprint submitted to J. Geom. Phys. \textit{The regular state in higher order gravity}.\\
\href{http://arxiv.org/abs/1212.2412}{arXiv:1212.2412}. 
\end{itemize}
} 
\end{minipage} 
\end{center} 
\vfill

\clearpage 


\pagestyle{empty} 

\null\vfill 

\begin{flushright}
\textit{To my wonderfull wife Eleni,\\ and my sweet kids Olga and Nikos...}
\end{flushright}

\vfill\vfill\vfill\vfill\vfill\vfill\null 

\clearpage 


\pagestyle{fancy} 

\lhead{\emph{Contents}} 
\tableofcontents 

\mainmatter 

\pagestyle{fancy} 



\chapter{Introduction} 

\label{Chapter1} 

\lhead{Chapter 1. \emph{Introduction}} 

Since 1998 with the discovery of the acceleration of the universe through observations of type Ia supernovae \cite{R98,P98,R99,P99,R01,T03,K03,R04,B04}, according to WMAP and PLANCK, approximately 77\% of its energy content is in an invisible and unclustered form called \emph{dark energy}. So far, in order to explain this cosmic acceleration three main classes of models have been proposed. The first is the cosmological constant $\Lambda$, the simplest possible form of dark energy, as it is constant both in space and time and leads to the standard $\Lambda$-CDM (Lambda cold dark matter) model of cosmology that provides a good account of the acceleration of the universe \cite{hlk13,AAA2013} and \cite{prl10}, Chap. 1. The second one is the dark energy, in which the fluid satisfies the equation of state: $p\approx -\rho$, where $p,\rho$ are the pressure and the energy density, respectively \cite{tf,abc06,Wet88,PR88,cst06,fth08,pad08}. Finally, the third class of models is the higher-order or modified $f(R)$ gravity \cite{Weyl18,whi84,ba-co88,maeda88,maeda89,Sch07,cots97,cct08,cdtt04,dk03,chi03,dick04,clba05,olmo05,cnot06,no06,shs06,bbpst06,esk06,na06,cse06,ftbm06,apt07}, where $f(R)$ is a nonlinear function of the scalar curvature, with which we shall deal extensively below.

Many works focused on the cosmological implications of higher-order $f(R)$ gravity since such models may give an alternative explanation and study of the acceleration effects observed in cosmology \cite{cap02,noB03,noD03,vol03,cfdett05}. In particular, we shall focus on quadratic, 
\be
f(R)=R+\ep R^2,
\label{f(R) start}
\ee
gravity theories, which were concieved as a mathematical alternative to Einstein's theory \cite{Buc70}. 

The field equations following from (\ref{f(R) start}) are of fourth-order, owing to the fact that they contain derivatives up to the fourth-order of the components of the metric $g_{ij}$ with respect to the space-time coordinates \cite{Far08,tt83}. Specifically, the term $R^2$ is responsible for the fourth-order character of the field equations, while the Einstein-Hilbert term $R$ contains derivatives up to the second-order.  Therefore, the theory of general relativity consitutes the simplest choice of the higher-order gravity theories, and corresponds to the simplest choice of the coefficient $\ep$ in (\ref{f(R) start}), $\ep=0$. Consequently, these two cosmological theories do not share the same geometric degrees of freedom. In particular, in comparison with general relativity, higher-order gravity theory has additional degrees of freedom.

A concern which arises with $f(R)$ gravity theory is associated with the corresponding initial value or Cauchy problem. A surface $\mathcal{M}_{t}$ is said to be a \emph{Cauchy surface} if every point of space-time is in the domain of dependence of $\mathcal{M}_{t}$ (\cite{grpart1}, Chap. 5). More precisely, the existence of a Cauchy surface ensures that there is a single slice of space-time such that information on that slice suffices to completely determine the physics throughout the entire space-time. Starting from an assignment of suitable Cauchy data on a Cauchy surface, the subsequent dynamical evolution of the system must be uniquely determined. Then, the initial value problem is \emph{well-formulated}. For instance, electromagnetism is a well-formulated field theory. Nevertheless, sometimes a well-formulated theory may not be adequate concerning its stability properties. It is important for the study of any theory to know if it is predictive in the sense that small perturbations in the initial data lead to small perturbations in the dynamics over all the space-time \cite{bo83}. This means that the structure of the theory must not be affected by small perturbations of the initial data. In the case that a problem is well-formulated and also preserves the causal structure of the theory we say it is \emph{well-posed} \cite{cv09,cc68}.

In general relativity, the Cauchy problem has intensively analysed \cite{sgr}, Chap. V and \cite{srel}, Chap. 30. However, in the case of scalar-tensor theories, the Cauchy problem has not been analysed with the same degree of completeness. In \cite{han72}, it was shown that any scalar-tensor theory corresponds to a higher-order gravity theory, and this fact was exploited so that the initial value problem can be studied. In $f(R)$ gravity, the choice of coordinates is decisive in order to infer whether the theory is well-formulated and possibly well-posed. For example, in \cite{noa83} through harmonic coordinates, the initial value problem of general relativity is proven to be well-posed. This means that general relativity is a stable theory. Further, in the same paper, the study of higher-order derivative gravity leads to the fact that this theory also has a well-posed initial value problem formulation. However, using (\ref{f(R) start}) in vacuum, one can perform a conformal transformation \cite{whi84,bc91,fgn99,rett96,c08}, and prove that the solutions in higher-order gravity reduce to solutions of Einstein's theory with a minimally coupled scalar field $\phi$ with self-interaction potential $V(\phi)$, where the de Sitter solution corresponds to a local extremum of $V(\phi)$ at $\phi=0$. But, in doing so, the two theories can not be considered as quantitavely the same. Besides, the degrees of freedom of these two theories are not the same and, for this reason, according to \cite{noa83}: `\textit{the initial value problem can be shown to be qualitatively the same as that of the conformally coupled scalar field}'. 

In this Thesis, the goal is to find the behaviour of the solutions of the field equations in higher-order $R + \ep R^2$ gravity theory by using formal series expansions of the metric $g_{ij}$. Prior to anything else, in order to study this behavior, we shall employ a \emph{synchronous reference system}, in which the metric necessarily satisfies the conditions \cite{bkl70,bkl82}:
\be
g_{00}=1
\label{start condition for g00},
\ee
and,
\be
g_{0\a}=0.
\label{start condition for g0a}
\ee
Hence, according to the $(3+1)$-decomposition \cite{mul11,sal06}, the line element is given by the expression,
\bq
ds^2 &=& dt^2 + g_{\a\b}dx^\a dx^\b \nonumber \\
&=& dt^2 - \g_{\a\b}dx^\a dx^\b.
\label{start relation for line element}
\eq
It is worth noting that in a synchronous reference frame the four-vector
\be
u^i = (dt/ds,dx^\a/ds),
\label{start relation for velocities}
\ee
tangent to the world lines $x^\a = \textrm{constant}$, has components
\be
u^0 = 1, \quad   u^\a = 0.
\label{start relation for velocities2}
\ee 
Therefore, the timelines are geodesics in the four dimensional spacetime and the four-vector $u^i$ satisfies the geodesic equations:
\be
\dfrac{du^i}{ds} + \G^i_{jk}u^j u^k = 0.
\label{geodesic equations}
\ee 
Based on the method of Landau-Lifshitz (see \cite{ll}, Chap. 11,14), we are interested in the forms,
\be
\g_{\a\b} = (\g_{\a\b})^{(0)} + t(\g_{\a\b})^{(1)} + t^2 (\g_{\a\b})^{(2)} + \cdots
\label{start relation for regular series}
\ee
and,
\be
\g_{\a\b} = t(\g_{\a\b})^{(1)} + t^2 (\g_{\a\b})^{(2)} + \cdots
\label{start relation for singular series}
\ee
where the $(\g_{\a\b})^{(n)}$, $n=0,1,2,...$ are functions of the space coordinates. These two forms of the $3$-metric $\g_{\a\b}$ are the simplest ones that we could consider. In fact, they also occur in solutions of the Friedmann equations both in general relativity and higher-order gravity in vacuum as well as in the presence of radiation as we will see extensively in Chapter \ref{Chapter4}. In particular, the exact solutions of the Friedmann equations in cases of vacuum and radiation for all values of the constant curvature $k$ $(k=-1,0,+1)$ serve as a motivation for the form of the $3$-metric $\g_{\a\b}$. The solution in empty space, where the pressure $p$ and the energy density $\rho$ are set equal to zero, corresponds to the form (\ref{start relation for regular series}), while the solution in case of radiation, in which we have $p=\rho/3$, corresponds to the form (\ref{start relation for singular series}).

There are many works which use these two forms or similar forms of the $3$-metric and study assorted properties about the behaviour of the geometric solutions of the field equations. In \cite{lk63}, there is a study of the arbitrary functions included in the solution of the Einstein's equations in the case that space is filled with matter with the equation of state $p=\rho/3$. It is shown that the solution has five less degrees of freedom compared with the general solution of the problem (see also \cite{cdlp94,dl95}). Because of our interest in radiation models of higher-order gravity, we give some details of this result in Appendix \ref{AppendixB}.

According to the standard cosmological model, a natural equation of state for the matter near the cosmological singularity is that of radiation. However, taking into consideration the development of the inflationary cosmological models \cite{star79,star80,guth81,lin82,lin83,star86,lin90}, as well as the appearance of brane and M theory cosmological models \cite{bf01,bf01sec}, one realizes that matter plays different roles in different stages of cosmological evolution and thus, the equation of state can assume different types. For this reason, in \cite{kks02} a generalization of the quasi-isotropic solution of the Einstein equations near cosmological singularity found by Lifshitz and Khalatnikov in 1960 for the case of the radiation-dominated Universe, was proposed through an expression of the spatial metric of the form,
\be
\g_{\a\b} = t^m (\g_{\a\b})^{(m)} + t^n (\g_{\a\b})^{(n)}, \quad m<n,
\label{kks02 metric}
\ee   
with fluid matter with equation of state $p=k\rho$.

In \cite{kkms03}, there is an analysis concerning the quasi-isotropic inhomogeneous solution of the Einstein equations near a cosmological singularity in the form of a series expansion in the synchronous reference system in the case of a two-fluid cosmological model. The inhomogeneous quasi-isotropic solution for the spatial metric is represented as,
\be
\g_{\a\b} = \sum_{n,m=0} {\g}^{(n,m)}_{\a\b}t^{2\g_0 + n(2-2\g_0) + m(2-\g_0 \a_1)},
\label{kkms03 metric}
\ee
where,
\be
\a_1 = 3(1+k_1),
\label{a1k1}
\ee
\be
\g_0 = \dfrac{2}{3(1+k_2)},
\label{g0k2}
\ee
with the equations of state $p_l = k_l \rho_l$, $l=1,2$ of the two barotropic fluids.

In \cite{ss87}, using the synchronous system of reference details are given about the general solution of the Einstein equations with a $\Lambda$ term. It is shown that the general solution asymptotically approaches de Sitter spacetime. In order to prove this result, the relation which is used is,
\be
\g_{\a\b} = exp(2Ht)h_{\a\b},
\label{gh relation}
\ee
where,
\be
h_{\a\b} = a_{\a\b} + \sum_{n=1}^\infty b^{(n)}_{\a\b}exp(-nHt),
\label{hab series}
\ee
where $H$ depends on the spatial coordinates $x^\a$ (see also \cite{mss88,ren06}).

In \cite{bct10}, there is a study of the general solution of the Einstein equations near a future sudden singularity \cite{bgt86,bar04}, in the case that the pressure $p$ and the energy density $\rho$ are not connected with an equation of state of the form $p=f(\rho)$, while the scale factor has the form,
\be
a(t) = 1 + Bt^q + C(t_s - t)^n,
\label{scale factor gscs}
\ee
where $B>0,q>0,n>0$ and $C$ are free constants and $0<t<t_s$. Setting $a(0)=0$, the scale factor becomes,
\be
a(t) = 1 + (\dfrac{t}{t_s})^q (a_s - 1) - (1 - \dfrac{t}{t_s})^n,
\label{scale factor gscs new}
\ee
where $a_s \equiv a(t_s)$ (see also \cite{bct13,cb07,bct12}). Taking into account the reduction of the Einstein equations,
\bq
3H^2 &=& \rho - \dfrac{3k}{a^2}, \nonumber \\
\label{hubble}
\dot{\rho} + 3H(\rho + p) &=& 0, \nonumber \\
\label{dot ener dens}
\dfrac{\ddot{a}}{a} &=& - \dfrac{\rho + 3p}{6},
\label{ddot a}
\eq
where $H=\dot{a}/a$ is the Hubble parameter, and restricting $1<n<2$ and $0<q\leq 1$, then for $t\rightarrow t_s$ we have that the solution (\ref{scale factor gscs new}) satisfies,
\be
a \rightarrow a_s (\textrm{finite}), \quad H \rightarrow H_s(\textrm{finite}) \quad \textrm{and} \quad \rho \rightarrow \rho_s(\textrm{finite}) >0, 
\label{aHrho}
\ee
with
\be
p \rightarrow \infty.
\label{infinite p}
\ee 
Using (\ref{start relation for line element}) and the linear asymptotic form of the solution (\ref{scale factor gscs new}), namely,
\be
a(t) = a_s + q(1 - a_s)(1 - \dfrac{t}{t_s}),
\label{linear scale factor}
\ee
the general series expansion is,
\be
\g_{\a\b} = a_{\a\b} + t b_{\a\b} + t^n c_{\a\b} + \cdots
\label{ser expan gscs}
\ee
(having also considered a further change of the time coordinate in order to place the future sudden singularity at $t=0$). From the series representation of the extrinsic curvature,
\be
K_{\a\b} = \pa_t \g_{\a\b},
\label{start ex cur}
\ee
we may then obtain series expansions of the components of the Ricci tensors,
\be
R^0_0 = -\dfrac{1}{2}\pa_t K - \dfrac{1}{4}K^\b_\a K^\a_\b,
\label{start R00 mixed}
\ee
\be
R^0_\a = \dfrac{1}{2} \left ( \nabla_\b K^\b_\a - \nabla_\a K \right ),
\label{start R0a mixed}
\ee
\be
R^\b_\a = -P^\b_\a - \dfrac{1}{4}KK^\b_\a,
\label{start Rba mixed}
\ee
where $P^\b_\a$ is the Ricci tensor associated with $\g_{\a\b}$. Taking into consideration the relations (\ref{start condition for g00}) and (\ref{start condition for g0a}), the Einstein equations,
\be
R^i_j - \dfrac{1}{2}\d^i_j R = 8\pi T^i_j,
\label{Eistein tensor}
\ee
where,
\be
T^i_j = 8\pi [(\rho + p)u^i u_j - p\d^i_j],
\label{stress energy tensor gscs}
\ee  
lead to $13$ independent functions. Subtracting the $4$ coordinates covariances, there remain $9$ independent arbitrary functions of the three space coordinates on each surface of constant time. This solution is quasi-isotropic and corresponds to a general solution of the problem as required by the structure of the initial value problem. In particular, one expects there will be $12$ data from $g_{\a\b}$ and its time derivative, plus $3$ free velocity components $u_\a$, plus $2$ from the pressure $p$ and the energy density $\rho$, giving a total $17$ independent data. However, due to the four constraint equations and the four diffeomorphism transformations, the number of the arbitrary functions is finally equal to $9$.

Ever since the fundamental realization that Einstein's equations can be equivalently studied as a geometric, nonlinear system of evolution equations with constraints, there has been a considerable stream of problems in general relativity  of a classical dynamical nature crucially depending on the notion of evolution from prescribed initial data (cf. \cite{ycb} and references therein). In particular, for the cosmological case, at least since the pioneering investigations of Lifshitz and Khalatnikov in relativistic cosmology \cite{lk63}, perturbative approaches to the question of genericity of cosmological solutions constructed asymptotically from given initial data,  have been advanced in various directions in general cosmological theory. This was the way to prove the existence of  vacuum, regular cosmological solutions in general relativity having the required number to qualify as general ones, and also the non existence of a general singular, radiation solution in the same context, cf. \cite{lk63,ll}, eventually leading to the realization that the initial state was of a more complex nature \cite{bkl70,bkl82,he}. These first results were based on an approximation technique that consisted of several steps, writing down a suitable expansion of the metric, substituting it to the field equations, and counting the number of free (or arbitrary) functions needed to make the whole scheme consistent.

The original perturbative approximation scheme mentioned above proved to be especially fertile, and in fact it is being exploited in various directions ever since. It was used by Starobinski to study solutions of the Einstein equations with a positive cosmological constant, especially with regard to the question of the asymptotic stability of de Sitter space used in inflationary scenarios \cite{star1,star2}. The formal series expansions were in addition, used more recently in \cite{kks02} to study the genericity question in relativistic cosmologies with a general equation of state $p=w\r$, where it was proved that, in order to be able to construct  a general singular solution initially, certain restrictions on the fluid parameter $w$ are needed. Further, the `ultrastiff' case was recently considered in a more rigorous way using Fuchsian techniques  in Ref. \cite{hs}. Formal series where also used in \cite{kkms03} to study the more involved problem of perturbing an FRW universe containing two suitable fluids, and also in \cite{bct10} where it was shown how to construct a sudden singularity with the  genericity properties of a general solution, a situation met previously only in the `no-hair' behaviour of inflation. Extensions  to certain higher-order gravity theories have also been considered in a formal series context, especially in connection with the stability of de Sitter space under generic perturbations in these theories, cf. \cite{ss87,mss88}.

It is interesting that the original approximation scheme of formal perturbation series has been further refined and applied in various situations in cosmology, cf. \cite{mss88,ste90,tom93} and references therein. More recently, Rendall has put the original formal series expansion techniques used by Starobinski \cite{star1} for the stability of de Sitter space, in a more rigorous basis and was able to prove various interesting theorems concerning function counting and formal series solutions of the Einstein equations with a positive cosmological constant in an `initial value problem' spirit, cf. \cite{ren1} (see also \cite{ren06}).

In this Thesis, we are interested in the possible genericity of \emph{regular} or \emph{singular} solutions defined by formal series expansions in the context of vacuum as well as radiation-dominated models in higher-order gravity derived from the analytic lagrangian $R+\epsilon R^2$ theory. In particular, we prove that the higher-order gravity field equations in vacuum admit a unique solution in the form of a regular formal power series expansion (\ref{start relation for regular series}) which contains the right number of free functions to qualify as a general solution of the system. This requires a careful function counting technique, and for this purpose it is necessary to develop a formulation of the theory as a system of evolution equations with constraints. However, if we use singular formal series expansion of the form (\ref{start relation for singular series}), the higher-order gravity field equations in vacuum do not provide the right number of free functions that correspond to the general solution. 

Repeating this process for the higher-order field equations with radiation, we conclude that neither in the case of regular, nor in the case of a singular formal power series expansion we obtain the right number of arbitrary functions in order to characterize the solution as a general one.   

A more detailed plan of this Thesis is as follows.\\
In Chapter \ref{Chapter2}, we use the ADM (Arnowitt-Deser-Misner) formalism as a step to study the evolution problem of higher-order gravity theories in vacuum as well as in the presence of a fluid. We exploit normal coordinates in which the spatial coordinates move with the fluid while the time coordinate is the proper time along the fluid world lines, and deal with the function-counting problem of the quadratic theory in vacuum and the corresponding function-counting problem in presence of a fluid. We split the field equations into constraints and evolution equations, using a (3+1)-splitting formulation and find the exact number of the degrees of freedom for the vacuum theory as well as the theory including a fluid.

In Chapter \ref{Chapter3}, we study the Cauchy-Kovalevskaya formulation of higher-order gravity theory in vacuum as well as in the presence of a fluid. We subject the higher-order gravity equations to the initial value problem of Cauchy as in the case of general relativity. We further give the proof of the Cauchy-Kovalevskaya property for both systems of evolution equations, and we state the conditions under which the initial value problem has an analytic solution. We also discuss the equivalence between general relativity plus a scalar field and higher-order gravity theory in vacuum from this point of view.

In Chapter \ref{Chapter4}, we introduce the perturbative formulation of higher-order gravity assuming a formal series representation of the spatial metric. We state the \textit{Friedmann equations} normal case letters of higher-order gravity theory and show that the exact solutions for the scale factor $a=a(t)$ of these equations in the cases of vacuum and radiation remain the same with the corresponding solutions in general relativity. We then develop the Landau-Lifschitz pertubative method in the vicinity of a point that is regular in the time and in the vicinity of a point that is singular in the time as well in this context.

In Chapter \ref{Chapter5}, we study the form of the solution of the gravitational field equations of higher-order gravity in vacuum as well as in the case of radiation, in the vicinity of a point that is regular in the time. We use the form (\ref{start relation for regular series}) for the $3$-metric and calculate the various quantities appearing in the higher-order field equations in terms of the initial data $(\g_{\a\b})^{(n)}$, $n=0,1,2,3,4$. Then we study the form of the solutions in vacuum and radiation. We also prove that the degrees of freedom in both cases remain the same provided that the order of the regular {3}-metric becomes greater than $4$. We further elaborate on the problem of choosing the initial data amongst the initial functions in both cases.

In Chapter \ref{Chapter6}, we study the form of the solution of the gravitational field equations of higher-order gravity in vacuum as well as in the case of radiation in the vicinity of a point that is singular in the time. We present the higher-order field equations by using the $3$-metric (\ref{start relation for singular series}), and we calculate the various quantities of these equations in terms of the initial data. We prove that the form (\ref{start relation for singular series}) cannot be admitted as a solution of the higher-order gravity equations in vacuum. We also find the form of the solution in the case of radiation and show that the degrees of freedom in case of radiation remain the same if the order of the singular {3}-metric becomes greater than $4$.

This Thesis has also two Appendices where we give information concerning the concepts that we use in this work. In Appendix \ref{AppendixA}, we give details regarding the solution of the Einstein equations in cases of vacuum and radiation by using the regular form (\ref{start relation for regular series}). We prove that this form corresponds to the general solution of the problem in case of vacuum, and in a particular solution in case of radiation. In Appendix \ref{AppendixB}, we also give details regarding the solution of the Einstein equations in cases of vacuum and radiation by using the singular form (\ref{start relation for singular series}). In particular, we show that the singular form of the $3$-metric does not give any kind of solution in case of vacuum but is admitted as a particular solution in case of radiation.



\chapter{(3+1)-formulation of higher-order gravity} 

\label{Chapter2} 

\lhead{Chapter 2. \emph{(3+1)-formulation of higher-order gravity}} 

In this Chapter, we introduce the ADM (Arnowitt-Deser-Misner) formalism as a necessary first step to the evolution problem of cosmological asymptotics in higher-order gravity theories in vacuum as well as in the presence of a fluid. In Section \ref{The Synchronous reference system}, we exploit normal coordinates in which the spatial coordinates move with the fluid, while the time coordinate is the proper time along the fluid world lines. In Section \ref{Vacuum field equations}, we deal with the function-counting problem of the quadratic theory in vacuum. After splitting the field equations into constraints and evolution equations, using the (3+1)-splitting formulation, we find the exact number of the degrees of freedom according to the vacuum theory. Finally, in Section \ref{Higher-order field equations in the presence of a fluid}, we introduce a fluid in the quadratic theory, and repeat the function-counting procedure for this important cosmic species.    


\section{The synchronous reference system}
\label{The Synchronous reference system}

In the first part of this section, we introduce an orthonormal frame that will allow us to view four-dimensional spacetime as a geometric object generated by the time evolution of spatial {3}-geometry (time-slice). This frame has important consequences on the mathematical description of various geometric and physical quantities of interest.


\subsection{Minkowski normal coordinates}
\label{Minkowski normal coordinates}

Let $\mathcal{V}$ be a connected, paracompact, Hausdorff and smooth manifold and $g_{ij}$ a Lorentzian metric of signature $(+,-,-,-)$ on $\mathcal{V}$. For a point $p=(x^0,x^1,x^2,x^3)$ in $\mathcal{V}$ we denote by $x^1, x^2, x^3$ the spatial coordinates, and by $x^0=t$ the time coordinate physically defined by an arbitrarily moving clock. The null cone in the tangent space 
$T_p \mathcal{V}$ of $\mathcal{V}$ at $p$ is then given by,  
\be
(x^0)^2-(x^1)^2-(x^2)^2-(x^3)^2=0.
\label{eq:Minkowski_equation}
\ee
We call the coordinates $x^0,x^1,x^2,x^3$ \emph{Minkowski normal coordinates} if in addition to Eq. (\ref{eq:Minkowski_equation}) the vector field $\partial_t$ is timelike and future-pointing.


\subsection{Proper time interval}
\label{Proper time interval}

The spacetime interval $ds$ between any two events with Minkowski normal coordinates $(t,x^\a)$ and 
$(t+dt,x^\a + dx^\a)$, $\a=1,2,3,$ satisfies the general expression,  
\be
ds^2=g_{ij}dx^idx^j 
=g_{00}dt^2 + 2g_{0\a}dtdx^\a + g_{\a\b}dx^\a dx^\b,
\label{eq:general_ds^2}
\ee
and setting $dx^\a=0$, we find that,
\be
ds=\sqrt{g_{00}}dt,
\label{eq:interval_proper_time}
\ee
so that,
\be
s=\int \sqrt{g_{00}}dt,
\label{eq:proper_time}
\ee
giving the proper time interval between any two finitely seperated events. We notice that the form (\ref{eq:general_ds^2}) yields: 
\be
g_{00}>0.
\label{pos g00}
\ee
We now proceed to find some interesting properties of the components of the {4}-metric $g_{ij}$ that we shall employ in Chapters \ref{Chapter5} and \ref{Chapter6}.


\subsection{Properties of the {4}-metric $g_{ij}$}
\label{Properties of the {4}-metric}

We start by introducing the notation,
\be
\g_{\a\b}=\dfrac{g_{0\a}g_{0\b}}{g_{00}}-g_{\a\b},
\label{eq:initial_g_{ab}}
\ee
which will be useful below to transform Eq. (\ref{eq:general_ds^2}) into a form that does not have cross product terms of the form $dx^\a dt$. The absence of these terms will allow us to work with a reference system in which synchronization of clocks located at different points in space is achieved. Since $g_{00}>0$, the line element (\ref{eq:general_ds^2}) can be written in the form,
\be
ds^2=\left( \sqrt{g_{00}}dt + \dfrac{g_{0\a}}{\sqrt{g_{00}}}dx^\a \right)^2 - \g_{\a\b}dx^\a dx^\b.
\label{eq:ds^2_g_00_g_0a_g_ab}
\ee
We now claim that the matrix $-\g_{\a\b}$ is the inverse of $g^{\a\b}$. Indeed, the identity,
\be
g^{ik}g_{kj}=\d^i_j,
\label{eq:identity_d^i_j}
\ee 
implies the following relations between the various components of $g_{ij}$. For the $0\a$-components of $g_{ij}$ it holds that,
\be
g^{\a\b}g_{\b 0} + g^{\a 0}g_{00}=0,
\label{eq:identity_i=a_j=0}
\ee
for the $\a\g$-components we have,
\be
g^{\a\b}g_{\b\g} + g^{\a 0}g_{0\g}=\d^\a_\g,
\label{eq:identity_i=a_j=g}
\ee
and for the $00$-component we find that,
\be
g^{0\b}g_{\b 0} + g^{00}g_{00}=1.
\label{eq:identity_i=0_j=0}
\ee
Solving for $g^{\a 0}$ from (\ref{eq:identity_i=a_j=0}) and substituting in (\ref{eq:identity_i=a_j=g}) using (\ref{eq:initial_g_{ab}}) we find,
\be
g^{\a\b}(-\g_{\b\g})=\d^\a_\g,
\label{eq:reciprocal_tensor_g_ab}
\ee
namely, that
\be
-\g^{\a\b}=g^{\a\b},
\label{eq:reciprocal_tensor_g^ab}
\ee
and the claim follows. 

We further claim that the quadratic form $\g_{\a\b}dx^\a dx^\b$ is positive definite, equal to a sum of squares of three linearly-independent {1}-forms (\cite{ycbcdwmII}, chap. V),
\be
\g_{\a\b}dx^\a dx^\b=\sum^3_{\k=1} (a^\k_\b dx^\b)^2,
\label{eq:g_ab_one_forms}
\ee
so that the matrix $\g_{\a\b}$ is positive definite. To see this, we consider two points A and B in the same time-slice with coordinates $x^\a$ and $x^\a+dx^\a$ respectively. Then, the required time for a light signal leaving point B, gets to point A and returns back to point B is apparently equal to twice the distance between the two given points. Since $ds$ is equal to zero, solving Eq. (\ref{eq:general_ds^2}) with respect to $dt$ and using the form (\ref{eq:initial_g_{ab}}), we find two distinct roots $dt_{(1)}$ and $dt_{(2)}$ where,
\be
dt_{(1)}=\dfrac{1}{g_{00}}\left(-g_{0\a}dx^\a - \sqrt{(g_{0\a}g_{0\b}-g_{\a\b}g_{00})dx^\a dx^\b}\right)
        =-\dfrac{g_{0\a}}{g_{00}}-\sqrt{\dfrac{1}{g_{00}}\g_{\a\b}dx^\a dx^\b}, 
\label{eq:root_dt1}
\ee
and
\be
dt_{(2)}=\dfrac{1}{g_{00}}\left(-g_{0\a}dx^\a + \sqrt{(g_{0\a}g_{0\b}-g_{\a\b}g_{00})dx^\a dx^\b}\right)
        =-\dfrac{g_{0\a}}{g_{00}}+\sqrt{\dfrac{1}{g_{00}}\g_{\a\b}dx^\a dx^\b}.
\label{eq:root_dt2}
\ee 
Since $g_{00}>0$ and $\dfrac{1}{g_{00}}\g_{\a\b}dx^\a dx^\b>0$, we finally conclude that,
\be
\g_{\a\b}dx^\a dx^\b>0,
\label{eq:quadratic_form_positive_definite}
\ee
from which the claim follows. Then, since all the three upper left determinants of $\g_{\a\b}$ must satisfy the conditions,  (\cite{str}, chap. 6),
\be
\g_{11}>0,\quad 
\det{\left(\begin{array}{cc}
\g_{11} & \g_{12} \\ 
\g_{21} & \g_{22}
\end{array}\right)}>0,\quad 
\det{(\g_{\a\b})}=\g>0,
\label{eq:determinant_g_ab}
\ee
taking into account that $g_{00}>0$, we obtain the following conditions for the various subdeterminants of $g_{ij}$ and the full determinant $\det{(g_{ij})}=g$ as well,
\be
\det{\left(\begin{array}{cc}
g_{00} & g_{01} \\ 
g_{10} & g_{11}
\end{array}\right)}<0,\quad 
\det{\left(\begin{array}{ccc}
g_{00} & g_{01}& g_{02} \\ 
g_{10} & g_{11}& g_{12}\\
g_{20} & g_{21}& g_{22}
\end{array}\right)}>0,\quad 
g<0,
\label{eq:determinant_g_ij}
\ee
where,
\be
g=-g_{00}\g.
\label{detg_ij_detg_ab}
\ee  
We can now use the various components of the {4}-metric $g_{ij}$ to give the conditions of clock synchronization of points located at different regions in same time-slice. These conditions of the {4}-metric give us the opportunity to choose a reference system that allows a complete synchronization of clocks.   


\subsection{Clock synchronization}
\label{Clock synchronization}

Using relations (\ref{eq:root_dt1}), (\ref{eq:root_dt2}), we note that the difference in the values of the time $t$ for two events infinitesimally near each other is given by,
\be
\Delta t=\dfrac{1}{2}(dt_{(1)}+dt_{(2)})=-\dfrac{g_{0\a}}{g_{00}}dx^\a.
\label{eq:Dt}
\ee
Then, it follows that the two clocks will be synchronized provided $g_{0\a}$ all vanish, and, we should regard these two events as simultaneous. Consequently, the exceptional cases of reference systems in which,
\be
g_{0\a} = 0,
\label{g0a=0}
\ee  
make it possible for a complete synchronization of the clocks.


\subsection{Cauchy-adapted frame}
\label{Cauchy-adapted frame}

We now introduce a splitted form of the {4}-metric $g_{ij}$ which will be used in what follows concerning the appropriate reference system of our analysis. For the spacetime $(\mathcal{V},g_{ij})$ where $\mathcal{V}=\mathbb{R}\times\mathcal{M}$, with $\mathcal{M}$ being an orientable 3-manifold, $g_{ij}$ is a Lorentzian metric, analytic and
with signature $(+,-,-,-)$, we consider a diffeomorphism $\phi:\mathcal{V}\longrightarrow \mathcal{M}\times I, I\subseteq \mathbb{R}$, such that, the submanifolds $\phi^{-1}(\{t\}\times \mathcal{M})=\mathcal{M}_{t}, t\in I$ are spacelike and the curves $\phi^{-1}(\{x\}\times I), x\in \mathcal{M}$ are timelike. Such a frame is called a \emph{Cauchy adapted frame} \cite{ycb,yc02,cots004}. A vector field $\partial/\partial t$ on $\mathcal{V}$ is defined by these curves and it can be decomposed into normal and parallel components relative to the slicing as follows:
\be
\partial_t=Nn+\overline{N}.
\label{eq:lapse_shift}
\ee
Here $N$ is a positive function on $\mathcal{V}$, $n$ is a future-directed, unit, normal field, and $\overline{N}$ is a vector field tangent to the slices $\mathcal{M}_{t}$. We call $N$ the \textit{lapse} function and $\overline{N}$ the \textit{shift} vector field. Then we set, 
\be
\overline{N}=N^\a \partial_\a,
\label{eq:shift vector field}
\ee
and due to the fact that $\partial_t$ is a timelike vector field on $\mathcal{V}$ we obtain,
\be
g_{00}=g(\partial_t,\partial_t)=N^2 + N^\a N_\a>0,
\label{eq:positive_g_00}
\ee
and
\be
g_{0 \a}=g(\partial_t, \partial_\a)=N_\a, \quad g^{0 \a}=N^\a/N^2.
\label{eq:shift}
\ee
Using (\ref{eq:positive_g_00}) and (\ref{eq:shift}), we find that Eq. (\ref{eq:ds^2_g_00_g_0a_g_ab}) becomes,
\be
ds^2=(N^{2} + N^\a N_\a)dt^{2} + 2N_\a dx^\a dt -(\g_{\a\b} - \frac{N_\a N_\b}{N^{2} + N^\a N_\a})dx^\a dx^\b,
\label{eq:ds^2_lapse_shift}
\ee
or,
\be
ds^2=N^{2}dt^{2} - (\g_{\a\b}-\frac{N_\a N_\b}{N^2 + N^\g N_\g})(dx^\a + N^\a dt)(dx^\b + N^\b dt).
\label{eq:ghj}
\ee


\subsection{Synchronous coordinates}
\label{Synchronous coordinates}

In this Thesis, we shall deal exclusively with the simplest choice, $N=1, N^\a=0$, which means that $g_{00}=1,g_{0\a}=0$ (the foliation away from $\mathcal{M}_{0}$ defined by the choice $N=1$ is called a \emph{geodesic slicing} \cite{gou}). In this case, one sometimes uses different jargon and speaks of a \emph{synchronous} \cite{ll}, or a \emph{geodesic normal} \cite{ycb, ss}, or a \emph{Gaussian normal} \cite{ns,mtw} system of local coordinates on $\mathcal{V}$ for this choice. In a synchronous reference system the timelines are geodesics of $\mathcal{V}$, while the hypersurfaces $\mathcal{M}_{t}$ with constant $t$ are orthogonal to these geodesics. Thus, from now on, for the {4}-metric $g_{ij}$, we are going to use the form, 
\be
ds^2=dt^2-\g_{\a\b}dx^\a dx^\b,
\label{eq:Synchronous_ds^2}
\ee 
where the time $t$ measures proper time. Then, for the Christoffel symbols,
\be
\G^i_{jk} = \dfrac{1}{2}g^{il}(\pa_k g_{lj} + \pa_j g_{lk} - \pa_l g_{jk}),
\label{Cs}
\ee
we obtain,
\be
\G^0_{00} = \G^\a_{00} = \G^0_{0\a} = 0,
\label{Cs2,3zeros}
\ee 
\be
\G^0_{\a\b} = \dfrac{1}{2}\pa_t \g_{\a\b},
\label{Cs1upzero}
\ee
and,
\be
\G^\a_{0\b} = \dfrac{1}{2}\g^{\a\g}\pa_t \g_{\g\b}.
\label{Cs1downzero}
\ee

Below and for the rest of this Chapter we first deal with the vacuum field equations of the higher order gravity theory and subsequently with the corresponding equations in the case of a fluid. In both cases, we shall study the function-counting problem of the theory.


\section{Vacuum field equations}
\label{Vacuum field equations}

We start with the vacuum field equations of the higher-order gravity theory derived from an analytic lagrangian $f(R)$, that is the equations \cite{tf,Sch07,cv},
\be
L_{ij}=f'(R)R_{ij} -\frac{1}{2}f(R)g_{ij} -\nabla_i\nabla_j f'(R) +g_{ij}\Box_g f'(R)=0.
\label{eq:FEs}
\ee
The Ricci tensor $R_{ij}$ and the scalar curvature $R$ are defined by,
\bq
R_{ij} &=& R^k_{ikj}, \\
\label{Rt}
R &=& g^{ij}R_{ij},
\label{scR}
\eq
where,
\be
R^i_{klm} = \pa_l \G^i_{km} - \pa_m \G^i_{kl} + \G^i_{nl}\G^n_{km} - \G^i_{nm}\G^n_{kl},
\label{Riemann}
\ee
and $\Box_g$ is the \textit{d' Alembertian} operator acting on a scalar function, namely,
\be
\Box_g f'(R) = g^{mn}\nabla_m\nabla_n f'(R).
\label{id Box}
\ee


\subsection{Synchronous equations}
\label{Synchronous equations}

We shall be concerned below with the quadratic theory 
\be
f(R)=R+\ep R^2, \quad \ep=\textrm{constant}
\label{quadratic eq}
\ee
in vacuum. The field equations (\ref{eq:FEs}) for the theory (\ref{quadratic eq}) in a Cauchy adapted frame split as follows:
\be
L_{00} :=(1+ 2\ep R)R_{00} -\frac{1}{2}(R+\ep R^2) +2\ep g^{\a\b} \nabla_\a \nabla_\b R =0,
\label{eq:Loo}
\ee
\be
L_{0\a} :=(1 +2\ep R)R_{0\a} -2\ep \nabla_0 \nabla_\a R =0,
\label{eq:Loa}
\ee
\be
L_{\a\b} := (1 +2\ep R)R_{\a\b} -\frac{1}{2}(R+\ep R^2) g_{\a\b} -2\ep \nabla_\a \nabla_\b R +2\ep g_{\a\b} \Box_g R =0.
\label{eq:Lab}
\ee
The terms, 
\[\nabla_\a \nabla_\b R, \quad \nabla_0 \nabla_\a R, \quad \Box_g R,  \]
appearing in Eq. (\ref{eq:Loo}), (\ref{eq:Loa}) and (\ref{eq:Lab}) satisfy, 
\be
\nabla_\a \nabla_\b R = \pa_\a (\pa_\b R) - \G^0_{\a\b}\pa_t R - \G^\g_{\a\b}\pa_\g R,
\label{nabla_a nabla_b R}
\ee
\be
\nabla_0 \nabla_\a R = \pa_\a (\pa_t R) - \G^\b_{\a 0}\pa_\b R,
\label{nabla_0 nabla_a R}
\ee
and,
\be
\Box_g R = {\pa_t}^2 R - \g^{\a\b}\nabla_\a \nabla_\b R.
\label{BoxR}
\ee
Thus (\ref{eq:Loo}) and (\ref{eq:Loa}) contain the first derivative of the scalar curvature $R$ with respect to the proper time $t$, while Eq. (\ref{eq:Lab}) the second derivative. This means that the order of these differential equations is not the same. To make it more clear, we define the extrinsic curvature $K_{\a\b}$ by the \emph{first variational equation}, 
\be
\partial_t \g_{\a\b} = K_{\a\b},
\label{first variational equation}
\ee
and we may then express the various curvatures in terms of $K_{\a\b}$.

The components of the Ricci tensor are then given by,
\be
R_{00} = -\frac{1}{2}\partial_t K  -\frac{1} {4}K_\a^\b K_\b^\a,
\label{eq:R00mixeddown}
\ee
\be
R_{0\a}  = \frac{1}{2} (\nabla_\b K^\b_\a - \nabla_\a K),
\label{eq:R0amixeddown}
\ee
and,
\be
R_{\a\b}= P_{\a\b} +\frac{1} {2} \partial_t K_{\a\b} +\frac{1} {4}K K_{\a\b} -\frac{1}{2}K^\g_\a K_{\b\g},
\label{eq:Rabmixeddown}
\ee
where 
\be
K = \g^{\a\b}K_{\a\b} = \textrm{tr}K_{\a\b},
\label{trK}
\ee
and $P_{\a\b}$ denotes the three-dimensional Ricci tensor associated with the {3}-metric $\g_{\a\b}$ on $\mathcal{M}_{t}$.
For the scalar curvature we find, by tracing equation (\ref{eq:R00mixeddown}) and (\ref{eq:Rabmixeddown}),
\be
R=-P-\partial_t K-\frac{1}{4}K^2-\frac{1}{4}K^\b_\a K^\a_\b,
\label{eq:Rscalar}
\ee
where $P$ is the trace of the tensor $P_{\a\b}$ with respect to the {3}-metric $\g_{\a\b}$. Because of Eq. (\ref{eq:Rscalar}), relations (\ref{nabla_a nabla_b R}), (\ref{nabla_0 nabla_a R}) contain third derivatives of the {3}-metric $\g_{\a\b}$ and (\ref{BoxR}) fourth derivatives of $\g_{\a\b}$. Therefore (\ref{eq:Loo}), (\ref{eq:Loa}) are third-order differential equations with respect to the proper time $t$ and, similarly, (\ref{eq:Lab}) is a fourth-order differential equation with respect to $t$. This result is going to be used in our analysis in chapters \ref{Chapter5} and \ref{Chapter6}. 

We are ready now to proceed with the (3+1)-splitting formulation of the quadratic theory in vacuum into evolution equations and constaints, in order to search for the true degrees of freedom of the $R+\ep R^2$ theory in vacuum.


\subsection{ADM equations}
\label{ADM equations}

Apart from the \emph{first variational equation},
\be
\partial_t \g_{\a\b} = K_{\a\b},
\label{eq:pg}
\ee
that has already been introduced above, we define the \emph{acceleration tensor} $D_{\a\b}$ through the \emph{second variational equation},
\be
\partial_t K_{\a\b} = D_{\a\b},
\label{eq:pK}
\ee
and further introduce the \emph{jerk tensor} (3rd-order derivatives) $W_{\a\b}$, through the \emph{jerk} equation,
\be
\partial_t D_{\a\b} = W_{\a\b}.
\label{eq:pD}
\ee
Compared to the case of general relativity, the need to define the two additional tensors $D_{\a\b}$ and $W_{\a\b}$ follows directly from the \textit{order} of the partial differential equations (\ref{eq:Loo}), (\ref{eq:Loa}) and (\ref{eq:Lab}).
The space tensors $D_{\a\b}$ and $W_{\a\b}$ are obviously symmetric, and owing to the fourth-order nature of the higher-order field equations (\ref{eq:Loo}), (\ref{eq:Loa}), (\ref{eq:Lab}), they play an important role in what follows. 

We imagine that as in the case of general relativity \cite{ycb}, evolution starts from some spesific initial data set satisfying some constraint equations, and proceeds to evolve these data using evolution in time. In higher-order gravity, we call any initial data set a $5$-plet $(\mathcal{M}_{t},\g_{\a\b},K_{\a\b},D_{\a\b},W_{\a\b})$, where the submanifolds $\mathcal{M}_{t}$ are spacelike and $(\g_{\a\b},K_{\a\b},D_{\a\b},W_{\a\b})$ are three-dimensional tensors.

Apart from the equations (\ref{eq:pg}) (velocity equation), (\ref{eq:pK}) (acceleration equation), (\ref{eq:pD}) (jerk equation), any initial data set $(\mathcal{M}_{t},\g_{\a\b},K_{\a\b},D_{\a\b},W_{\a\b})$ must further satisfy Eq. (\ref{eq:Lab}). This results in the following evolution equation:
\begin{eqnarray}
&&\partial_t W =\frac{1}{6\ep}( \frac{1}{2} P +\frac{1}{8} K^2  -\frac{5}{8} K^{\a\b} K_{\a\b} +D ) + \nonumber \\
&&\frac{1}{6}[P^2 +\frac{1}{4}P K^2 -\frac{1}{4}P K^{\a\b} K_{\a\b} +\frac{1}{32} K^4 -\frac{1}{16}K^2 K^{\a\b} K_{\a\b}  -\nonumber \\
&&6K K^\a_\b K^\b_\g K^\g_\a -\frac{99}{32} (K^{\a\b} K_{\a\b})^2 +27 K^\a_\b K^\b_\g K^\g_\d K^\d_\a +9 K K^{\a\b} D_{\a\b}  -  \nonumber \\
&&57 K^\a_\b K^\b_\g D^\g_\a  +\frac{13}{2}D K^{\a\b} K_{\a\b} -\frac{7}{2}D^2 +15 D^{\a\b} D_{\a\b}  -3K W   +\nonumber \\
&&15K^{\a\b} W_{\a\b}  -6\partial_t (\partial_t P) -\nonumber \\
&&4\g^{\a\b} \nabla_\a \nabla_\b (-P -D +\frac{3}{4} K^{\a\b} K_{\a\b} -\frac{1}{4}K^2)],
\label{eq:pW}
\end{eqnarray}
called here the \emph{snap} equation.

Using the \emph{acceleration tensor}  $D_{\a\b}$, the Ricci tensor splittings given above in Eqns. (\ref{eq:R00mixeddown}), (\ref{eq:R0amixeddown}) and (\ref{eq:Rabmixeddown}) become,
\be
R_{00} = -\frac{1}{2}D +\frac{1} {4}K_\a^\b K_\b^\a,
\label{eq:R00mixeddownnew}
\ee
\be
R_{0\a}  = \frac{1}{2} (\nabla_\b K^\b_\a - \nabla_\a K),
\label{eq:R0amixeddownnew}
\ee
\be
R_{\a\b}= P_{\a\b} +\frac{1} {2} D_{\a\b} +\frac{1} {4}K K_{\a\b} -\frac{1}{2}K^\g_\a K_{\b\g},
\label{eq:Rabmixeddownnew}
\ee
while for the scalar curvature we obtain,
\be
R = -P -D -\frac{1}{4}K^2 +\frac{3}{4}K^\a_\b K^\b_\a,
\label{eq:scalarR}
\ee
where 
\be
P = \g^{\a\b}P_{\a\b} = \textrm{tr}P_{\a\b}, 
\label{traceP}
\ee
and,
\be
D = \g^{\a\b}D_{\a\b} = \textrm{tr}D_{\a\b}.
\label{traceD}
\ee


\subsection{Constraints}
\label{Constraints}

We continue by substituting these forms (\ref{eq:R00mixeddownnew}), (\ref{eq:R0amixeddown}), (\ref{eq:Rabmixeddownnew}) and (\ref{eq:scalarR}) into the identities (\ref{eq:Loo}), (\ref{eq:Loa}) to find the following equations, which, in a manner analogous with the situation in general relativity, we call \emph{constraints}:

\emph{Hamiltonian Constraint}
\begin{eqnarray}
\mathcal{C}_0: &=& \frac{1}{2}P  +\frac{1}{8}K^2 -\frac{1}{8}K^{\a\b} K_{\a\b} + \nonumber \\
&&\ep[-\frac{1}{2}P^2  -\frac{1}{4}P K^2 +\frac{1}{4}P K^{\a\b} K_{\a\b}  -\frac{1}{32}K^4 +\frac{1}{16}K^2 K^{\a\b} K_{\a\b} + \nonumber\\
&&\frac{3}{32}(K^{\a\b} K_{\a\b})^2 -\frac{1}{2}D K^{\a\b} K_{\a\b} +\frac{1}{2}D^2     - \nonumber\\
&&2\g^{\a\b} \nabla_\a \nabla_\b(-P - \frac{1}{4}K^2 +\frac{3}{4}K^{\g\d} K_{\g\d}  -D)] =0,
\label{eq:hamiltonian}
\end{eqnarray}

\emph{Momentum Constraints}
\begin{eqnarray}
\mathcal{C}_\a : &=& \frac{1}{2} (\nabla_\b K^\b_\a - \nabla_\a K) + \nonumber \\
&&\ep[(-P  -\frac{1}{4}K^2 +\frac{3}{4}K^\d_\g K^\g_\d -D) (\nabla_\b K^\b_\a - \nabla_\a K)  - \nonumber \\
&&\nabla_\a (-2\partial_t P + K K^{\g\d} K_{\g\d} -3 K^\b_\g K^\g_\d K^\d_\b - \nonumber \\
&&K D +5K^{\g\d} D_{\g\d} -2W)]=0.
\label{eq:momentum}
\end{eqnarray}
We conclude that the initial data $(\g_{\a\b},K_{\a\b},D_{\a\b},W_{\a\b})$ cannot be chosen arbitrarily but they must satisfy  the constraint equations (\ref{eq:hamiltonian}) and (\ref{eq:momentum}) on each time-slice $\mathcal{M}_{t}$.


\subsection{Function-counting}
\label{Function-counting}

Subtracting four diffeomorphism transformations, three of them due to the possibility of arbitrary transformations of the three space coordinates and one more due to the arbitrariness in choosing the initial hypersurface for setting up the synchronous reference system, we find that the number of the arbitrary functions which have to be specified initially is equal to $24-4-4=16$. 

This also agrees with the number that Barrow finds in \cite{ba}. If we denote by $F$ the number of the non-interacting fluids and by $S$ the number of the scalar fields of the theory, then for the description of higher-order gravity cosmologies, we need $16 + 4F + 2S$ independent arbitrary functions of space. To make clearer, in a $4$-dimensional spacetime the initial data of each fluid require four functions to be specified and, thus, the $F$ non-interacting fluids require $4F$ functions to describe them in general. In addition to that, to specify each one of the $S$ non-interacting scalar fields, $\phi_i$, $i=1,2,...,S$, being presented with self interaction potentials $V(\phi_i)$, we need two more functions $\phi_j$ and $\pa_t \phi_j$, namely that $2S$ such functions in general. Taking into account that in addition to $\g_{\a\b}$, $\pa_t \g_{\a\b}$ we must further specify ${\pa_t}^2 \g_{\a\b}$ and ${\pa_t}^3 \g_{\a\b}$ then, the final number of the independent arbitrary functions of higher-order gravity is equal to $16 + 4F + 2S$. Therefore, in the vacuum case, where $F=0,S=0$, we end up with $16$ degrees of freedom.

We can consider the four evolution equations (\ref{eq:pg}), (\ref{eq:pK}), (\ref{eq:pD}) and (\ref{eq:pW}) as the higher-order gravity analogues of the ADM equations of general relativity in vacuum. These equations together with the constraints  (\ref{eq:hamiltonian}) and (\ref{eq:momentum}) describe the \emph{time development} $(\mathcal{V},g_{ij})$ of any initial data set $(\mathcal{M}_{t},\g_{\a\b},K_{\a\b},D_{\a\b},W_{\a\b})$ in higher-order gravity theories in vacuum, with the $4$-metric $g_{ij}$ satisfying Eq. (\ref{eq:Synchronous_ds^2}).\\
We can now proceed to study the ADM formulation problem in higher-order gravity in the presence of a fluid.


\section{Higher-order field equations in the presence of a fluid}
\label{Higher-order field equations in the presence of a fluid}

Our starting point is the field equations of the higher-order gravity theory derived from an analytic lagrangian $f(R)$, that is the equations
\be
L_{ij}=f'(R)R_{ij} -\frac{1}{2}f(R)g_{ij} -\nabla_i\nabla_j f'(R) +g_{ij}\Box_g f'(R)=8\pi G T_{ij},
\label{eq:FEs_rad}
\ee
where $G$ is Newton's gravitational constant and $T_{ij}$ is the energy-momentum tensor satisfying the conservations laws of energy and momentum, given by the equations of motion,
\be
\nabla_i T^i_j=0.
\label{eq:conservation laws}
\ee
Equations (\ref{eq:conservation laws}) follow directly from (\ref{eq:FEs_rad}) because of the fact that (\cite{edd}, chap. IV),
\be
\nabla_i L^i_j = 0.
\label{clL}
\ee 


\subsection{Field equations of the quadratic theory $R+\ep R^2$ in the presence of a fluid}
\label{Field equations of the quadratic theory in the presence of a fluid}

Similarly to the case of vacuum, we shall be concerned below with the quadratic theory $f(R)=R+\ep R^2$. The field equations (\ref{eq:FEs_rad}) in a \emph{Cauchy adapted frame} split as follows:
\be
L_{00} = 8\pi G T_{00},
\label{eq:Loo_rad}
\ee
\be
L_{0\a} = 8\pi G T_{0 \a},
\label{eq:Loa_rad}
\ee
\be
L_{\a\b} = 8\pi G T_{\a\b},
\label{eq:Lab_rad}
\ee
where $L_{00},L_{0\a},L_{\a\b}$ are given by Eqns. (\ref{eq:Loo}), (\ref{eq:Loa}) and (\ref{eq:Lab}) respectively and the components of the Ricci tensors $R_{00}, R_{0\a}, R_{\a\b}$ and the scalar curvature $R$ satisfy the relations (\ref{eq:R00mixeddownnew}), (\ref{eq:R0amixeddown}), (\ref{eq:Rabmixeddownnew}) and (\ref{eq:scalarR}) respectively.
Thus, to apply the (3+1)-splitting formulation of this theory into evolution equations and constaints in order to find the true degrees of freedom of the $(R+\ep R^2)$-fluid theory, we first need splitting relations for the fluid energy-momentum tensor.

\subsection{Energy-Momentum tensor}
\label{Energy-Momentum tensor}

We consider the tangent vector,
\be
u^i=\left(\dfrac{dt}{ds},\dfrac{dx^\a}{ds}\right)=(u^0,u^\a),
\label{eq:velocity}
\ee
to a timelike curve parametrized by the proper time $s$, the \emph{$4$-velocity} of the curve with unit length satisfying the identity,
\be
1=u_i u^i= u_0 u^0 + u_\a u^\a.
\label{eq:velocities identity}
\ee
The $4$-velocity $u^i$ is used for a covariant description of continuous matter distributions given by the energy-momentum tensor $T_{ij}$.
In particular, a cosmological \emph{perfect fluid} is defined to be a continuous distribution of matter with energy-momentum tensor of the form,
\be
T^i_j=(\rho+p)u^i u_j-p\d^i_j,
\label{eq:general perfect fluid}
\ee
where $\rho$ is the energy density and $p$ is the pressure. 
Using the equation of state,
\be
p=w\rho,
\label{eq:general equation of state}
\ee
where $w\in \mathbb{R}$ is the equation of state parameter, and taking into account that in a synchronous reference system we have $u_0=u^0$, the components of the energy-momentum tensor $T^i_j$ become,
\bq
T^0_0 &=& \rho(1+w){u_0}^2 -w\rho,
\label{T^0_0 general} \\
T^0_\a &=& \rho(1+w) u_\a u_0,
\label{T^0_a general} \\
T^\b_\a &=& \rho(1+w)u^\b u_\a -w\rho \d^\b_\a.
\label{T^b_a general}
\eq
Also we have the trace,
\be
T=\textrm{tr}T_{ij}=\rho (1-3w).
\label{T general}
\ee
The equations of motion (\ref{eq:conservation laws}) yield the following equations,
\be
[(1+w){u_0}^2 -w]\nabla_0 \rho + 2(1+w)\rho u_0 \nabla_0 u_0 + (1+w)g^{\a\b}\nabla_\a (\rho u_\b u_0)=0,
\label{eq of motion 0}
\ee
\be
(1+w)\nabla_0 (\rho u_\b u_0) + (1+w)\nabla_\a (\rho u^\a u_\b) - w\d^\a_\b \nabla_\a \rho=0.
\label{eq of motion a}
\ee

\subsection{Function counting of $R+\ep R^2$ in the presence of a fluid}
\label{Function counting in the presence of a fluid}

According to the field equations (\ref{eq:Loo_rad}) and (\ref{eq:Loa_rad}), we find the following equations, the constraints for the higher-order-gravity-matter theory: 

\emph{Hamiltonian Constraint}
\begin{eqnarray}
\mathcal{C}_0 : &=& \frac{1}{2}P  +\frac{1}{8}K^2 -\frac{1}{8}K^{\a\b} K_{\a\b} + \nonumber \\
&&\ep[-\frac{1}{2}P^2  -\frac{1}{4}P K^2 +\frac{1}{4}P K^{\a\b} K_{\a\b}  -\frac{1}{32}K^4 +\frac{1}{16}K^2 K^{\a\b} K_{\a\b} + \nonumber\\
&&\frac{3}{32}(K^{\a\b} K_{\a\b})^2 -\frac{1}{2}D K^{\a\b} K_{\a\b} +\frac{1}{2}D^2     - \nonumber\\
&&2\g^{\a\b} \nabla_\a \nabla_\b(-P - \frac{1}{4}K^2 +\frac{3}{4}K^{\g\d} K_{\g\d}  -D)] =8\pi G T_{00},
\label{eq:hamiltonian_rad}
\end{eqnarray}

\emph{Momentum Constraints}
\begin{eqnarray}
\mathcal{C}_\a : &=& \frac{1}{2} (\nabla_\b K^\b_\a - \nabla_\a K) + \nonumber \\
&&\ep[(-P  -\frac{1}{4}K^2 +\frac{3}{4}K^\d_\g K^\g_\d -D) (\nabla_\b K^\b_\a - \nabla_\a K)  - \nonumber \\
&&\nabla_\a (-2\partial_t P + K K^{\g\d} K_{\g\d} -3 K^\b_\g K^\g_\d K^\d_\b - \nonumber \\
&&K D +5K^{\g\d} D_{\g\d} -2W)]=8\pi G T_{0\a}.
\label{eq:momentum_rad}
\end{eqnarray}
Apart from the equations (\ref{eq:pg}) (velocity equation), (\ref{eq:pK}) (acceleration equation), (\ref{eq:pD}) (jerk equation), any initial data set $(\mathcal{M}_{t},\g_{\a\b},K_{\a\b},D_{\a\b},W_{\a\b})$ together with the quantities $p,\rho,u^i$ must further satisfy Eq. (\ref{eq:Lab_rad}). This results the \emph{snap} equation with matter,
\begin{eqnarray}
&&\partial_t W =\frac{1}{6\ep}(8\pi G T^\a_\a + \frac{1}{2} P +\frac{1}{8} K^2  -\frac{5}{8} K^{\a\b} K_{\a\b} +D ) + \nonumber \\
&&\frac{1}{6}[P^2 +\frac{1}{4}P K^2 -\frac{1}{4}P K^{\a\b} K_{\a\b} +\frac{1}{32} K^4 -\frac{1}{16}K^2 K^{\a\b} K_{\a\b}  -\nonumber \\
&&6K K^\a_\b K^\b_\g K^\g_\a -\frac{99}{32} (K^{\a\b} K_{\a\b})^2 +27 K^\a_\b K^\b_\g K^\g_\d K^\d_\a +9 K K^{\a\b} D_{\a\b}  -  \nonumber \\
&&57 K^\a_\b K^\b_\g D^\g_\a  +\frac{13}{2}D K^{\a\b} K_{\a\b} -\frac{7}{2}D^2 +15 D^{\a\b} D_{\a\b}  -3K W   +\nonumber \\
&&15K^{\a\b} W_{\a\b}  -6\partial_t (\partial_t P) -\nonumber \\
&&4\g^{\a\b} \nabla_\a \nabla_\b (-P -D +\frac{3}{4} K^{\a\b} K_{\a\b} -\frac{1}{4}K^2)].
\label{eq:pW with T}
\end{eqnarray}
We end up with a dynamical system consisting of $30$ arbitrary functions, the $24$ \emph{initial data} $(\g_{\a\b},K_{\a\b},D_{\a\b},W_{\a\b})$ together with the $6$ functions $p,\rho,u^i$ satisfying evolution equations (\ref{eq:pg}), (\ref{eq:pK}), (\ref{eq:pD}) and (\ref{eq:pW with T}) as well as the four constraint equations (\ref{eq:hamiltonian_rad}) and (\ref{eq:momentum_rad}) on each slice $\mathcal{M}_{t}$. The quantities $p,\rho$ satisfy the equation of state (\ref{eq:general equation of state}), and the velocities the identity (\ref{eq:velocities identity}).

In addition to that, subtracting 4 diffeomorphism transformations, we find that the number of the arbitrary functions which have to be specified initially is equal to $30-4-1-1-4=20$ (from the initial $30$ functions we have to subtract in turn $4$ from the constraint equations, $1$ from the equation of state, $1$ from the identity for the velocities and more $4$ from the diffeomorphism transformations). Of course, this is the number we expected to find and it is also consistent with the number that Barrow finds in \cite{ba} for $F=1,S=0$. 

We can consider the four evolution equations (\ref{eq:pg}), (\ref{eq:pK}), (\ref{eq:pD}) and (\ref{eq:pW with T}) as the higher-order gravity-matter analogues of the ADM equations of general relativity with matter fields. These equations together with the constraints  (\ref{eq:hamiltonian_rad}) and (\ref{eq:momentum_rad}), the equation of state (\ref{eq:general equation of state}) and the identity (\ref{eq:velocities identity}) describe the
\emph{time development} $(\mathcal{V},g_{ij})$ of any initial data set $(\mathcal{M}_{t},\g_{\a\b},K_{\a\b},D_{\a\b},W_{\a\b})$ together with the quantities $p,\rho,u^i$ in higher order gravity theories. 

In order to study the problems of regularity-singularity in higher-order gravity in the cases of vacuum and various matter contributions, we must first verify that the dynamical systems we introduced in both cases have the Cauchy-Kovalevskaya property. This property will ensure that the corresponding evolution problems are \emph{well-formulated} in the sense that starting from any assignment of initial data, the subsequent dynamical evolution is uniquely determined.  

\chapter{Cauchy-Kovalevskaya formulation in higher-order gravity} 

\label{Chapter3} 


\lhead{Chapter 3. \emph{Cauchy-Kovalevskaya formulation in higher-order gravity}} 
 
In this Chapter, we study the Cauchy-Kovalevskaya formulation in higher-order gravity theory in vacuum as well as in the presence of a fluid. In Section \ref{The initial value problem of Cauchy in higher-order gravity}, we gain further insight into the mathematical content of higher-order gravity equations by applying them to the initial value problem of Cauchy as in the case of general relativity \cite{w}. In Section \ref{Cauchy-Kovalevskaya property}, we give the proof of the Cauchy-Kovalevskaya property for both systems of evolution equations, while in Section \ref{Analytic solution of the local Cauchy problem}, we state the conditions under which the initial value problem has a solution which is analytic. We also discuss the equivalence between general relativity plus a scalar field and higher-order gravity theory in vacuum from this viewpoint.


\section{The initial value problem of Cauchy in higher-order gravity}
\label{The initial value problem of Cauchy in higher-order gravity}

In the previous Chapter, we found the number of the arbitrary functions that any general solution of the field equations of higher-order gravity must contain. The \textit{initial value} or the \textit{Cauchy} problem of higher-order gravity is the problem of studying evolution using these equations, under given initial conditions.


\subsection{Initial-value formulation}
\label{Initial-value formulation}

Suppose that we are given $g_{ij}$, $\pa_t g_{ij}$, ${\pa_t}^2 g_{ij}$ and ${\pa_t}^3 g_{ij}$ everywhere on the hypersurface $\mathcal{M}_{t}$. Then in order to compute $g_{ij}$, $\pa_t g_{ij}$, ${\pa_t}^2 g_{ij}$ and ${\pa_t}^3 g_{ij}$ at a time $t + dt$, we will need to be able to extract from the field equations a formula for ${\pa_t}^4 g_{ij}$. By continuing this process, we can compute these tensors for all $x^\a$ and $t$.

From the left-hand side of the field equations,
\be
L^{ij} = \left( R^{ij} - \dfrac{1}{2}g^{ij}R \right) + \ep \left( 2RR^{ij} - \dfrac{1}{2}g^{ij}R^2 - 2\nabla^i \nabla^j R + 2g^{ij}\Box_g R \right),
\label{fecon}
\ee
and taking into consideration the identity, 
\be
\nabla_i L^{ij} = 0,
\label{id Lcon}
\ee
we obtain,
\be
\pa_t L^{0j} = -\pa_\a L^{\a j} - \G^i_{il}L^{lj} - \G^j_{il}L^{il}.
\label{id Lcon t}
\ee 
The right-hand side of Eq. (\ref{id Lcon t}) contains time derivatives up to the fourth order. Thus $L^{0j}$ contains no time derivatives higher than ${\pa_t}^3$. Of course, we have already this result in Section \ref{Synchronous equations}, but here we present this argument with a more transparent way more suitable to the initial value problem.

Apparently, from (\ref{id Lcon t}), the $(00)$-component and $(0\a)$-components do not provide us with any information about the time evolution of the higher-order equations of the gravitational field. Besides, the equations,
\bq
L^{00} &=& 8\pi G T^{00}, 
\label{L00con} \\
L^{0\a} &=& 8\pi G T^{0\a},
\label{Loacon}
\eq
constitute constraints on the initial data at $t$, and so only the remaining six equations, namely,
\be
L^{\a\b} = 8\pi G T^{\a\b},
\label{Labcon}
\ee
include such time derivatives. 

To sum up, Eq. (\ref{Labcon}) determines the six ${\pa_t}^4 g^{\a\b}$, but leaves four left over derivatives, namely, ${\pa_t}^4 g^{0j}$ indeterminate. It would be convenient to apply four coordinate transformations, but unfortunately, even though $g_{ij}$ would remain the same at $t$ it alters everywhere else.

This difficulty could be overcome by imposing four coordinate conditions that fix the coordinate system. In the next two subsections, and solely for the analysis of the Cauchy problem we are going to use \textit{harmonic coordinates}, as a particularly convenient choice of approach.


\subsection{Harmonic coordinates}
\label{Harmonic coordinates}     

We start with the \emph{harmonic coordinate conditions},
\be
g^{jk}\G^i_{jk} = 0,
\label{hccG}
\ee 
namely that,
\be
\dfrac{1}{2}g^{jk}g^{il}(\pa_k g_{lj} + \pa_j g_{lk} - \pa_l g_{jk}) = 0.
\label{hccGanal}
\ee
Taking into account that,
\be
\G^i_{ij} = \dfrac{1}{2}g^{il}\pa_j g_{li},
\label{G^i_ij}
\ee
and the fact that,
\be
g^{il}\pa_j g_{li} = \pa_j lng = \dfrac{2}{\sqrt{g}}\pa_j \sqrt{g},
\label{id sqrtg}
\ee
the \emph{harmonic coordinate conditions} read,
\be
\pa_j (\sqrt{g}g^{ij}) = 0.
\label{hcc}
\ee
This last condition is crusial in the analysis of the Cauchy problem.


\subsection{The Cauchy problem}
\label{The Cauchy problem}
 
The \emph{harmonic coordinate conditions} (\ref{hcc}) tell us that,
\be
\pa_t (\sqrt{g}g^{i0}) = -\pa_\a (\sqrt{g}g^{i\a}).
\label{hccanal}
\ee
Differentiating (\ref{hccanal}) with respect to time, we obtain,
\be
{\pa_t}^2 (\sqrt{g}g^{i0}) = -\pa_\a \left(\pa_t (\sqrt{g}g^{i\a}) \right).
\label{hccanalsec}
\ee
Then for the third derivative we have,
\be
{\pa_t}^3 (\sqrt{g}g^{i0}) = -\pa_\a \left({\pa_t}^2 (\sqrt{g}g^{i\a}) \right),
\label{hccanalsec}
\ee
and finally,
\be
{\pa_t}^4 (\sqrt{g}g^{i0}) = -\pa_\a \left({\pa_t}^3 (\sqrt{g}g^{i\a}) \right).
\label{hccanalsec}
\ee
Consequently, the six equations (\ref{Labcon}) together with the four equations (\ref{hccanalsec}) suffice to determine the fourth time derivatives of all $g_{ij}$ and thus, the initial value problem of Cauchy can be solved.

We are now ready to study the Cauchy problem when the initial data are analytic functions. The first thing we need is to prove that the systems of the evolution equations of higher-order gravity (both in the case of vacuum and in the presence of a fluid) which have already been presented in Chapter \ref{Chapter2}, constitute  Cauchy-Kovalevskaya type systems.


\section{Cauchy-Kovalevskaya property}
\label{Cauchy-Kovalevskaya property}

Our starting point are the `dangerous' terms of the systems, namely, terms that may include time derivatives. We prove below that these terms are eventually `purely spatial', in all cases. At first, we show this in the case of vacuum.  


\subsection{Proof of the Cauchy-Kovalevskaya property in vacuum}
\label{Proof of the Cauchy-Kovalevskaya property in vacuum}

We present details of the proof that the system consisting of the four evolution equations (\ref{eq:pg}), (\ref{eq:pK}), (\ref{eq:pD}) and (\ref{eq:pW}) constitutes a Cauchy-Kovalevskaya type system. Indeed, the only `dangerous' terms in these equations are the three terms,
\be
\partial_t P,\quad\partial_t(\partial_t P), \quad\nabla_\alpha \nabla_\beta (-P -D +\frac{3}{4}K^{\gamma\delta} K_{\gamma\delta}-\frac{1}{4}K^2),
\label{eq:dangerous_terms}
\ee
present in the constraints (\ref{eq:hamiltonian}), (\ref{eq:momentum}) and in the snap equation (\ref{eq:pW}). The spatial connection coefficients are given by the obvious formula,
\be
\Gamma^\mu_{\alpha\beta}=\frac{1}{2}\gamma^{\mu\epsilon}(\partial_\beta \gamma_{\alpha\epsilon} +\partial_\alpha \gamma_{\beta\epsilon} -\partial_\epsilon \gamma_{\alpha\beta}).
\label{gammas}
\ee
We then have,
\be\partial_t \Gamma^\mu_{\alpha\beta}= -\frac{1}{2}K^{\mu\epsilon}(\partial_\beta \gamma_{\alpha\epsilon} +\partial_\alpha \gamma_{\beta\epsilon} -\partial_\epsilon \gamma_{\alpha\beta}) +\frac{1}{2}\gamma^{\mu\epsilon}(\partial_\beta K_{\alpha\epsilon} +\partial_\alpha K_{\beta\epsilon}- \partial_\epsilon K_{\alpha\beta}),
\label{dtgamma}     
\ee
and,
\begin{eqnarray}
\partial_t^2 \Gamma^\mu_{\alpha\beta}&=& - K^{\mu\epsilon}(\partial_\beta K_{\alpha\epsilon} +\partial_\alpha K_{\beta\epsilon}- \partial_\epsilon K_{\alpha\beta})\nonumber\\ &-&\frac{1}{2}(D^{\mu\epsilon}-2K^{\mu\eta} K^\epsilon_\eta)(\partial_\beta \gamma_{\alpha\epsilon} +\partial_\alpha \gamma_{\beta\epsilon} -\partial_\epsilon \gamma_{\alpha\beta}) \nonumber\\ &+&\frac{1}{2}\gamma^{\mu\epsilon}(\partial_\beta D_{\alpha\epsilon} +
\partial_\alpha D_{\beta\epsilon}- \partial_\epsilon D_{\alpha\beta}).
\label{ddtgamma}
\end{eqnarray}
We also set,
\be 
\G_{\e\a\b}=\frac{1}{2}(\pa_\b \g_{\a\e} +\pa_\a \g_{\b\e} -\pa_\e \g_{\a\b}),
\label{eq:big_Gamma}
\ee
so that 
\be
\G^\m_{\a\b}=\g^{\m\e} \G_{\e\a\b},
\label{GupGdown}
\ee
and further we set,
\be 
Z_{\e\a\b}=\frac{1}{2}(\pa_\b K_{\a\e} +\pa_\a K_{\b\e} -\pa_\e K_{\a\b}), 
\label{big_Zeta}   
\ee
and,
\be 
H_{\e\a\b}=\frac{1}{2}(\pa_\b D_{\a\e} +\pa_\a D_{\b\e} -\pa_\e D_{\a\b}). 
\label{big_Heta}     
\ee
Then we find that the time derivatives of these `fully covariant symbols' satisfy,
\bq
\pa_t \G_{\e\a\b} &=& Z_{\e\a\b},\\
\pa_t Z_{\e\a\b} &=& H_{\e\a\b}.
\eq
Hence, we conclude that the first time derivatives of the spatial connection coefficients depend only on $(\g_{\a\b},K_{\a\b})$ and their first \emph{spatial} derivatives, while the second  time derivatives of the spatial connection coefficients depend only on $(\g_{\a\b},K_{\a\b},D_{\a\b})$ and their first \emph{spatial} derivatives.

Now, the spatial Ricci tensor is given by,
\be 
P_{\a\b}=\pa_\m \G^\m_{\a\b} -\pa_\b \G^\m_{\a\m} +\G^\m_{\a\b}\G^\ep_{\m\ep} -\G^\m_{\a\ep}\G^\ep_{\b\m},
\label{eq:sp_Ricci_ten_big}
\ee
and so its time derivative is calculated to be,
\be 
\partial_t P_{\alpha\beta}= \partial_\mu (\partial_t \Gamma^\mu_{\alpha\beta}) -\partial_\beta (\partial_t \Gamma^\mu_{\alpha\mu}) +  \partial_t \Gamma^\mu_{\alpha\beta} \Gamma^\epsilon_{\mu\epsilon} +\Gamma^\mu_{\alpha\beta} \partial_t \Gamma^\epsilon_{\mu\epsilon} -\partial_t \Gamma^\mu_{\alpha\epsilon} \Gamma^\epsilon_{\beta\mu} -\Gamma^\mu_{\alpha\epsilon} \partial_t \Gamma^\epsilon_{\beta\mu}.  
\label{eq:partial_sp_Ricci_ten_big}
\ee
Then, for the spatial scalar curvature,
\be 
P=\g^{\a\b}P_{\a\b} ,
\label{sp_scalar_curv}
\ee
we find that,
\be 
\partial_t P=-K^{\alpha\beta} P_{\alpha\beta}+\gamma^{\alpha\beta}\partial_t P_{\alpha\beta},
\label{eq:partial_sp_scalar_curv}
\ee
and, therefore, the first of the dangerous terms finally reads:
\begin{eqnarray}
\pa_t P&=&K^{\a\b}(-\pa_\m \g^{\m\ep}\G_{\ep\a\b} - \g^{\m\ep}\pa_\m\G_{\ep\a\b} +\pa_\b \g^{\m\ep}\G_{\ep\a\m} + \g^{\m\ep}\pa_\b\G_{\ep\a\m} - \g^{\m\ep}\g^{\z\e}\G_{\ep\a\b}\G_{\e\m\z}  \nonumber\\
&+&\g^{\m\e}\g^{\ep\z}\G_{\e\a\ep}\G_{\z\b\m})   + \g^{\a\b}[-\pa_\m K^{\m\ep}\G_{\ep\a\b} - K^{\m\ep}\pa_\m\G_{\ep\a\b} +\pa_\m \g^{\m\ep}Z_{\ep\a\b} + \g^{\m\ep}\pa_\m Z_{\ep\a\b}  \nonumber\\
&-& \pa_\b K^{\m\ep}\G_{\ep\a\m} - K^{\m\ep}\pa_\b\G_{\ep\a\m}   + \pa_\b \g^{\m\ep}Z_{\ep\a\m} + \g^{\m\ep}\pa_\b Z_{\ep\a\m} \nonumber\\
&+& \g^{\z\e}\G_{\e\m\z}(-K^{\m\ep}\G_{\ep\a\b} +\g^{\m\ep}Z_{\ep\a\b})
+\g^{\m\z}\G_{\z\a\b}(-K^{\ep\e}\G_{\e\m\ep} +\g^{\ep\e}Z_{\e\m\ep}) \nonumber\\
&-& \g^{\ep\z}\G_{\z\b\m}(-K^{\m\e}\G_{\e\a\ep} +\g^{\m\e}Z_{\e\a\ep})
-\g^{\m\z}\G_{\z\a\ep}(-K^{\ep\e}\G_{\e\b\m} +\g^{\ep\e}Z_{\e\b\m})].
\label{eq:d_t_P}
\end{eqnarray}
This result means that the first dangerous term is `purely spatial'. It also implies that the third dangerous term is also purely spatial, for it is calculated to be of the form,
\begin{eqnarray}
 \nabla_\alpha \nabla_\beta (-P -D +\frac{3}{4}K^{\gamma\delta} K_{\gamma\delta}-\frac{1}{4}K^2)&=& \partial_\alpha[\partial_\beta (-P -D +\frac{3}{4}K^{\gamma\delta} K_{\gamma\delta}-\frac{1}{4}K^2)] \nonumber\\ &-&\Gamma^\mu_{\alpha\beta}\partial_\mu (-P -D +\frac{3}{4}K^{\gamma\delta} K_{\gamma\delta}-\frac{1}{4}K^2)\nonumber \\
&-&\frac{1}{2}K_{\alpha\beta}(-\partial_t P-W +\frac{5}{2}K^{\alpha\beta}D_{\alpha\beta} \nonumber\\
&-&\frac{3}{2}K^{\alpha\gamma}K^\beta_\gamma K_{\alpha\beta} -\frac{1}{2}KD \nonumber \\
&+&\frac{1}{2}K K^{\alpha\beta} K_{\alpha\beta}),
\label{eq:third_dangerous_term}
\end{eqnarray}
means that trouble could only have arisen from the first dangerous term which however as we showed above, is purely spatial.

Lastly, since
\be
\partial_t(\partial_t P)=-(D^{\alpha\beta}-2K^{\alpha\gamma}K^\beta_\gamma)P_{\alpha\beta} - 2K^{\alpha\beta}\partial_t P_{\alpha\beta} +\gamma^{\alpha\beta}\partial_t(\partial_t P_{\alpha\beta}), \ee
and,
\bq
\partial_t^2 P_{\alpha\beta}&=&\partial_\mu (\partial_t^2 \Gamma^\mu_{\alpha\beta}) -\partial_\beta(\partial_t^2 \Gamma^\mu_{\alpha\mu}) +  \partial_t^2 \Gamma^\mu_{\alpha\beta} \Gamma^\epsilon_{\mu\epsilon} +\Gamma^\mu_{\alpha\beta} \partial_t^2 \Gamma^\epsilon_{\mu\epsilon} - \partial_t^2 \Gamma^\mu_{\alpha\epsilon} \Gamma^\epsilon_{\beta\mu} \nonumber\\
&-&  \partial_t^2 \Gamma^\epsilon_{\beta\mu}\Gamma^\mu_{\alpha\epsilon} +2 \partial_t \Gamma^\mu_{\alpha\beta}\partial_t\Gamma^\epsilon_{\mu\epsilon} -
2 \partial_t \Gamma^\mu_{\alpha\epsilon} \partial_t\Gamma^\epsilon_{\beta\mu},
\label{eq:d_t^2_P}
\eq
we find that the second dangerous term depends on $(\g_{\a\b},K_{\a\b},D_{\a\b})$ and its first and second spatial derivatives, namely, it has the form:
\begin{eqnarray}
\pa_t(\pa_t P)&=&(D^{\a\b}-2K^{\a\g}K^\b_\g)(-\pa_\m \g^{\m\ep}\G_{\ep\a\b} - \g^{\m\ep}\pa_\m\G_{\ep\a\b} +\pa_\b \g^{\m\ep}\G_{\ep\a\m} + \g^{\m\ep}\pa_\b\G_{\ep\a\m}\nonumber\\
&-& \g^{\m\ep}\g^{\z\e}\G_{\ep\a\b}\G_{\e\m\z} +\g^{\m\e}\g^{\ep\z}\G_{\e\a\ep}\G_{\z\b\m})\nonumber\\
&-&2K^{\a\b}[-\pa_\m K^{\m\ep}\G_{\ep\a\b} - K^{\m\ep}\pa_\m\G_{\ep\a\b} +\pa_\m \g^{\m\ep}Z_{\ep\a\b} + \g^{\m\ep}\pa_\m Z_{\ep\a\b} - \pa_\b K^{\m\ep}\G_{\ep\a\m}  \nonumber\\
&-& K^{\m\ep}\pa_\b\G_{\ep\a\m} + \pa_\b \g^{\m\ep}Z_{\ep\a\m} + \g^{\m\ep}\pa_\b Z_{\ep\a\m} \nonumber\\
&+& \g^{\z\e}\G_{\e\m\z}(-K^{\m\ep}\G_{\ep\a\b} +\g^{\m\ep}Z_{\ep\a\b})
+\g^{\m\z}\G_{\z\a\b}(-K^{\ep\e}\G_{\e\m\ep} +\g^{\ep\e}Z_{\e\m\ep}) \nonumber\\
&-& \g^{\ep\z}\G_{\z\b\m}(-K^{\m\e}\G_{\e\a\ep} +\g^{\m\e}Z_{\e\a\ep})
-\g^{\m\z}\G_{\z\a\ep}(-K^{\ep\e}\G_{\e\b\m} +\g^{\ep\e}Z_{\e\b\m})] \nonumber\\
&+& \g^{\a\b}\{-\pa_\m(D^{\m\ep}-2K^{\m\g}K^\ep_\g)\G_{\ep\a\b} - (D^{\m\ep}-2K^{\m\g}K^\ep_\g)\pa_\m\G_{\ep\a\b}\nonumber\\
&-& 2\pa_\m K^{\m\ep}Z_{\ep\a\b}-2K^{\m\ep}\pa_\m Z_{\ep\a\b} + \pa_\m \g^{\m\ep}H_{\ep\a\b} + \g^{\m\ep}\pa_\m H_{\ep\a\b}\nonumber\\
&+& \pa_\b(D^{\m\ep}-2K^{\m\g}K^\ep_\g)\G_{\ep\a\m} + (D^{\m\ep}-2K^{\m\g}K^\ep_\g)\pa_\b\G_{\ep\a\m}\nonumber\\
&+& 2\pa_\b K^{\m\ep}Z_{\ep\a\m} + 2K^{\m\ep}\pa_\b Z_{\ep\a\m} - \pa_\b \g^{\m\ep}H_{\ep\a\m} - \g^{\m\ep}\pa_\b H_{\ep\a\m}\nonumber\\
&+& \g^{\z\e}[-(D^{\m\ep}-2K^{\m\g}K^\ep_\g)\G_{\ep\a\b}-2K^{\m\ep}Z_{\ep\a\b} + \g^{\m\ep}H_{\ep\a\b}]\G_{\e\m\z} \nonumber\\
&+& \g^{\m\z}[-(D^{\ep\e}-2K^{\ep\g}K^\e_\g)\G_{\e\m\ep}-2K^{\ep\e}Z_{\e\m\ep} + \g^{\ep\e}H_{\e\m\ep}]\G_{\z\a\b} \nonumber\\
&-& \g^{\ep\z}[-(D^{\m\e}-2K^{\m\g}K^\e_\g)\G_{\e\a\ep}-2K^{\m\e}Z_{\e\a\ep} + \g^{\m\e}H_{\e\a\ep}]\G_{\z\b\m}\nonumber \\
&-& \g^{\m\z}[-(D^{\ep\e}-2K^{\ep\g}K^\e_\g)\G_{\e\b\m}-2K^{\ep\e}Z_{\e\b\m} + \g^{\ep\e}H_{\e\b\m}]\G_{\z\a\ep}\nonumber \\
&+& 2(-K^{\m\ep}\G_{\ep\a\b}+\g^{\m\ep}Z_{\ep\a\b})(-K^{\z\e}\G_{\e\m\z}+\g^{\z\e}Z_{\e\m\z})\nonumber\\
&-& 2(-K^{\m\e}\G_{\e\a\ep}+\g^{\m\e}Z_{\e\a\ep})(-K^{\ep\z}\G_{\z\b\m}+\g^{\ep\z}Z_{\z\b\m})\}.
\label{eq:d_t_d_t_P}
\end{eqnarray}
This shows that the evolution equations form a Cauchy-Kovalevskaya system. Based on this result, we will perform a pertubative analysis of our system of higher order equations, namely, the evolution equations (\ref{eq:pg}), (\ref{eq:pK}), (\ref{eq:pD}) and (\ref{eq:pW}) together with the constraints  (\ref{eq:hamiltonian}) and (\ref{eq:momentum}).


\subsection{The Cauchy-Kovalevskaya property in the presence of a fluid}
\label{The Cauchy-Kovalevskaya property in the presence of a fluid}

In order to verify the function-counting argument of the higher-order gravity-matter theory ($20$ arbitrary functions), we also need to prove that the field equations are well defined, at least in the analytic case.
To be concrete, we need to prove that these equations are of the Cauchy-Kovalevskaya type. In doing so, we must show that there are no time derivatives in the constraints (\ref{eq:hamiltonian_rad}) and (\ref{eq:momentum_rad}) and the derivatives of the unknowns are (through the evolution equations (\ref{eq:pg}), (\ref{eq:pK}), (\ref{eq:pD}) and (\ref{eq:pW with T}))  analytic functions of the coordinates, the unknowns, and their first and second \emph{space} derivatives. The only `dangerous' terms in the dynamical system presently are again the three terms met previously, namely, 
\be
\partial_t P,\quad\partial_t(\partial_t P), \quad\nabla_\alpha \nabla_\beta (-P -D +\frac{3}{4}K^{\gamma\delta} K_{\gamma\delta}-\frac{1}{4}K^2),
\label{eq:dangerous_terms_rad}
\ee
present in the constraints Eqns. (\ref{eq:hamiltonian_rad}), (\ref{eq:momentum_rad}) and in the snap equation (\ref{eq:pW with T}). However, this is already shown in previous subsection \ref{Proof of the Cauchy-Kovalevskaya property in vacuum}.


\section{Analytic solution of the local Cauchy problem}
\label{Analytic solution of the local Cauchy problem}

In order to verify the function-counting argument of Chapter \ref{Chapter2} ($16$ arbitrary functions for the vacuum, $20$ arbitrary functions in case of a fluid), we further need to prove that the evolution equations (\ref{eq:pg}), (\ref{eq:pK}), (\ref{eq:pD}) and (\ref{eq:pW}) together with the constraints for each case are well defined in the sense that there is a well-defined Cauchy problem, at least for the analytic case.


\subsection{Analytic solution in vacuum}
\label{Analytic solution in vacuum}

In case of vacuum we have proved that the evolution and constraint equations are of the Cauchy-Kovalevskaya type. From this conclusion and the Cauchy-Kovalevskaya theorem, we are directly led to the following result, local Cauchy problem (analytic case cf. \cite{ycb}):
 \begin{theorem}
 For $N=1,N^\a=0$, if we prescribe analytic initial data $(\g_{\a\b},K_{\a\b},\\
 D_{\a\b},W_{\a\b})$ on some initial slice $\mathcal{M}_{0}$, then there exists a neighborhood of $\mathcal{M}_{0}$ in $\mathbb{R}\times\mathcal{M}$ such that the evolution equations (\ref{eq:pg}), (\ref{eq:pK}), (\ref{eq:pD}) and (\ref{eq:pW}) have an analytic solution in this neighborhood consistent with these data. This analytic solution is the development of  the prescribed initial data on  $\mathcal{M}_{0}$ if and only if these initial data satisfy the constraints.
  \end{theorem}
For the last part of this theorem there is an alternative justification if we use the conformal equivalence theorem of higher-order gravity theories \cite{ba-co88}, that is working in the Einstein-frame representation, and a theorem on symmetric hyperbolic systems, cf. \cite{ycb}, pp. 150-1, and also \cite{ay}. Indeed, in the Einstein frame representation, the theory under discussion is general relativity plus a self-interacting scalar field, and so the Einstein equations,
\be
R^i_j - \dfrac{1}{2}\d^i_j R = 8\pi G T^i_j,
\label{Einstein eq}
\ee
give a system of the form,
\be
R^0_0 = -\frac{1}{2}\partial_t K -\frac{1} {4}K_\a^\b K_\b^\a = 8\pi G(T^0_0 -\frac{1}{2}T),
\label{eq:RooT}
\ee
\be
R^0_\a  = \frac{1}{2} (\nabla_\b K^\b_\a - \nabla_\a K) = 8\pi G T^0_\a,
\label{eq:RoaT}
\ee
\be
R^\b_\a = -P^\b_\a - \frac{1}{2\sqrt{\g}} \partial_t(\sqrt{\g}K^\b_\a) = 8\pi G (T^\b_\a -\frac{1}{2}\d^\b_\a T),
\label{eq:RbaT}
\ee
and the wave equation $\nabla^i \nabla_i \phi -V'(\phi)=0$  for $\phi$ having the particular scalar field potential given in  \cite{ba-co88}. This system is of the form given in Thm. 4.1 of \cite{ycb}, p. 150, from which the result follows.

From the conformal transformation theorem above, we notice that there is an equivalence between our higher-order gravity theory and general relativity plus a self-interacting scalar field which satisfies the wave equation on $\mathcal{M}\times I, I\subseteq \mathbb{R}$. Therefore in the conformal frame we indeed end up with six arbitrary functions, four of them are from the geometric part and two more from the wave equation of the scalar field. Furthermore, through the conformal transformation \cite{c08},
\be
\phi=\ln (1+2\epsilon R), 
\label{con transf}
\ee
the scalar field $\phi$ is exactly defined to be directly related to the scalar curvature $R$,  and so we see that the function-counting resulting in the number 6 may be interpreted as giving the number of arbitrary functions associated with the conformal picture of the theory in the Einstein frame. 

We note in passing that this result explains a completely different function-counting that is usually associated with higher-order gravity theories, cf. \cite{star1,star2}, namely, that using the trace equation for the scalar curvature $R$ we get two initial data functions, and, these, together with the additional four other arbitrary functions coming from general relativity, result in the theory having finally six functions for the general solution in the analytic case. However, it may not seem completely correct to count the metric functions as independent functions together with the initial data corresponding to the trace equation for the scalar curvature and its first derivative, which, in turn, depend on the metric and its derivatives. Besides, six is also the number for the higher-order equations in the original frame rather than the Einstein frame representation, assumes that the conformal transformations is not-singular. This counting method leaves completely open the counting of the arbitrary functions when the conformal factor $\Omega$, where,
\be
\Omega^2 = e^\phi,
\label{con factor}
\ee
is equal to zero \cite{bc,mss88}.
 
In order to confirm that the right number of arbitrary functions of this same theory in the original Jordan frame is 16, the only thing left to show is the proof of the Cauchy-Kovalevskaya property of the system of the four evolution equations.


\subsection{Analytic solution in the presence of a fluid}
\label{Analytic solution in the presence of a fluid}

Similarly with the results obtained in subsection \ref{Analytic solution in vacuum} and the Cauchy-Kovalevskaya theorem, we are led to the local Cauchy problem (analytic case) for the case of a fluid:
\begin{theorem}
For $N=1,N^\a=0$, if we prescribe analytic initial data $(\g_{\a\b},K_{\a\b}, \\
D_{\a\b},W_{\a\b})$ together with the data $(p,\rho,u^i)$ satisfying Eqns. (\ref{eq:general equation of state}), (\ref{eq of motion 0}), (\ref{eq of motion a}) and the identity (\ref{eq:velocities identity}) on some initial slice $\mathcal{M}_{0}$, then there exists a neighborhood of $\mathcal{M}_{0}$ in $\mathbb{R}\times\mathcal{M}$ such that the evolution equations (\ref{eq:pg}), (\ref{eq:pK}), (\ref{eq:pD}) and (\ref{eq:pW with T}) have an analytic solution in this neighborhood consistent with these data. This analytic solution is the development of  the prescribed initial data on  $\mathcal{M}_{0}$ if and only if these initial data satisfy the constraints.
\end{theorem}  
We are now ready to apply in the next Chapter a perturbative formalism of higher-order gravity in a similar manner to that presented in \cite{ll} in the cases of vacuum and radiation in general relativity. In particular, we are going to study both cases by using a formal series representation of the spatial metric. In fact, for the analysis of the vacuum and radiation problem, we will consider two different formal series of the spatial metric for each case. The first one is a regular series representation of the spatial metric and the second is not a regular, but a singular formal representation of the spatial metric. In this Thesis, such a perturbative analysis will play a decisive role and in fact will generalize the results holding in general relativity to the framework of higher-order gravity. Speciffically, including the results concerning the Einstein theory, we are going to deal with eight different problems (two of them have been fully analyzed by Landau and Lifschitz in \cite{ll}):
\begin{itemize}

\item general relativity in cases of vacuum and radiation (with regular and singular formal series) and,  
\item higher-order $R+\ep R^2$ gravity in cases of vacuum and radiation (with regular and singular formal series).
\end{itemize}


\chapter{Perturbative formulation of higher-order gravity} 

\label{Chapter4} 

\lhead{Chapter 4. \emph{Perturbative formulation of higher-order gravity}} 

In this Chapter, we introduce the perturbative formulation in higher-order gravity assuming a formal series representation of the spatial metric. In Section \ref{Friedmann Equations}, we find an exact solution of the scale factor $a=a(t)$ that is provided by the corresponding \textit{Friedmann equations} of higher-order gravity theory in cases of vacuum and radiation. In Section \ref{Perturbative analysis with a regular formal series representation of the spatial metric}, we use the Landau-Lifschitz pertubative method \cite{ll} in the vicinity of a point that is regular in the time. In Section \ref{Perturbative analysis with a singular formal series representation of the spatial metric}, we use the same method in the vicinity of a point that is not regular, but singular in the time.


\section{Friedmann Equations}
\label{Friedmann Equations}

In this Section we present the Friedmann equations of General Relativity as well as the corresponing generalized Friedmann equations of higher-order gravity and we study the connection between their solutions.



\subsection{Solutions of Friedmann equations}
\label{Solutions of Friedmann Equations}

In General Relativity and in Friedmann-Robertson-Walker universes, using the Robertson-Walker metric (\cite{hel}, chap. 14 and \cite{clba05}),
\be
{ds}^2 = {dt}^2 - {a^2}(t)\left[ \dfrac{{dr}^2}{1-kr^2} + r^2({d\theta}^2 + {\sin}^2 \theta {d\phi}^2) \right],
\label{rwmetric}
\ee
we find that the off-diagonal components of the Ricci tensor are equal to zero and the diagonal components are given by,
\bq
R_{00} &=& -3\dfrac{\ddot{a}}{a},\\
\label{R00sf}
R_{11} &=& \dfrac{a\ddot{a}+2{\dot{a}}^2 +2k}{1-kr^2},\\
\label{R11sf}
R_{22} &=& \dfrac{a\ddot{a}+2{\dot{a}}^2 +2k}{r^2},\\
\label{R22sf}
R_{33} &=& \dfrac{a\ddot{a}+2{\dot{a}}^2 +2k}{r^2 {\sin}^2 \theta}.
\label{R33sf}
\eq
Also for the scalar curvature we obtain,
\be
R = -6\left( \dfrac{\ddot{a}}{a} + \dfrac{{\dot{a}}^2}{a^2} + \dfrac{k}{a^2} \right),
\label{Rsf}
\ee
where dot denotes differentiation with respect to $t$. Then the Friedmann equations for the energy density and the pressure are respectively (\cite{lppt}, chap. 19),
\be
\dfrac{8\pi G\rho}{3} = \dfrac{\dot{a}^2 + k}{a^2},
\label{Friedmann eq GR density}
\ee
and,
\be
8\pi G p=-2\dfrac{\ddot{a}}{a}- \dfrac{\dot{a}^2 +k}{a^2},
\label{Friedmann eq GR pressure}
\ee
where $k$ is the constant curvature, with values $-1,0,+1$ correspondingly to hyperbolic, flat and spherical geometries (see \cite{har}, chap. 18).

In case of an empty space $(p=0,\rho=0)$, the general solution is,
\be
a = a_0 + a_1 t,
\label{gses}
\ee
where,
\be
{a_1}^2 = -k,
\label{a1}
\ee
for $k=-1,0$.

Further, in radiation models ($w=\dfrac{1}{3}$, see \cite{grpart2}, chap. 11), that is,
\be
p=\dfrac{\rho}{3},
\label{radiation eq}
\ee 
the exact solutions of the Friedmann equations for $k=-1,0,+1$ are respectively (\cite{on}, chap. 12):
\be
a=\sqrt{c}[(1+\dfrac{t}{\sqrt{c}})^2 -1]^{1/2}=(2\sqrt{c}t+t^2)^{1/2},
\label{sol rad GR k=-1}
\ee 
\be
a=(4c)^{1/4}t^{1/2},
\label{sol rad GR k=0}
\ee 
\be
a=\sqrt{c}[1-(1-\dfrac{t}{\sqrt{c}})^2]^{1/2}=(2\sqrt{c}t-t^2)^{1/2},
\label{sol rad GR k=+1}
\ee 
where, 
\be
c=\dfrac{8\pi G \rho a^4}{3}
\label{eq c}
\ee
and,
\be
\rho a^4=\textrm{constant},
\label{eq ra^4}
\ee
which expresses the conservation of mass-energy (\ref{eq:conservation laws}). 


\subsection{Generalized Friedmann Equations}
\label{Generalized Friedmann Equations}

Working similarly with the lagrangian $f(R)=R + \ep R^2$ and the field equations (\ref{eq:FEs_rad}), apart from the Eqns. (\ref{R00sf})-(\ref{Rsf}), we find that,
\bq
\pa_t R &=& -6\left( \dfrac{\dddot{a}}{a} + \dfrac{\dot{a}\ddot{a}}{a^2} - 2\dfrac{{\dot{a}}^3}{a^3} - 2k\dfrac{\dot{a}}{a^3} \right), \\
\label{pa_t Rsf}
{\pa_t}^2 R &=& -6\left( \dfrac{a^{(4)}}{a} + \dfrac{{\ddot{a}}^2}{a^2} - 8\dfrac{{\dot{a}}^2 \ddot{a}}{a^3} + 6\dfrac{{\dot{a}}^4}{a^4} - 2k\dfrac{\ddot{a}}{a^3} + 6k\dfrac{{\dot{a}}^2}{a^4} \right),
\label{pa_t2 Rsf}
\eq
as well as,
\bq
\nabla_1 \nabla_1 R &=& \dfrac{6}{1-kr^2}\left( \dot{a} \dddot{a} + \dfrac{{\dot{a}}^2 \ddot{a}}{a} - 2\dfrac{{\dot{a}}^4}{a^2} - 2k\dfrac{{\dot{a}}^2}{a^2} \right), \\
\label{nabla1nabla1Rsf}
\nabla_2 \nabla_2 R &=& 6r^2 \left( \dot{a} \dddot{a} + \dfrac{{\dot{a}}^2 \ddot{a}}{a} - 2\dfrac{{\dot{a}}^4}{a^2} - 2k\dfrac{{\dot{a}}^2}{a^2} \right), \\
\label{nabla2nabla2Rsf}
\nabla_3 \nabla_3 R &=& 6r^2\sin{\theta} \left( \dot{a} \dddot{a} + \dfrac{{\dot{a}}^2 \ddot{a}}{a} - 2\dfrac{{\dot{a}}^4}{a^2} - 2k\dfrac{{\dot{a}}^2}{a^2} \right), \\
\label{nabla3nabla3Rsf}
\nabla_\a \nabla_\b R &=& 0, \quad \textrm{if} \quad \a \neq \b.
\label{nablaanablabRsf}
\eq
Then, the $00$-component \cite{ckt} and $\a \a$-components lead us to the generalized Friedmann equations for the energy density $\rho$ and the pressure $p$, which respectively are:
\be
\dfrac{8\pi G\rho}{3}=\dfrac{k+\dot{a}^2}{a^2} +6\ep \left( 2\dfrac{\dot{a}\dddot{a}}{a^2} +2\dfrac{\dot{a}^2 \ddot{a}}{a^3} -\dfrac{\ddot{a}^2}{a^2} -3\dfrac{\dot{a}^4}{a^4} -2k\dfrac{\dot{a}^2}{a^4} +\dfrac{k^2}{a^4} \right),
\label{gen Friedmann eq density}
\ee
and,
\be
8\pi G p=-2\dfrac{\ddot{a}}{a}- \dfrac{\dot{a}^2 +k}{a^2} +6\ep \left( -2\dfrac{a^{(4)}}{a} -4\dfrac{\dot{a} \dddot{a}}{a^2} -3\dfrac{\ddot{a}^2}{a^2} +12\dfrac{\dot{a}^2 \ddot{a}}{a^3} -3\dfrac{\dot{a}^4}{a^4} +4k\dfrac{\ddot{a}}{a^3} -2k\dfrac{\dot{a}^2}{a^4} +\dfrac{k^2}{a^4} \right),
\label{gen Friedmann eq pressure}
\ee
where the general relativity case may be recovered when $\ep \rightarrow 0$. 
The question that naturally emerge is if, in cases of vacuum and radiation, the corresponding solutions (\ref{gses}) and (\ref{sol rad GR k=-1}), (\ref{sol rad GR k=0}), (\ref{sol rad GR k=+1}) of the Friedmann equations, for $k=-1,0,+1$ respectively, are exact solutions of the generalized Friedmann equations (\ref{gen Friedmann eq density}) and (\ref{gen Friedmann eq pressure}) as well. To answer this question, we notice that each one of Eqns. (\ref{gen Friedmann eq density}) and (\ref{gen Friedmann eq pressure}) consists of two parts. The first parts of those are not multiplied by the coefficient $6\ep$ and, in fact, are equal to the quantities $\dfrac{8\pi G\rho_{GR}}{3}$ and $8\pi G p_{GR}$ respectively.

This means that in both cases $(p=\rho=0,p=\rho/3)$ the corresponding solutions (\ref{gses}) and (\ref{sol rad GR k=-1}), (\ref{sol rad GR k=0}), (\ref{sol rad GR k=+1}), for the appropriate $k$ each time, make these parts identically equal. In addition to that, substituting $a$ from the solutions (\ref{gses}) and (\ref{sol rad GR k=-1}), (\ref{sol rad GR k=0}), (\ref{sol rad GR k=+1}), for the corresponding $k$ each time, to the second parts of (\ref{gen Friedmann eq density}) and (\ref{gen Friedmann eq pressure}), which are the parts, 
\beq
2\dfrac{\dot{a}\dddot{a}}{a^2} +2\dfrac{\dot{a}^2 \ddot{a}}{a^3} -\dfrac{\ddot{a}^2}{a^2} -3\dfrac{\dot{a}^4}{a^4} -2k\dfrac{\dot{a}^2}{a^4} +\dfrac{k^2}{a^4} , 
\eeq
and,
\beq
-2\dfrac{a^{(4)}}{a} -4\dfrac{\dot{a} \dddot{a}}{a^2} -3\dfrac{\ddot{a}^2}{a^2} +12\dfrac{\dot{a}^2 \ddot{a}}{a^3} -3\dfrac{\dot{a}^4}{a^4} +4k\dfrac{\ddot{a}}{a^3} -2k\dfrac{\dot{a}^2}{a^4} +\dfrac{k^2}{a^4}, 
\eeq
we immediately find that these parts are identically equal to zero.

Therefore, we conclude that, in cases of vacuum and radiation, relations (\ref{gses}) and (\ref{sol rad GR k=-1}), (\ref{sol rad GR k=0}), (\ref{sol rad GR k=+1}), for the corresponding $k$ each time, are also exact solutions of the generalized Friedmann equations. Also, we note that relations (\ref{eq c}) and (\ref{eq ra^4}) are still valid, due to the fact that they arise from the conservation laws (\ref{eq:conservation laws}). This result is very important concerning the appropriate form of the spatial metric $\g_{\a\b}$ that we are going to use in what follows. 

In fact, we are going to choose a form of the $3$-metric $\g_{\a\b}$, that is consistent with the consideration that on largest scales the galaxy distribution appears to be homogeneous and isotropic \cite{wald} and so with the exact solutions of the generalized Friedmann equations.


\section{Perturbative analysis with a regular formal series representation of the spatial metric}
\label{Perturbative analysis with a regular formal series representation of the spatial metric}

In this Section, taking into account that the scale factor $a$ given by Eq. (\ref{gses}) is an exact solution of the generalized Friedmann equations (\ref{gen Friedmann eq density}) and (\ref{gen Friedmann eq pressure}), we are going to use the Landau-Lifschitz pertubative method \cite{ll} in higher order gravity, in the vicinity of o point that is regular in the time. In particular, we assume a regular formal series representation of the spatial metric of the form,
\be
\g_{\a\b}= \g^{(0)}_{\a\b} +\g^{(1)}_{\a\b}\;t + \g^{(2)}_{\a\b}\;t^2 + \g^{(3)}_{\a\b}\;t^3 + \g^{(4)}_{\a\b}\;t^4 + \cdots,
\label{eq:3dimmetric}
\ee
where the $ \g^{(0)}_{\a\b} , \g^{(1)}_{\a\b} , \g^{(2)}_{\a\b} , \g^{(3)}_{\a\b} , \g^{(4)}_{\a\b},\cdots$ are functions of the space coordinates. 


\subsection{Regular formal series representation}
\label{Regular formal series representation}

In the first step of this analysis, after taking into consideration the order of the higher order gravity equations in cases of vacuum and radiation that we have dealt with, we shall be interested only in the part of the formal series shown, that is up to order four. Because of this, we shall drop the dots at the end of the various expressions to simplify the overall appearance\footnote{Differentiation of such formal series with respect to either space or the time variables is defined term by term, whereas multiplication of two such expressions results when the various terms are multiplied and terms of same powers of $t$ are taken together.}. Nevertheless, in the next two Chapters, we will show that, even if the order of the formal series is greater than four, then the results are qualitatively the same as that of the series (\ref{eq:3dimmetric}). Thus before substitution to the evolution and constraint higher order equations, we note that the expression  (\ref{eq:3dimmetric}) contains  $30$ degrees of freedom ($6$ of each spatial matrix $\g^{(n)}_{\a\b},n=0,\cdots,4$). Note that setting $\g^{(0)}_{\a\b}=\d_{\a\b} $ and $\g^{(n)}_{\a\b}=0,n>0,$ we have Minkowski space included here as an exact solution of the equations, and so our perturbation analysis covers this case too.

We set,
\be
 \g^{(0)}_{\a\b}= a_{\a\b}, \,\,  \g^{(1)}_{\a\b}=b_{\a\b}, \,\,   \g^{(2)}_{\a\b}= c_{\a\b}, \, \,   \g^{(3)}_{\a\b}= d_{\a\b}, \,\,
     \g^{(4)}_{\a\b}= e_{\a\b},
     \label{free}
\ee
where the data $a_{\a\b} ,b_{\a\b} ,c_{\a\b} ,d_{\a\b} ,e_{\a\b}$, are arbitrary, nontrivial analytic functions of the space coordinates. 
To proceed, we shall need the expression of the formal expansion of the reciprocal tensor $\g^{\a\b}$. To find this, we use the identity,
\be
\g_{\a\b}\g^{\b\g}=\d_\a^\g.
\label{eq:identity of gamma_ab}
\ee
Then the various coefficients $ \g^{(\mu)\,\a\b},\mu=0,\cdots,4,$ of $t$ in the expansion $\g^{\a\b}=\sum_{n=0}^{\infty}\g^{(n)\,\a\b}t^n$ are found to be:
\bq
 \g^{\a\b}&=& a^{\a\b} - b^{\a\b}t + \left(b^{\a\g}\;b_\g^\b - c^{\a\b}\right)t^2 +
 \left(-d^{\a\b} + b^{\a\g}\;c_\g^\b - b^{\a\g}\;b_\g^\d\;b_\d^\b +
 c^{\a\g}\;b_\g^\b\right)t^3   \nonumber     \\
&+& \left(-e^{\a\b} + b^{\a\g}d_\g^\b - b^{\a\g}b_\g^\d\;c_\d^\b + c^{\a\g}\;c_\g^\b + d^{\a\g}\;b_\g^\b - b^{\a\g}\; c_\g^\d b_\d^\b + b^{\a\g}b_\g^\d b_\d^\ep b_\ep^\b \right. \nonumber \\
&-& \left. c^{\a\g}\;b_\g^\d b_\d^\b\right)t^4. 
\label{eq:3diminvmetric}
\eq
 Note that $a_{\a\b}a^{\b\g}=\d_\a^\g$ and the indices of $b_{\a\b},c_{\a\b},d_{\a\b},e_{\a\b}$ are raised by $a^{\a\b}$.
For any tensor $X$, using the formal expansion  ($\ref{eq:3dimmetric}$), we can recursively calculate the coefficients in the expansion
\be
X_{\a\b}= X^{(0)}_{\a\b} +X^{(1)}_{\a\b}\;t + X^{(2)}_{\a\b}\;t^2 + X^{(3)}_{\a\b}\;t^3 + X^{(4)}_{\a\b}\;t^4.
\label{x tensor}
\ee
In particular, we can write down a general iterated formula for the $n$-th order term, $X^{(n)}_{\a\b}$. For instance, for the extrinsic curvature $K_{\a\b}$, we write,
 \be
K_{\a\b}= K^{(0)}_{\a\b} +K^{(1)}_{\a\b}\;t + K^{(2)}_{\a\b}\;t^2 + K^{(3)}_{\a\b}\;t^3.
\label{K tensor}
\ee
We now calculate all the essential quantities that we are going to use in the next Chapters in order to compare the number of the arbitrary functions that the formal series provide with those we found after the function-counting analysis in Chapter \ref{Chapter2}.


\subsection{Regular formal series representation of $K_{\a\b},D_{\a\b},W_{\a\b}$}
\label{Regular formal series representation of K,D,W}

In terms of the data $a_{\a\b},b_{\a\b},c_{\a\b},d_{\a\b},e_{\a\b}$, we have explicitly,
\be
K_{\a\b}=\partial_t\g_{\a\b}=b_{\a\b} + 2c_{\a\b}t + 3d_{\a\b}t^2 + 4e_{\a\b}t^3,
\label{eq:Kab}
\ee
and for the mixed components we obtain,
\bq
K_\b^\a&=&\g^{\a\g}K_{\g\b}=b_\b^\a + (2c_\b^\a - b^{\a\g}b_{\g\b})t + (3d_\b^\a - 2b^{\a\g}c_{\g\b} + b^{\a\d} b_\d^\g  b_{\g\b} - c^{\a\g}b_{\g\b})t^2   \nonumber   \\
&+& (4e_\b^\a - 3b^{\a\g}d_{\g\b} + 2b^{\a\d}b_\d^\g c_{\g\b} - 2c^{\a\g}c_{\g\b} - d^{\a\g}b_{\g\b} + b^{\a\d}c_\d^\g b_{\g\b} - b^{\a\d}b_\d^\ep b_\ep^\g b_{\g\b} \nonumber\\
&+&  c^{\a\d}b_\d^\g b_{\g\b})t^3.
\label{eq:Kmixed}
\eq
Further, taking into account the third inequality of (\ref{eq:determinant_g_ab}), the mean curvature,
\be
K=K_\a^\a=\g^{\a\b} \partial_t\g_{\a\b}=\partial_t  ln(\g),
\label{mean curvature}
\ee
is given by the form,
\bq
K&=&b + (2c - b_\b^\a b_\a^\b)t + (3d - 3b_\b^\a c_\a^\b + b^{\a\b}b_\b^\g b_{\g\a})t^2
\nonumber\\
&+& (4e -4b_\b^\a d_\a^\b + 4b^{\a\b}c_\b^\g b_{\g\a} - b^{\a\b}b_\b^\g b_\g^\d b_{\d\a}
-2 c_\b^\a c_\a^\b)t^3.
\label{eq:trK}
\eq
For completeness, we also give the expressions of the coefficients $K^{(n),\,\a\b}$ of the fully contravariant symbols,
\bq
K^{\a\b}&=&b^{\a\b} -2(b^{\a\g}b^\b_\g -c^{\a\b})t -3(-d^{\a\b} +b^{\a\g}c^\b_\g -b^{\a\g}b^\d_\g b^\b_\d +c^{\a\g}b^\b_\g)t^2
\nonumber\\
&-& 4(-e^{\a\b} +b^{\a\g}d^\b_\g -b^{\a\g}b^\d_\g c^\b_\d +c^{\a\g}c^\b_\g +d^{\a\g}b^\b_\g -b^{\a\g}c^\d_\g b^\b_\d\nonumber\\ &+&b^{\a\g}b^\d_\g b^\ep_\d b^\b_\ep -c^{\a\g}b^\d_\g b^\b_\d)t^3.
\label{eq:K^ab}
\eq
Using these forms, we can find the various components of the acceleration and jerk tensors to the required order. We have that the perturbation of the acceleration tensor in terms of the prescribed data is given by the form:
\bq
D_{\a\b}=\pa_t K_{\a\b}=2c_{\a\b} +6d_{\a\b}t +12e_{\a\b}t^2.
\label{eq:Dab}
\eq
Further we find,
\bq
D^\a_\b=\g^{\a\g}D_{\g\b}=2c^\a_\b +2(3d^\a_\b -b^\a_\g c^\g_\b)t +2(6e^\a_\b -3b^\a_\g d^\g_\b +b^\a_\d b^\d_\g c^\g_\b -c^\a_\g c^\g_\b)t^2,
\label{eq:Dmixed}
\eq
and,
\bq
D=2c +2(3d -b^\a_\b c^\b_\a)t +2(6e -3b^\a_\b d^\b_\a +b^\g_\b b^\b_\a c^\a_\g -c^\a_\b c^\b_\a)t^2,
\label{eq:trD}
\eq
where the trace is given by,
\be
D=D^\a_\a=\g^{\a\b} \pa_t K_{\a\b}.
\label{eq:trace_D}
\ee
For the fully contravariant components, we find,
\bq
D^{\a\b}&=& 2c^{\a\b} +2(3d^{\a\b} -b^{\a\g} c^\b_\g -b^{\d\b} c^\a_\d)t \nonumber\\
&+& 2(6e^{\a\b} -3b^{\a\g} d^\b_\g -3b^{\d\b} d^\a_\d -2c^{\a\g} c^\b_\g +b^{\a\d} b^\g_\d c^\b_\g +b^{\d\b} b^\a_\g c^\g_\d
\nonumber \\
&+&b^{\d\g} b^\b_\g c^\a_\d)t^2.
\label{eq:D^ab}
\eq
Lastly, the jerk perturbation series is found to be,
\bq
W_{\a\b}=\pa_t D_{\a\b}= 6d_{\a\b} +24e_{\a\b}t,
\label{eq:Wab}
\eq
so that,
\bq
W= 6d +6(4e -b^\a_\b d^\b_\a)t,
\label{eq:trW}
\eq
where,
\be
W=W^\a_\a=\g^{\a\b} \pa_t D_{\a\b}.
\label{eq:trace_W}
\ee


\subsection{Regular formal series representation of the Ricci tensor}
\label{Regular formal series representation of the Ricci tensor}

The various components of the Ricci curvature are more complicated when expressed in terms of the asymptotic data $a,b,c,d,e$ and we give them below. Using Eqns. (\ref{eq:R00mixeddownnew}), (\ref{eq:R0amixeddownnew}) and (\ref{eq:Rabmixeddownnew}) we find, 
\bq
R^0_0 &=& -\frac{1}{2}D +\frac{1} {4}K_\a^\b K_\b^\a =
\left(-c + \frac{1}{4}b_\a^\b b_\b^\a\right) +
\left(-3d + 2b_\a^\b c_\b^\a - \frac{1}{2}b_\a^\b b_\g^\a b_\b^\g\right)t \nonumber  \\
&+& \left(-6e +\frac{9}{2}b_\a^\b d_\b^\a -\frac{7}{2}b_\a^\b b_\g^\a c_\b^\g + \frac{3} {4}b_\g^\b
b_\a^\g b_\d^\a b_\b^\d + 2c_\a^\b c_\b^\a\right)t^2,
\label{eq:R00mixed}   
\eq
and,
\bq
R^0_\a &=& \frac{1}{2} (\nabla_\b K^\b_\a - \nabla_\a K)=
\frac{1}{2}\left(\nabla_\b b^\b_\a - \nabla_\a b\right) +
\left(\nabla_\b c^\b_\a - \nabla_\a c -\frac{1} {2}\nabla_\b (b_\g^\b b_\a^\g) +
\frac{1}{2}\nabla_\a (b_\g^\b b_\b^\g)\right)t   \nonumber   \\
&+& \left[\frac{3}{2}(\nabla_\b d^\b_\a - \nabla_\a d) - \nabla_\b (b^\b_\g c^\g_\a) +
\frac{3}{2}\nabla_\a (b^\b_\g c^\g_\b) - \frac{1}{2}\nabla_\b (c^\b_\g b^\g_\a)
+\frac{1}{2}\nabla_\b (b^\b_\d b^\d_\g b^\g_\a) \right. \nonumber \\
&-& \left. \frac{1}{2}\nabla_\a (b^\d_\b b^\b_\g b^\g_\d)\right]t^2
\nonumber \\
&+& \left[2(\nabla_\b e^\b_\a -\nabla_\a e) -\frac{3}{2}\nabla_\b (b^\b_\g d^\g_\a)
+2\nabla_\a (b^\b_\g d^\g_\b) -\frac{1}{2}\nabla_\b (d^\b_\g b^\g_\a) -
\nabla_\b (c^\b_\g c^\g_\a) +\nabla_\a (c^\b_\g c^\g_\b)    \right.\nonumber \\
&+& \left.\nabla_\b (b^\b_\d b^\d_\g c^\g_\a)
+\frac{1}{2}\nabla_\b (b^\b_\d c^\d_\g b^\g_\a)
+\frac{1}{2}\nabla_\b (c^\b_\d b^\d_\g b^\g_\a) -2\nabla_\a (b^\d_\b c^\b_\g b^\g_\d) \right.\nonumber\\
&-&\left.\frac{1}{2}\nabla_\b (b^\b_\d b^\d_\ep b^\ep_\g b^\g_\a) +
\frac{1}{2}\nabla_\a (b^\d_\b b^\b_\g b^\g_\ep b^\ep_\d)\right] t^3,
\label{eq:R0amixed}
\eq
and also,
\bq
R^\b_\a &=&
-P^\b_\a -\frac{1} {4} KK^\b_\a
-\frac{1} {2}D^\b_\a +\frac{1}{2}K^\b_\g K^\g_\a= \left(-(P^\b_\a)^{(0)}
-c^\b_\a + \frac{1} {2}b^\b_\g b^\g_\a -\frac{1} {4}b^\b_\a b\right) \nonumber \\
&+& \left(-3d^\b_\a + 2b^\b_\g c^\g_\a - b^\b_\d b^\d_\g b^\g_\a + c^\b_\g b^\g_\a -
\frac{1}{2}b^\b_\a c +\frac{1} {4}b^\d_\g b^\g_\d b^\b_\a -
\frac{1}{2}c^\b_\a b +\frac{1} {4}b^\b_\g b^\g_\a b-(P^\b_\a)^{(1)}\right)t    \nonumber     \\
&+& \left(-6e^\b_\a +\frac{9} {2}b^\b_\g d^\g_\a -3b^\b_\d b^\d_\g c^\g_\a +3c^\b_\g c^\g_\a
 +\frac{3} {2}d^\b_\g b^\g_\a -\frac{3} {2}b^\b_\d c^\d_\g b^\g_\a +\frac{3} {2}b^\b_\d b^\d_\ep
 b^\ep_\g b^\g_\a -\frac{3} {2}c^\b_\d b^\d_\g b^\g_\a \right.   \nonumber  \\
&-& \frac{3} {4}d^\b_\a b +\frac{1} {2}b^\b_\g
 c^\g_\a b -\frac{1} {4}b^\b_\d b^\d_\g b^\g_\a b +\frac{1} {4}c^\b_\g b^\g_\a b -c^\b_\a c +
 \frac{1} {2}c^\b_\a b^\d_\g b^\g_\d +\frac{1} {2}b^\b_\g b^\g_\a c
 -\frac{1} {4}b^\b_\g b^\g_\a b^\d_\ep b^\ep_\d                \nonumber    \\
&-& \left.\frac{3} {4}b^\b_\a d +\frac{3} {4}b^\b_\a b^\d_\g c^\g_\d -
 \frac{1} {4}b^\b_\a b^\d_\ep b^\ep_\g b^\g_\d-(P^\b_\a)^{(2)}\right) t^2,
\label{eq:Rabmixed}
\eq
where $P_{\a\b}$ is the Ricci tensor associated with $\g_{\a\b}$. In fact, starting with the expression of the Ricci tensor,
\be
R_{ij} = R^l_{ilj} = \pa_l \G^l_{ij} - \pa_j \G^l_{il} + \G^l_{kl}\G^k_{ij} - \G^l_{kj}\G^k_{il},
\label{Ricci tensor, Riemann}
\ee
we find that the expression of the three-dimensional Ricci tensor is,
\be
P_{\a\b} = \pa_\m \G^\m_{\a\b} - \pa_\b \G^\m_{\a\m} + \G^\m_{\a\b}\G^\ep_{\m\ep} - \G^\m_{\a\ep}\G^\ep_{\b\m}.
\label{three-dimensional Ricci tensor}
\ee 
Due to Eqns. (\ref{eq:3dimmetric}) and (\ref{eq:3diminvmetric}), we find the form of the Christoffel symbols of the second kind, which is,
\be
\G^\m_{\a\b} = (\G^\m_{\a\b})^{(0)} + t (\G^\m_{\a\b})^{(1)} + t^2 (\G^\m_{\a\b})^{(2)} + t^3 (\G^\m_{\a\b})^{(3)} + t^4 (\G^\m_{\a\b})^{(4)}.
\label{Cskreg}
\ee 
Therefore, the form of the three-dimensional Ricci tensor and the corresponding mixed three-dimensional Ricci tensor becomes,
\be
P_{\a\b} = (P_{\a\b})^{(0)} + t (P_{\a\b})^{(1)} + t^2 (P_{\a\b})^{(2)} + t^3 (P_{\a\b})^{(3)} + t^4 (P_{\a\b})^{(4)},
\label{three-dimensional Ricci tensor regular}
\ee
and,
\be
P^\b_\a = \g^{\b\m}P_{\m\a} = (P^\b_\a)^{(0)} + t (P^\b_\a)^{(1)} + t^2 (P^\b_\a)^{(2)} + t^3 (P^\b_\a)^{(3)} + t^4 (P^\b_\a)^{(4)}.
\label{mixed three-dimensional Ricci tensor regular}
\ee
We also find,
\be
P = \g^{\a\b}P_{\a\b} = P^{(0)} + t P^{(1)} + t^2 P^{(2)} + t^3 P^{(3)} + t^4 P^{(4)}.
\label{mixed three-dimensional Ricci tensor regular}
\ee
Then, the scalar curvature becomes an expression of the form,
\be
R = R^{(0)} + R^{(1)} t + R^{(2)} t^2,
\label{eq:Rshort}
\ee
and explicitly we have,
\bq
R&=&-P -\frac{1}{4}K^2 +\frac{3}{4}K^\b_\a K^\a_\b -D \nonumber\\
&=& \left(-P^{(0)} -2c +\frac{3}{4}b^\b_\a b^\a_\b -\frac{1} {4}b^2\right) +
\left(-bc -6d +5b^\b_\a c^\a_\b -\frac{3}{2}b^\b_\a b^\a_\g b^\g_\b +
\frac{1}{2}b^\b_\a b^\a_\b b-P^{(1)}\right)t   \nonumber  \\
&+& \left(-\frac{3}{2}bd -c^2 -12e +\frac{21}{2}b^\b_\a d^\a_\b +5c^\b_\a c^\a_\b -
\frac{19}{2}b^\b_\a b^\a_\g c^\g_\b +\frac{9}{4}b^\b_\g b^\g_\a b^\a_\d b^\d_\b +
\frac{3} {2}b^\b_\a c^\a_\b b\right.     \nonumber  \\
&+& \left. b^\b_\a b^\a_\b c -\frac{1} {2}b^\g_\b b^\b_\a b^\a_\g b -
    \frac{1} {4}(b^\b_\a b^\a_\b)^2-P^{(2)}\right) t^2.
\label{eq:trR}
\eq


\subsection{Constraints and snap equation}
\label{Constraints and snap equation}

In terms of $a_{\a\b},b_{\a\b},c_{\a\b},d_{\a\b},e_{\a\b}$, the hamiltonian constraint (\ref{eq:hamiltonian}) becomes,
\bq
\mathcal{C}_0 &=& \frac{1}{2}P^{(0)} -\frac{1}{8}b^\b_\a b^\a_\b +\frac{1}{8}b^2 
+\ep \left \{2(-P^{(0)} -2c +\frac{3}{4}b^\b_\a b^\a_\b -\frac{1} {4}b^2)(-c + \frac{1}{4}b^\d_\g b^\g_\d) \right. \nonumber \\ &-& \left. \frac{1}{2}(-P^{(0)} -2c +\frac{3}{4}b^\b_\a b^\a_\b -\frac{1} {4}b^2)^2 \right. \nonumber \\
&+& \left. b(-bc -6d +5b^\b_\a c^\a_\b -\frac{3}{2}b^\b_\a b^\a_\g b^\g_\b +\frac{1}{2}b^\b_\a b^\a_\b b -P^{(1)}) \right. \nonumber \\
&+& \left. a^{\a\b}\left[-2\partial_\a \left( \partial_\b(-P^{(0)} -2c +\frac{3}{4}b^\d_\g b^\g_\d -\frac{1} {4}b^2) \right) \right. \right. \nonumber \\
&+& \left. \left. a^{\m\ep}A_{\a\b\ep}\partial_\m(-P^{(0)} -2c +\frac{3}{4}b^\d_\g b^\g_\d -\frac{1} {4}b^2)\right]\right \} \nonumber \\
&+& t\left \{ \frac{1}{2}P^{(1)}-\frac{1}{4}b^\b_\a b^\a_\b b+\frac{1}{4}b^\b_\a b^\a_\g b^\g_\b -\frac{1}{2}b^\b_\a c^\a_\b +\frac{1}{2}bc \right. \nonumber \\
&+&\left. \ep \left \{ 2(-P^{(0)} -2c +\frac{3}{4}b^\ep_\d b^\d_\ep -\frac{1} {4}b^2)(-3d + 2b_\a^\b c_\b^\a - \frac{1}{2}b_\a^\b b_\g^\a b_\b^\g) \right. \right. \nonumber \\
&+& \left. \left. 2(-bc -6d +5b^\b_\a c^\a_\b -\frac{3}{2}b^\b_\a b^\a_\g b^\g_\b +
\frac{1}{2}b^\b_\a b^\a_\b b-P^{(1)})(-c + \frac{1}{4}b^\ep_\d b^\d_\ep) \right. \right. \nonumber \\
&-& \left. \left. (-P^{(0)} -2c +\frac{3}{4}b^\ep_\d b^\d_\ep -\frac{1} {4}b^2)(-bc -6d +5b^\b_\a c^\a_\b -\frac{3}{2}b^\b_\a b^\a_\g b^\g_\b + \frac{1}{2}b^\b_\a b^\a_\b b-P^{(1)}) \right. \right. \nonumber \\
&+& \left. \left. 2c(-bc -6d +5b^\b_\a c^\a_\b -\frac{3}{2}b^\b_\a b^\a_\g b^\g_\b + \frac{1}{2}b^\b_\a b^\a_\b b-P^{(1)}) \right. \right. \nonumber \\
&+& \left. \left. 2b(-\frac{3}{2}bd -c^2 -12e +\frac{21}{2}b^\b_\a d^\a_\b +5c^\b_\a c^\a_\b - \frac{19}{2}b^\b_\a b^\a_\g c^\g_\b +\frac{9}{4}b^\b_\g b^\g_\a b^\a_\d b^\d_\b + \frac{3} {2}b^\b_\a c^\a_\b b \right. \right.    \nonumber  \\
&+& \left. \left. b^\b_\a b^\a_\b c -\frac{1} {2}b^\g_\b b^\b_\a b^\a_\g b - \frac{1} {4}(b^\b_\a b^\a_\b)^2-P^{(2)}) \right. \right. \nonumber \\
&+& \left. \left. a^{\a\b} \left[ -2\partial_\a \left( \partial_\b(-bc -6d +5b^\d_\g c^\g_\d -\frac{3}{2}b^\ep_\d b^\d_\g b^\g_\ep + \frac{1}{2}b^\d_\g b^\g_\d b-P^{(1)})\right) \right. \right. \right. \nonumber \\
&+& \left. \left. \left. 2E^\m_{\a\b} \pa_\m (-P^{(0)} -2c +\frac{3}{4}b^\d_\g b^\g_\d -\frac{1} {4}b^2) \right. \right. \right. \nonumber \\
&+& \left. \left. \left. a^{\m\ep}A_{\a\b\ep} \pa_\m (-bc -6d +5b^\d_\g c^\g_\d -\frac{3}{2}b^\d_\g b^\g_\z b^\z_\d +
\frac{1}{2}b^\d_\g b^\g_\d b-P^{(1)}) \right] \right. \right. \nonumber \\
&+& \left. \left. b^{\a\b} \left[ 2\partial_\a \left( \partial_\b(-P^{(0)} -2c +\frac{3}{4}b^\d_\g b^\g_\d -\frac{1} {4}b^2) \right) \right. \right. \right. \nonumber \\
&-& \left. \left. \left. b_{\a\b}(-bc -6d +5b^\d_\g c^\g_\d -\frac{3}{2}b^\ep_\d b^\d_\g b^\g_\ep + \frac{1}{2}b^\d_\g b^\g_\d b-P^{(1)}) \right. \right. \right. \nonumber \\
&-& \left. \left. \left. a^{\m\ep}A_{\a\b\ep} \pa_\m (-P^{(0)} -2c +\frac{3}{4}b^\d_\g b^\g_\d -\frac{1} {4}b^2) \right] \right \} \right \} \nonumber \\
&=& 0.
\label{eq:L00mixed}
\eq
From (\ref{eq:momentum}), we calculate the momentum constraints in the form,
\bq
\mathcal{C}_\a &=& \frac{1}{2}(\nabla_\b b^\b_\a-\nabla_\a b) 
+\ep \left[(-P_0 -2c +\frac{3}{4}b^\b_\a b^\a_\b -\frac{1} {4}b^2)(\nabla_\b b^\b_\a-\nabla_\a b) \right. \nonumber \\
&-& \left. 2\partial_\a(-bc -6d +5b^\b_\a c^\a_\b -\frac{3}{2}b^\b_\a b^\a_\g b^\g_\b 
+\frac{1}{2}b^\b_\a b^\a_\b b -P_1) \right. \nonumber \\
&+& \left. b^\b_\a \partial_\b(-P_0 -2c +\frac{3}{4}b^\b_\a b^\a_\b -\frac{1} {4}b^2) \right] \nonumber \\
&+& t \left \{ (\nabla_\b c^\b_\a - \nabla_\a c) -\frac{1} {2}\nabla_\b (b_\g^\b b_\a^\g) 
+ \frac{1} {2}\nabla_\a (b_\g^\b b_\b^\g) \right.  \nonumber \\
&+& \left. \ep \left \{ 2(-P^{(0)} -2c +\frac{3}{4}b^\ep_\d b^\d_\ep -\frac{1} {4}b^2)\left[(\nabla_\b c^\b_\a - \nabla_\a c) 
-\frac{1} {2}\nabla_\b (b_\g^\b b_\a^\g) + \frac{1} {2}\nabla_\a (b_\g^\b b_\b^\g)\right] \right. \right. \nonumber \\
&+& \left. \left. (-bc -6d +5b^\d_\g c^\g_\d -\frac{3}{2}b^\ep_\d b^\d_\g b^\g_\ep +
\frac{1}{2}b^\d_\g b^\g_\d b-P^{(1)})(\nabla_\b b^\b_\a - \nabla_\a b)  \right. \right. \nonumber \\
&-& \left. \left. 4\partial_\a (-\frac{3}{2}bd -c^2 -12e +\frac{21}{2}b^\g_\b d^\b_\g +5c^\g_\b c^\b_\g -
\frac{19}{2}b^\d_\g b^\g_\b c^\b_\d +\frac{9}{4}b^\ep_\d b^\d_\g b^\g_\b b^\b_\ep + \frac{3} {2}b^\g_\b c^\b_\g b \right. \right. \nonumber  \\
&+& \left. \left. b^\g_\b b^\b_\g c -\frac{1} {2}b^\d_\g b^\g_\b b^\b_\d b - \frac{1} {4}(b^\g_\b b^\b_\g)^2-P^{(2)}) \right. \right. \nonumber \\
&+& \left. \left. b^\b_\a \partial_\b(-bc -6d +5b^\d_\g c^\g_\d -\frac{3}{2}b^\ep_\d b^\d_\g b^\g_\ep + \frac{1}{2}b^\d_\g b^\g_\d b-P^{(1)}) \right. \right. \nonumber \\
&+& \left. \left. (2c^\b_\a -b^\b_\g b^\g_\a) \partial_\b (-P^{(0)} -2c +\frac{3}{4}b^\ep_\d b^\d_\ep -\frac{1} {4}b^2) \right \} \right \} \nonumber \\
&=& 0.
\label{eq:L0amixed}
\eq
Finally, the snap equation (\ref{eq:pW}) gives,
\bq
&&\frac{1}{2}P^{(0)}+2c-\frac{5}{8}b^\b_\a b^\a_\b+\frac{1}{8}b^2 \nonumber \\
&+&\ep\left\{2[-P^{(0)}+(-2c+\frac{3}{4}b^\d_\g b^\g_\d-\frac{1}{4}b^2)][-P^{(0)}+(-c+\frac{1}{2}b^\b_\a b^\a_\b-\frac{1}{4}b^2)]\right.\nonumber\\
&-&\left. \frac{3}{2}[-P^{(0)}+(-2c+\frac{3}{4}b^\d_\g b^\g_\d-\frac{1}{4}b^2)]^2 +6[-2P^{(2)}-3bd-2c^2-24e+21b^\g_\d d^\d_\g \right. \nonumber\\
&+& \left.10c^\g_\d c^\d_\g-19b^\ep_\d b^\d_\g c^\g_\ep+\frac{9}{2}b^\d_\g b^\g_\ep b^\ep_\zeta b^\zeta_\d+3b^\g_\d c^\d_\g b +2b^\g_\d b^\d_\g c-b^\g_\d b^\d_\ep b^\ep_\g b-\frac{1}{2}(b^\g_\d b^\d_\g)^2]\right.\nonumber\\
&+&\left.4a^{\a\b}[\partial_\a \left(\partial_\b (-P^{(0)} -2c +\frac{3}{4}b^\d_\g b^\g_\d
-\frac{1} {4}b^2)\right) \right. \nonumber \\
&-& \left. \frac{1}{2}b_{\a\b}(-bc -6d +5b^\b_\a c^\a_\b -\frac{3}{2}b^\b_\a b^\a_\g b^\g_\b +
\frac{1}{2}b^\b_\a b^\a_\b b-P^{(1)}) \right. \nonumber \\
&-& \left. (\G^\m_{\a\b})^{(0)}\partial_\m (-P^{(0)} -2c +\frac{3}{4}b^\d_\g b^\g_\d
-\frac{1} {4}b^2)]\right\} \nonumber \\
&=& 0.
\label{eq:trL}
\eq


\section{Perturbative analysis with a singular formal series representation of the spatial metric}
\label{Perturbative analysis with a singular formal series representation of the spatial metric}

Similarly to the previous Section, taking into account that the scale factor $a$ given by Eqns. (\ref{sol rad GR k=-1}), (\ref{sol rad GR k=0}), (\ref{sol rad GR k=+1}) is an exact solution of the generalized Friedmann equations (\ref{gen Friedmann eq density}) and (\ref{gen Friedmann eq pressure}) for the corresponding $k$ each time, we are also going to use the Landau-Lifschitz pertubative method in higher order gravity, in the vicinity of o point that is not regular, but singular in the time. In particular, we note that the exact solutions of the generalized Friedmann equations in cases of $k=-1,0,+1$, lead us to the fact that the dependence of the scale factor $a=a(t)$ when $t\rightarrow 0$ goes over the $k=0$ solution, and consequently we have,
\be
a\propto t^{1/2}.
\label{a prop t^1/2}
\ee
Thus, in accordance with (\ref{a prop t^1/2}), we assume a formal series representation of the spatial metric of the form:
\bq
\g_{\a\b} &=& \g^{(1)}_{\a\b}t + \g^{(2)}_{\a\b}t^2 + \g^{(3)}_{\a\b}t^3 + \g^{(4)}_{\a\b}t^4 + \cdots \nonumber \\
          &=& a_{\a\b}t + b_{\a\b}t^2 + c_{\a\b}t^3 + d_{\a\b}t^4 + \cdots,
\label{spatial metric rad}
\eq 
where the $\g^{(1)}_{\a\b}=a_{\a\b} , \g^{(2)}_{\a\b}=b_{\a\b} , \g^{(3)}_{\a\b}=c_{\a\b} , \g^{(4)}_{\a\b}=d_{\a\b},\cdots$ are functions of the space coordinates.


\subsection{Singular formal series representation}
\label{Singular formal series representation}

Taking into consideration the order of the higher order gravity equations in cases of vacuum and radiation that we have dealt with in Chapter \ref{Chapter2}, we shall be interested too in the part of the formal series shown, that is up to order four. Because of this, we shall also drop the dots at the end of the various expressions in order to simplify the overall appearance. Certainly, in the next two Chapters, we also prove that, even if the order of the formal series is greater than four, then the results are qualitatively the same as that of the series (\ref{spatial metric rad}). Thus before substitution to the evolution and constraint higher order equations, we note that the expression  (\ref{spatial metric rad}) contains  $24$ degrees of freedom ($6$ of each spatial matrix $\g^{(n)}_{\a\b},n=1,\cdots,4$).

We note that the data $a_{\a\b} ,b_{\a\b} ,c_{\a\b} ,d_{\a\b}$, are arbitrary, nontrivial analytic functions of the space coordinates.

To proceed, we find the expression of the formal expansion of the reciprocal tensor $\g^{\a\b}$ using the identity (\ref{eq:identity of gamma_ab}). Then the various coefficients $ \g^{(\mu)\,\a\b},\mu=1,\cdots,4,$ of $t$ in the expansion
$\g^{\a\b} = \sum_{n=-1}^{\infty}(\g^{\a\b})^{(n)} t^n$ are found to be:
\be
\g^{\a\b} = \dfrac{1}{t}a^{\a\b} - b^{\a\b} + t(-c^{\a\b} + b^{\a\g}b^\b_\g) +
t^2(-d^{\a\b} + b^{\a\g}c^\b_\g - b^{\a\g}b^\d_\g b^\b_\d + c^{\a\g}b^\b_\g). 
\label{eq:3diminvmetric_rad}
\ee
Note that $a_{\a\b}a^{\b\g}=\d_\a^\g$ and the indices of $b_{\a\b},c_{\a\b},d_{\a\b}$ are raised by $a^{\a\b}$.
For any tensor $X$, using the formal expansion  ($\ref{spatial metric rad}$), we can recursively calculate the coefficients in the expansion
\be
X_{\a\b} = \sum X^{(n)}_{\a\b}\;t^n,
\label{X tensor}
\ee
where the values of $n$ depend on the tensor.


\subsection{Singular formal series representation of $K_{\a\b},D_{\a\b},W_{\a\b}$}
\label{Singular formal series representation of K,D,W}

In terms of the data $a_{\a\b},b_{\a\b},c_{\a\b},d_{\a\b}$, we have explicitly,
\be
K_{\a\b} = \partial_t \g_{\a\b}= a_{\a\b} + 2tb_{\a\b} + 3t^2c_{\a\b} + 4t^3d_{\a\b},
\label{eq:Kab_rad}
\ee
and for the mixed components we obtain,
\be
K_\b^\a = \dfrac{1}{t}\d^\b_\a + b^\b_\a + t(2c^\b_\a - b^\b_\g b^\g_\a) +
t^2(3d^\b_\a - 2b^\b_\g c^\g_\a - c^\b_\g b^\g_\a + b^\b_\g b^\g_\d b^\d_\a).
\label{eq:Kmixed_rad}
\ee
Additionally, taking into account the third inequality of (\ref{eq:determinant_g_ab}), namely that,
\be
\g > 0,
\label{positive g_rad}
\ee
the mean curvature,
\be
K=K_\a^\a=\g^{\a\b} \partial_t\g_{\a\b}=\partial_t  ln(\g),
\label{mean curvature_rad}
\ee
is given by the form,
\be
K = \dfrac{3}{t} + b + t(2c - b^\b_\a b^\a_\b) + t^2(3d - 3b^\b_\a c^\a_\b + b^\b_\g b^\g_\a b^\a_\b),
\label{eq:trK_rad}
\ee
and for completeness, we also give the expressions of the coefficients $(K^{\a\b})^{(n)}$ of the fully contravariant symbols,
\be
K^{\a\b} = \dfrac{1}{t^2}a^{\a\b} + (c^{\a\b} - b^{\b\g}b^\a_\g) +
t(2d^{\a\b} - 2b^\b_\g c^{\g\a} - 2b^{\a\g}c^\b_\g + 2b^{\a\g}b^\d_\g b^\b_\d).
\label{eq:K^ab_rad}
\ee
Using these forms, we can find the various components of the acceleration and jerk tensors to the required order. We have that the perturbation of the acceleration tensor in terms of the prescribed data is given by the form:
\be
D_{\a\b} = \pa_t K_{\a\b} = 2b_{\a\b} + 6c_{\a\b}t + 12d_{\a\b}t^2.
\label{eq:Dab_rad}
\ee
Further, for the mixed components we find,
\be
D^\b_\a = \frac{2}{t}b^\b_\a + 2(3c^\b_\a -b^\b_\g b^\g_\a) +
2t(6d^\b_\a -3b^\b_\g c^\g_\a -c^\b_\g b^\g_\a 
+b^\b_\d b^\d_\g b^\g_\a),
\label{eq:Dmixed_rad}
\ee
and,
\be
D = \frac{2}{t}b + 2(3c -b^\b_\a b^\a_\b) + 2t(6d -4b^\b_\a c^\a_\b + b^\b_\g b^\g_\a b^\a_\b),
\label{eq:trD_rad}
\ee
where the trace is given by,
\be
D=D^\a_\a=\g^{\a\b} \pa_t K_{\a\b}.
\label{eq:trace_D_rad}
\ee
Also, for the fully contravariant components, we find,
\be
D^{\a\b} = \frac{2}{t^2}b^{\a\b} + \frac{2}{t}(3c^{\a\b} - 2b^\b_\g b^{\a\g}) +
2(6d^{\a\b} - 4b^\b_\g c^{\a\g} - 4b^\a_\g c^{\g\b} + 3b^{\a\g} b^\d_\g b^\b_\d).
\label{eq:D^ab_rad}
\ee
Lastly, the jerk perturbation series is found to be,
\be
W_{\a\b} = \pa_t D_{\a\b} = 6c_{\a\b} + 24d_{\a\b}t,
\label{eq:Wab}
\ee
so that,
\be
W = \dfrac{6}{t}c + 6(4d - b^\b_\a c^\a_\b),
\label{eq:trW_rad}
\ee
where,
\be
W=W^\a_\a=\g^{\a\b} \pa_t D_{\a\b}.
\label{eq:trace_W_rad}
\ee
So far, we have fully calculate all the terms that appear in higher-order gravity apart from the three-dimensional Ricci tensor $P_{\a\b}$ and it's trace $P=trP_{\a\b}$. 


\subsection{Singular formal series representation of $P_{\a\b}$}
\label{Singular formal series representation of K}

In order to determine $P_{\a\b}$, we start with the expression of the Ricci tensor, that is,
\be
R_{ij} = R^l_{ilj} = \pa_l \G^l_{ij} - \pa_j \G^l_{il} + \G^l_{kl}\G^k_{ij} - \G^l_{kj}\G^k_{il}.
\label{Ricci tensor}
\ee
Thus, the expression of the three-dimensional Ricci tensor is,
\be
P_{\a\b} = \pa_\m \G^\m_{\a\b} - \pa_\b \G^\m_{\a\m} + \G^\m_{\a\b}\G^\ep_{\m\ep} - \G^\m_{\a\ep}\G^\ep_{\b\m},
\label{three-dimensional Ricci tensor 2}
\ee
where the Christoffel symbols of the second kind,
\be
\G^\m_{\a\b} = \dfrac{1}{2}\g^{\m\ep}(\pa_\b \g_{\a\ep} + \pa_\a \g_{\b\ep} - \pa_\ep \g_{\a\b}),
\label{Christoffel symbols}
\ee
using the series (\ref{spatial metric rad}) become,
\bq
\G^\m_{\a\b} &=& \dfrac{1}{2}\big[\dfrac{1}{t}a^{\m\ep} - b^{\m\ep} + t(-c^{\m\ep} + b^{\m\g}b^\ep_\g) +
t^2(-d^{\m\ep} + b^{\m\g}c^\ep_\g + c^{\m\g}b^\ep_\g - b^{\m\g}b^\d_\g b^\ep_\d)\big] \nonumber \\
& \times & \left[t(\pa_\b a_{\a\ep} + \pa_\a a_{\b\ep} - \pa_\ep a_{\a\b}) +
t^2(\pa_\b b_{\a\ep} + \pa_\a b_{\b\ep} - \pa_\ep b_{\a\b}) \right. \nonumber \\
&+& \left. t^3(\pa_\b c_{\a\ep} + \pa_\a c_{\b\ep} - \pa_\ep c_{\a\b}) +
t^4(\pa_\b d_{\a\ep} + \pa_\a d_{\b\ep} - \pa_\ep d_{\a\b})\right].
\label{spatial Christoffel symbols}
\eq
In order to simplify our work, we set,
\be
A_{\a\b\ep} = \pa_\b a_{\a\ep} + \pa_\a a_{\b\ep} - \pa_\ep a_{\a\b},
\label{Aabe}
\ee
\be
B_{\a\b\ep} = \pa_\b b_{\a\ep} + \pa_\a b_{\b\ep} - \pa_\ep c_{\a\b},
\label{Babe}
\ee
\be
C_{\a\b\ep} = \pa_\b c_{\a\ep} + \pa_\a c_{\b\ep} - \pa_\ep c_{\a\b},
\label{Cabe}
\ee
\be
D_{\a\b\ep} = \pa_\b d_{\a\ep} + \pa_\a d_{\b\ep} - \pa_\ep d_{\a\b}.
\label{Dabe}
\ee
Therefore, relation (\ref{spatial Christoffel symbols}) becomes,
\be
\G^\m_{\a\b} = \dfrac{1}{2}(\dfrac{1}{t}a^{\m\ep} - b^{\m\ep} + t(\g^{\m\ep})^{(1)} + t^2(\g^{\m\ep})^{(2)})
\times(tA_{\a\b\ep} + t^2 B_{\a\b\ep} + t^3 C_{\a\b\ep} + t^4 D_{\a\b\ep}),
\label{spatial Christoffel symbols ABCD}
\ee
and so,
\bq
\G^\m_{\a\b} &=& \dfrac{1}{2}a^{\m\ep}A_{\a\b\ep} + \dfrac{1}{2}(a^{\m\ep}B_{\a\b\ep} - b^{\m\ep}A_{\a\b\ep})t
+ \dfrac{1}{2}(a^{\m\ep}C_{\a\b\ep} - b^{\m\ep}B_{\a\b\ep} + \g^{(1)\m\ep}A_{\a\b\ep})t^2 \nonumber \\
&+& \dfrac{1}{2}(a^{\m\ep}D_{\a\b\ep} - b^{\m\ep}C_{\a\b\ep} + (\g^{\m\ep})^{(1)}B_{\a\b\ep}
+ (\g^{\m\ep})^{(2)}A_{\a\b\ep})t^3. 
\label{spatial Christoffel symbols ABCD series}
\eq
We further set,
\be
\widetilde{\G}^\m_{\a\b} = \frac{1}{2}a^{\m\ep} A_{\a\b\ep},
\label{eq:Gtildemab}
\ee
\be
E^\m_{\a\b} = \frac{1}{2}(a^{\m\ep} B_{\a\b\ep} - b^{\m\ep} A_{\a\b\ep}),
\label{eq:Emab}
\ee
\be
F^\m_{\a\b} = \frac{1}{2}(a^{\m\ep} C_{\a\b\ep} - b^{\m\ep} B_{\a\b\ep} + e^{\m\ep} A_{\a\b\ep}),
\label{eq:Fmab}
\ee
\be
G^\m_{\a\b} = \frac{1}{2}(a^{\m\ep} D_{\a\b\ep} - b^{\m\ep} C_{\a\b\ep} + e^{\m\ep} B_{\a\b\ep} + f^{\m\ep} A_{\a\b\ep}),
\label{eq:Gmab}
\ee
and finally for the Christoffel symbols we obtain,
\be
\G^\m_{\a\b} = \widetilde{\G}^\m_{\a\b} + tE^\m_{\a\b} + t^2 F^\m_{\a\b} + t^3 G^\m_{\a\b}.
\label{finally Christoffel symbols}
\ee
Moreover, from (\ref{three-dimensional Ricci tensor 2}) using (\ref{finally Christoffel symbols}) we find,
\bq
P_{\a\b} &=& (\pa_\m \widetilde{\G}^\m_{\a\b} + t \pa_\m E^\m_{\a\b} + t^2 \pa_\m F^\m_{\a\b} + t^3 \pa_\m G^\m_{\a\b})
- (\pa_\b \widetilde{\G}^\m_{\a\m} + t \pa_\b E^\m_{\a\m} + t^2 \pa_\b F^\m_{\a\m} + t^3 \pa_\b G^\m_{\a\m}) \nonumber \\
&+& (\widetilde{\G}^\m_{\a\b} + tE^\m_{\a\b} + t^2 F^\m_{\a\b} + t^3 G^\m_{\a\b})
\times (\widetilde{\G}^\ep_{\m\ep} + tE^\ep_{\m\ep} + t^2 F^\ep_{\m\ep} + t^3 G^\ep_{\m\ep}) \nonumber \\
&-& (\widetilde{\G}^\m_{\a\ep} + tE^\m_{\a\ep} + t^2 F^\m_{\a\ep} + t^3 G^\m_{\a\ep})
\times (\widetilde{\G}^\ep_{\b\m} + tE^\ep_{\b\m} + t^2 F^\ep_{\b\m} + t^3 G^\ep_{\b\m}),
\label{three-dimensional Ricci tensor GEFG}
\eq
and further setting,
\be
\widetilde{P}_{\a_\b} = \pa_\m \widetilde{\G}^\m_{\a\b} - \pa_\b \widetilde{\G}^\m_{\a\m}
+ \widetilde{\G}^\m_{\a\b}\widetilde{\G}^\ep_{\m\ep} - \widetilde{\G}^\m_{\a\ep}\widetilde{\G}^\ep_{\b\m},
\label{eq:Ptildeab}
\ee
\be
H_{\a\b} = \pa_\m E^\m_{\a\b} - \pa_\b E^\m_{\a\m} + \widetilde{\G}^\m_{\a\b}E^\ep_{\m\ep}
+ \widetilde{\G}^\ep_{\m\ep}E^\m_{\a\b} -  \widetilde{\G}^\m_{\a\ep}E^\ep_{\b\m} - \widetilde{\G}^\ep_{\b\m}E^\m_{\a\ep},
\label{eq:Hab}
\ee
\be
I_{\a\b} = \pa_\m F^\m_{\a\b} - \pa_\b F^\m_{\a\m} + \widetilde{\G}^\m_{\a\b}F^\ep_{\m\ep}
+ \widetilde{\G}^\ep_{\m\ep}F^\m_{\a\b} - \widetilde{\G}^\m_{\a\ep}F^\ep_{\b\m} - \widetilde{\G}^\ep_{\b\m}F^\m_{\a\ep}
+ E^\m_{\a\b}E^\ep_{\m\ep} - E^\m_{\a\ep}E^\ep_{\b\m},
\label{eq:Iab}
\ee
\bq
J_{\a\b} &=& \pa_\m G^\m_{\a\b} - \pa_\b G^\m_{\a\m} +\widetilde{\G}^\m_{\a\b}G^\ep_{\m\ep}
+ \widetilde{\G}^\ep_{\m\ep}G^\m_{\a\b} - \widetilde{\G}^\m_{\a\ep}G^\ep_{\b\m} - \widetilde{\G}^\ep_{\b\m}G^\m_{\a\ep} 
\nonumber \\ 
&+& E^\m_{\a\b}F^\ep_{\m\ep} - E^\m_{\a\ep}F^\ep_{\b\m} + F^\m_{\a\b}E^\ep_{\m\ep} - F^\m_{\a\ep}E^\ep_{\b\m},
\label{eq:Jab}
\eq
we can write the three-dimensional Ricci tensor in the form,
\be
P_{\a\b} = \widetilde{P}_{\a\b} + tH_{\a\b} + t^2 I_{\a\b} + t^3 J_{\a\b}.
\label{finally three-dimensional Ricci tensor}
\ee
For the mixed three-dimensional Ricci tensor we have,
\be
P^\b_\a = \g^{\b\m}P_{\m\a} = (\dfrac{1}{t}a^{\b\m} - b^{\b\m} + t\g^{(1)\b\m} + t^2 \g^{(2)\b\m})
\times (\widetilde{P}_{\m_\a} + tH_{\m\a} + t^2 I_{\m\a} + t^3 J_{\m\a}),
\label{mixed Pab}
\ee
and setting,
\be
\k^\b_\a = a^{\b\m}H_{\m\a} - b^{\b\m}\widetilde{P}_{\m\a},
\label{eq:kab}
\ee
\be
\l^\b_\a = a^{\b\m}I_{\m\a} - b^{\b\m}H_{\m\a} + (\g^{\b\m})^{(1)}\widetilde{P}_{\m\a},
\label{eq:greekLab}
\ee
\be
\m^\b_\a = a^{\b\m}J_{\m\a} - b^{\b\m}I_{\m\a} + (\g^{\b\m})^{(1)}H_{\m\a} + (\g^{\b\m})^{(2)} \widetilde{P}_{\m\a}, 
\label{eq:Mab}
\ee
we find that the form of the mixed three-dimensional Ricci tensor is,
\be
P^\b_\a = \dfrac{1}{t}\widetilde{P}^\b_\a + \k^\b_\a + t\l^\b_\a + t^2 \m^\b_\a.
\label{final mixed Pab}
\ee
Finally, we obtain,
\be
P = P^\a_\a = \dfrac{1}{t}\widetilde{P} + \k + t\l + t^2 \m,
\label{final P}
\ee
where obviously,
\be
\k = \d^\a_\b \k^\b_\a,
\label{eq:k}
\ee
\be
\l = \d^\a_\b \l^\b_\a
\label{eq:greekL}
\ee
\be
\m = \d^\a_\b \m^\b_\a.
\label{eq:M}
\ee 
Using the above results, we can write the various components of the Ricci curvature and its trace expressed in terms of the asymptotic data $a_{\a\b},b_{\a\b},c_{\a\b},d_{\a\b}$. 


\subsection{Singular formal series representation of the Ricci tensor}
\label{Singular formal series representation of the Ricci tensor}

Using Eqns. (\ref{eq:R00mixeddownnew}), (\ref{eq:R0amixeddownnew}) and (\ref{eq:Rabmixeddownnew}) and the results of the previous Subsection concerning the quantities $P_{\a\b}$, the various components of the Ricci curvature become,
\be
R^0_0 = \frac{3}{4t^2} - \frac{1}{2t}b + (-2c +\frac{3}{4}b^\b_\a b^\a_\b) +
(-\frac{9}{2}d + \frac{7}{2}b^\b_\a c^\a_\b - b^\b_\g b^\g_\a b^\a_\b)t,
\label{eq:R00mixed_rad}
\ee
\bq
R^0_\a &=& \frac{1}{2}(\nabla_\b b^\b_\a - \nabla_\a b) + [(\nabla_\b c^\b_\a -\nabla_\a c) 
-\frac{1}{2}\nabla_\b (b^\b_\g b^\g_\a) + \frac{1}{2}\nabla_\a (b^\b_\g b^\g_\b)]t \nonumber \\
&+& \left[\frac{3}{2}(\nabla_\b d^\b_\a - \nabla_\a d) - \nabla_\b (b^\b_\g c^\g_\a) - \frac{1}{2}\nabla_\b (c^\b_\g b^\g_\a)
+ \frac{3}{2}\nabla_\a (b^\b_\g c^\g_\b) + \frac{1}{2}\nabla_\b(b^\b_\g b^\g_\d b^\d_\a) \right. \nonumber \\ 
&-& \left. \frac{1}{2}\nabla_\a(b^\b_\g b^\g_\d b^\d_\b) \right]t^2,
\label{eq:R0amixed_rad}
\eq
\bq
R^\b_\a &=& -\frac{1}{4t^2}\d^\b_\a - \frac{1}{4t}(4\widetilde{P}^\b_\a + 3b^\b_\a +b\d^\b_\a) 
+ (-\frac{5}{2}c^\b_\a + \frac{5}{4}b^\b_\g b^\g_\a - \frac{1}{4}bb^\b_\a - \frac{1}{2}c\d^\b_\a 
+ \frac{1}{4}b^\d_\g b^\g_\d \d^\b_\a - \k^\b_\a) \nonumber \\
&+& (-\frac{21}{4}d^\b_\a + \frac{7}{2}b^\b_\g c^\g_\a + \frac{7}{4}c^\b_\g b^\g_\a - \frac{7}{4}b^\b_\g b^\g_\d b^\d_\a
- \frac{1}{2}bc^\b_\a + \frac{1}{4}bb^\b_\g b^\g_\a - \frac{1}{2}cb^\b_\a 
+ \frac{1}{4}b^\d_\g b^\g_\d b^\b_\a - \frac{3}{4}d\d^\b_\a  \nonumber \\
&+& \frac{3}{4}b^\d_\g c^\g_\d \d^\b_\a - \frac{1}{4}b^\d_\g b^\g_\ep b^\ep_\d \d^\b_\a - \l^\b_\a)t,
\label{eq:Rabmixed_rad}
\eq
and the  scalar curvature becomes an expression of the form,
\be
R = \dfrac{1}{t}R^{(-1)} + R^{(0)}  + R^{(1)} t,
\label{eq:Rshortsin}
\ee
where explicitly we have,
\bq
R &=& -\frac{1}{t}(\widetilde{P} + 2b) + (-6c -\frac{1}{4}b^2 + \frac{11}{4}b^\b_\a b^\a_\b - \k) \nonumber \\
&+& (-12d -bc + 11b^\b_\a c^\a_\b - \frac{7}{2}b^\b_\g b^\g_\a b^\a_\b + \frac{1}{2}bb^\b_\a b^\a_\b - \l)t.
\label{Rscalar_rad}
\eq
 


\subsection{Constraints and snap equation}
\label{Constraints and snap equation2}

In terms of $a_{\a\b},b_{\a\b},c_{\a\b},d_{\a\b}$, the hamiltonian constraint (\ref{eq:hamiltonian_rad}) becomes,
\bq
\mathcal{C}_0 &=& \dfrac{1}{t^3}\left( \dfrac{3}{2}\ep(\widetilde{P} + 2b) \right) \nonumber \\
&+& \dfrac{1}{t^2} \left \{ \dfrac{3}{4} + \ep \left[-\dfrac{1}{2}(\widetilde{P}^2 - 4b^2)
+ \dfrac{3}{2}(-6c - \dfrac{1}{4}b^2 + \dfrac{11}{4}b^\b_\a b^\a_\b - \k)
 \right. \right. \nonumber \\
&+& \left. \left.  2a^{\a\b}\pa_\a \left(\pa_\b (\widetilde{P} + 2b)\right) -2a^{\a\b}\widetilde{\G}^\m_{\a\b}\pa_\m(\widetilde{P} + 2b) \right] \right \} \nonumber \\
&+& \dfrac{1}{t} \left \{ \dfrac{1}{2}(\widetilde{P}+b) + \ep \left[ (\widetilde{P} + 2b)(\dfrac{1}{4}b^\b_\a b^\a_\b - \dfrac{1}{4}b^2 - \k)
- b(-6c -\frac{1}{4}b^2 + \frac{11}{4}b^\b_\a b^\a_\b - \k) \right. \right. \nonumber \\
&+& \left. \left.\dfrac{9}{2}(-12d -bc + 11b^\b_\a c^\a_\b - \frac{7}{2}b^\b_\g b^\g_\a b^\a_\b + \frac{1}{2}bb^\b_\a b^\a_\b - \l)
- 2b^{\a\b}\pa_\a \left(\pa_\b(\widetilde{P}+2b)\right) \right. \right. \nonumber \\
&-& \left. \left. 2a^{\a\b}\pa_\a \left(\pa_\b (-6c -\frac{1}{4}b^2 + \frac{11}{4}b^\d_\g b^\g_\d - \k)\right)
+ 2b^{\a\b}\widetilde{\G}^\m_{\a\b}\pa_\m(\widetilde{P} + 2b) \right. \right. \nonumber \\
&+& \left. \left. 2a^{\a\b}\widetilde{\G}^\m_{\a\b}\pa_\m (-6c -\frac{1}{4}b^2 + \frac{11}{4}b^\d_\g b^\g_\d - \k)
- 2a^{\a\b}E^\m_{\a\b}\pa_\m (\widetilde{P} + 2b) \right] \right \} \nonumber \\
&=& 8\pi G T^0_0.
\label{C0rad all}
\eq
From (\ref{eq:momentum_rad}), the momentum constraints become,
\bq
\mathcal{C}_\a &=& \dfrac{1}{t^2}\left(-3\ep \pa_\a (\widetilde{P} + 2b)\right) \nonumber \\
&+& \dfrac{1}{t} \left \{ \ep \left[-(\widetilde{P} + 2b)(\nabla_\b b^\b_\a - \nabla_\a b)
+ \pa_\a (-6c -\frac{1}{4}b^2 + \frac{11}{4}b^\d_\g b^\g_\d - \k) \right. \right. \nonumber \\
&-& \left. \left. b^\b_\a \pa_\b (\widetilde{P} + 2b) \right] \right \} \nonumber \\
&+& t \left \{ \frac{1}{2}(\nabla_\b b^\b_\a -\nabla_\a b) + \ep \left[-2(\widetilde{P} + 2b)\big[(\nabla_\b c^\b_\a -\nabla_\a c) 
-\frac{1}{2}\nabla_\b (b^\b_\g b^\g_\a) +\frac{1}{2}\nabla_\a (b^\b_\g b^\g_\b)\big] \right. \right. \nonumber \\
&+& \left. \left. (-6c -\frac{1}{4}b^2 + \frac{11}{4}b^\d_\g b^\g_\d - \k)(\nabla_\b b^\b_\a -\nabla_\a b) \right. \right. \nonumber \\
&-& \left. \left. \pa_\a (-12d -bc + 11b^\d_\g c^\g_\d - \frac{7}{2}b^\b_\g b^\g_\d b^\d_\b + \frac{1}{2}bb^\d_\g b^\g_\d - \l)
\right. \right. \nonumber \\
&+& \left. \left. b^\b_\a \pa_\b (-6c -\frac{1}{4}b^2 + \frac{11}{4}b^\d_\g b^\g_\d - \k)
- (2c^\b_\a - b^\b_\g b^\g_\a)\pa_\b (\widetilde{P} + 2b) \right] \right \} \nonumber \\
&=& 8\pi G T^0_\a 
\label{Ca rad all}
\eq
Finally, the snap equation (\ref{eq:pW with T}) gives,
\bq
&& \dfrac{1}{t^3}\left(-\dfrac{9}{2}\ep (\widetilde{P} + 2b)\right) \nonumber \\
&+& \dfrac{1}{t^2} \left \{ -\dfrac{3}{4} + \ep \left[(\widetilde{P} + 2b)(\dfrac{1}{2}\widetilde{P} + 2b)
-\dfrac{3}{2} (-6c -\frac{1}{4}b^2 + \frac{11}{4}b^\b_\a b^\a_\b - \k) \right. \right. \nonumber \\
&-& \left. \left. 2a^{\a\b}\pa_\a \left(\pa_\b(\widetilde{P} + 2b)\right) + 2a^{\a\b}\widetilde{\G}^\m_{\a\b}\pa_\m(\widetilde{P} + 2b)
+6 a^{\a\b}\pa_\a \left(\pa_\b(\widetilde{P} + 2b)\right) \right. \right. \nonumber \\
&-& \left. \left. 6 a^{\a\b}\widetilde{\G}^\m_{\a\b}\pa_\m(\widetilde{P} + 2b) \right] \right \} \nonumber \\
&+& \dfrac{1}{t} \left \{ -\dfrac{1}{2}\widetilde{P} + \dfrac{3}{2}b 
+ \ep \left[\dfrac{9}{2}(-12d - bc + 11b^\b_\a c^\a_\b - \dfrac{7}{2}b^\g_\b b^\b_\a b^\a_\g
+ \dfrac{1}{2}bb^\b_\a b^\a_\b - \l) \right. \right. \nonumber \\
&+& \left. \left. (\widetilde{P} + 2b)(-6c + \dfrac{9}{4}b^\b_\a b^\a_\b - \dfrac{1}{4}b^2 - \k) \right. \right. \nonumber \\
&-& \left. \left. (2\widetilde{P} + 3b)(-6c - \dfrac{1}{4}b^2 + \dfrac{11}{4}b^\b_\a b^\a_\b - \k)
- 6 a^{\a\b}E^\m_{\a\b}\pa_\m (\widetilde{P} + 2b) \right. \right. \nonumber \\
&+& \left. \left. 2b^{\a\b}\pa_\a \left(\pa_\b(\widetilde{P} + 2b) \right)
- 2a^{\a\b}\widetilde{\G}^\m_{\a\b}\pa_\m (-6c - \dfrac{1}{4}b^2 + \dfrac{11}{4}b^\d_\g b^\g_\d - \k) \right. \right. \nonumber \\
&-& \left. \left. 2b^{\a\b}\widetilde{\G}^\m_{\a\b}\pa_\m (\widetilde{P} + 2b)
+ 2a^{\a\b}E^\m_{\a\b}\pa_\m (\widetilde{P} + 2b) \right. \right. \nonumber \\
&-& \left. \left. 6 a^{\a\b}\pa_\a \left(\pa_\b (-6c - \dfrac{1}{4}b^2 + \dfrac{11}{4}b^\d_\g b^\g_\d - \k) \right) \right. \right. \nonumber \\
&-& \left. \left. 6b^{\a\b}\pa_\a \left(\pa_\b (\widetilde{P} + 2b) \right)
+ 6 a^{\a\b}\widetilde{\G}^\m_{\a\b}\pa_\m (-6c - \dfrac{1}{4}b^2 + \dfrac{11}{4}b^\d_\g b^\g_\d - \k) \right. \right. \nonumber \\
&+& \left. \left. 6 b^{\a\b}\widetilde{\G}^\m_{\a\b}\pa_\m (\widetilde{P} + 2b)
+ 2a^{\a\b}\pa_\a \left(\pa_\b(-6c - \dfrac{1}{4}b^2 + \dfrac{11}{4}b^\d_\g b^\g_\d - \k) \right) \right] \right \} \nonumber \\
&=& 8\pi G T^\a_\a 
\label{snap rad all}
\eq
We are now ready to write our field equations in terms of the initial data $a_{\a\b},b_{\a\b},c_{\a\b},\\ d_{\a\b}$. This will allow us to count how many of the initial data are free. 

\chapter{Regular universes in higher-order gravity} 

\label{Chapter5} 

\lhead{Chapter 5. \emph{Regular universes in higher-order gravity}} 

In this Chapter, we study the form of the solution of the gravitational field equations of higher-order gravity in vacuum, as well as in the case of radiation, in the vicinity of a point that is regular in the time. In Section \ref{Analytic field equations through regular formal series expansion}, by using the $3$-metric (\ref{eq:3dimmetric}), we present the higher-order field equations and we calculate the various quantities of these equations in terms of the initial data $a_{\a\b},b_{\a\b},c_{\a\b},d_{\a\b},e_{\a\b}$. In Section \ref{Regular vacuum field equations}, by using (\ref{eq:3dimmetric}), we study the form of the solution of higher-order gravity equations in vacuum. In Section \ref{Regular field equations in case of radiation}, we also use the same {3}-metric to search for the form of the solution in the case of radiation. Then, in Section \ref{Degrees of freedom general}, we show that the degrees of freedom in both cases remain the same, if the order of the regular {3}-metric becomes greater than $4$. Finally, in Section \ref{Choice of the initial data}, we present the choice of the initial data between the initial functions of the two cases, taking into account the corresponding function-counting problems of Chapter \ref{Chapter2}. 


\section{Analytic field equations through regular formal series expansion}
\label{Analytic field equations through regular formal series expansion}

In this Section we present the vacuum field equations of higher-order gravity theory derived from the analytic lagrangian $f(R)=R+\ep R^2$, that is the Eqns. (\ref{eq:Loo}), (\ref{eq:Loa}) and (\ref{eq:Lab}), through regular formal series expansion (\ref{eq:3dimmetric}) and find the form of their solution. We also present the corresponding field equations in the case of radiation by using the same series.


\subsection{Series form of field equations}
\label{Series form of field equations}

Taking into consideration the order of the higher-order differential equations in vacuum, namely that Eqns. (\ref{eq:Loo}), (\ref{eq:Loa}) are third-order differential equations with respect to the proper time $t$ and Eq. (\ref{eq:Lab}) is fourth-order differential equation with respect to $t$ (see Section \ref{Synchronous equations}), we find it convenient to use the following notation:
\be
\mathcal{C}_0    =  (L^0_0)^{(0)} + t(L^0_0)^{(1)},  
\label{eq:C0}
\ee
\be
\mathcal{C}_\a   =  (L^0_\a)^{(0)} + t(L^0_\a)^{(1)}, 
\label{eq:Ca}
\ee
and,
\be
L^\b_\a  =  (L^\b_\a)^{(0)}, 
\label{eq:LAB}
\ee
in order to simplify our further work. We also use the same notation in the case of a fluid, because the order of the differential equations (\ref{eq:Loo_rad}), (\ref{eq:Loa_rad}) and (\ref{eq:Lab_rad}) is correspondingly the same with the order of vacuum differential equations. Naturally, the five terms $(L^0_0)^{(0)},(L^0_0)^{(1)},(L^0_\a)^{(0)},(L^0_\a)^{(1)},(L^\b_\a)^{(0)}$ appearing in (\ref{eq:C0}), (\ref{eq:Ca}) and (\ref{eq:LAB}) are not the same in both cases. However, they have the same dependence on the matrices $a_{\a\b},b_{\a\b},c_{\a\b},d_{\a\b},e_{\a\b}$. For this reason, we present the dependence for both cases once.


\subsection{Calculation of the series terms}
\label{Calculation of the series terms}

In terms of $a_{\a\b},b_{\a\b},c_{\a\b},d_{\a\b},e_{\a\b}$, the quantities appearing in these last three equations, by using the relations (\ref{eq:R00mixed}), (\ref{eq:R0amixed}), (\ref{eq:Rabmixed}) and (\ref{eq:trR}), become,
\bq
(L^0_0)^{(0)}&=&\frac{1}{2}P^{(0)} -\frac{1}{8}b^\b_\a b^\a_\b +\frac{1}{8}b^2 
+\ep \left \{2(-P^{(0)} -2c +\frac{3}{4}b^\b_\a b^\a_\b -\frac{1} {4}b^2)(-c + \frac{1}{4}b^\d_\g b^\g_\d) \right. \nonumber \\ &-& \left. \frac{1}{2}(-P^{(0)} -2c +\frac{3}{4}b^\b_\a b^\a_\b -\frac{1} {4}b^2)^2 \right. \nonumber \\
&+& \left. b(-bc -6d +5b^\b_\a c^\a_\b -\frac{3}{2}b^\b_\a b^\a_\g b^\g_\b +\frac{1}{2}b^\b_\a b^\a_\b b -P^{(1)}) \right. \nonumber \\
&+& \left. a^{\a\b}\left[-2\partial_\a \left( \partial_\b(-P^{(0)} -2c +\frac{3}{4}b^\d_\g b^\g_\d -\frac{1} {4}b^2) \right) \right. \right. \nonumber \\
&+& \left. \left. a^{\m\ep}A_{\a\b\ep}\partial_\m(-P^{(0)} -2c +\frac{3}{4}b^\d_\g b^\g_\d -\frac{1} {4}b^2)\right]\right \},
\label{analzeroL00mixed}
\eq
\bq
(L^0_0)^{(1)} &=& \frac{1}{2}P^{(1)}-\frac{1}{4}b^\b_\a b^\a_\b b+\frac{1}{4}b^\b_\a b^\a_\g b^\g_\b -\frac{1}{2}b^\b_\a c^\a_\b +\frac{1}{2}bc \nonumber \\
&+& \ep \left \{ 2(-P^{(0)} -2c +\frac{3}{4}b^\ep_\d b^\d_\ep -\frac{1} {4}b^2)(-3d + 2b_\a^\b c_\b^\a - \frac{1}{2}b_\a^\b b_\g^\a b_\b^\g) \right. \nonumber \\
&+& \left. 2(-bc -6d +5b^\b_\a c^\a_\b -\frac{3}{2}b^\b_\a b^\a_\g b^\g_\b +
\frac{1}{2}b^\b_\a b^\a_\b b-P^{(1)})(-c + \frac{1}{4}b^\ep_\d b^\d_\ep) \right. \nonumber \\
&-& \left. (-P^{(0)} -2c +\frac{3}{4}b^\ep_\d b^\d_\ep -\frac{1} {4}b^2)(-bc -6d +5b^\b_\a c^\a_\b -\frac{3}{2}b^\b_\a b^\a_\g b^\g_\b + \frac{1}{2}b^\b_\a b^\a_\b b-P^{(1)}) \right. \nonumber \\
&+& \left. 2c(-bc -6d +5b^\b_\a c^\a_\b -\frac{3}{2}b^\b_\a b^\a_\g b^\g_\b + \frac{1}{2}b^\b_\a b^\a_\b b-P^{(1)}) \right. \nonumber \\
&+& \left. 2b(-\frac{3}{2}bd -c^2 -12e +\frac{21}{2}b^\b_\a d^\a_\b +5c^\b_\a c^\a_\b - \frac{19}{2}b^\b_\a b^\a_\g c^\g_\b +\frac{9}{4}b^\b_\g b^\g_\a b^\a_\d b^\d_\b + \frac{3} {2}b^\b_\a c^\a_\b b \right.    \nonumber  \\
&+& \left. b^\b_\a b^\a_\b c -\frac{1} {2}b^\g_\b b^\b_\a b^\a_\g b - \frac{1} {4}(b^\b_\a b^\a_\b)^2-P^{(2)}) \right. \nonumber \\
&+& \left. a^{\a\b} \left[ -2\partial_\a \left( \partial_\b(-bc -6d +5b^\d_\g c^\g_\d -\frac{3}{2}b^\ep_\d b^\d_\g b^\g_\ep + \frac{1}{2}b^\d_\g b^\g_\d b-P^{(1)})\right) \right. \right. \nonumber \\
&+& \left. \left. 2E^\m_{\a\b} \pa_\m (-P^{(0)} -2c +\frac{3}{4}b^\d_\g b^\g_\d -\frac{1} {4}b^2) \right. \right. \nonumber \\
&+& \left. \left. a^{\m\ep}A_{\a\b\ep} \pa_\m (-bc -6d +5b^\d_\g c^\g_\d -\frac{3}{2}b^\d_\g b^\g_\z b^\z_\d +
\frac{1}{2}b^\d_\g b^\g_\d b-P^{(1)}) \right] \right. \nonumber \\
&+& \left. b^{\a\b} \left[ 2\partial_\a \left( \partial_\b(-P^{(0)} -2c +\frac{3}{4}b^\d_\g b^\g_\d -\frac{1} {4}b^2) \right) \right. \right. \nonumber \\
&-& \left. \left. b_{\a\b}(-bc -6d +5b^\d_\g c^\g_\d -\frac{3}{2}b^\ep_\d b^\d_\g b^\g_\ep + \frac{1}{2}b^\d_\g b^\g_\d b-P^{(1)}) \right. \right. \nonumber \\
&-& \left. \left. a^{\m\ep}A_{\a\b\ep} \pa_\m (-P^{(0)} -2c +\frac{3}{4}b^\d_\g b^\g_\d -\frac{1} {4}b^2) \right] \right \},
\label{analoneL00mixed}
\eq  
and,
\bq
(L^0_\a)^{(0)} &=& \frac{1}{2}(\nabla_\b b^\b_\a-\nabla_\a b) 
+\ep \left[(-P_0 -2c +\frac{3}{4}b^\b_\a b^\a_\b -\frac{1} {4}b^2)(\nabla_\b b^\b_\a-\nabla_\a b) \right. \nonumber \\
&-& \left. 2\partial_\a(-bc -6d +5b^\b_\a c^\a_\b -\frac{3}{2}b^\b_\a b^\a_\g b^\g_\b 
+\frac{1}{2}b^\b_\a b^\a_\b b -P_1) \right. \nonumber \\
&+& \left. b^\b_\a \partial_\b(-P_0 -2c +\frac{3}{4}b^\b_\a b^\a_\b -\frac{1} {4}b^2) \right],
\label{analzeroL0amixed}
\eq  
\bq
(L^0_\a)^{(1)} &=& (\nabla_\b c^\b_\a - \nabla_\a c) -\frac{1} {2}\nabla_\b (b_\g^\b b_\a^\g) 
+ \frac{1} {2}\nabla_\a (b_\g^\b b_\b^\g)  \nonumber \\
&+& \ep \left \{ 2(-P^{(0)} -2c +\frac{3}{4}b^\ep_\d b^\d_\ep -\frac{1} {4}b^2)\left[(\nabla_\b c^\b_\a - \nabla_\a c) 
-\frac{1} {2}\nabla_\b (b_\g^\b b_\a^\g) + \frac{1} {2}\nabla_\a (b_\g^\b b_\b^\g)\right] \right. \nonumber \\
&+& \left. (-bc -6d +5b^\d_\g c^\g_\d -\frac{3}{2}b^\ep_\d b^\d_\g b^\g_\ep +
\frac{1}{2}b^\d_\g b^\g_\d b-P^{(1)})(\nabla_\b b^\b_\a - \nabla_\a b)  \right. \nonumber \\
&-& \left. 4\partial_\a (-\frac{3}{2}bd -c^2 -12e +\frac{21}{2}b^\g_\b d^\b_\g +5c^\g_\b c^\b_\g -
\frac{19}{2}b^\d_\g b^\g_\b c^\b_\d +\frac{9}{4}b^\ep_\d b^\d_\g b^\g_\b b^\b_\ep + \frac{3} {2}b^\g_\b c^\b_\g b \right. \nonumber  \\
&+& \left. b^\g_\b b^\b_\g c -\frac{1} {2}b^\d_\g b^\g_\b b^\b_\d b - \frac{1} {4}(b^\g_\b b^\b_\g)^2-P^{(2)}) \right. \nonumber \\
&+& \left. b^\b_\a \partial_\b(-bc -6d +5b^\d_\g c^\g_\d -\frac{3}{2}b^\ep_\d b^\d_\g b^\g_\ep + \frac{1}{2}b^\d_\g b^\g_\d b-P^{(1)}) \right. \nonumber \\
&+& \left. (2c^\b_\a -b^\b_\g b^\g_\a) \partial_\b (-P^{(0)} -2c +\frac{3}{4}b^\ep_\d b^\d_\ep -\frac{1} {4}b^2) \right \},
\label{analoneL0amixed}
\eq 
and also,
\bq
(L^\b_\a)^{(0)}&=&-(P^\b_\a)^{(0)} -c^\b_\a +\frac{1} {2}b^\b_\g b^\g_\a -\frac{1} {4}b^\b_\a b -\frac{1}{2}\d^\b_\a(-P^{(0)}-2c+\frac{3}{4}b^\b_\a b^\a_\b -\frac{1}{4}b^2) \nonumber \\
&+&\ep \left \{2(-P^{(0)} -2c +\frac{3}{4}b^\b_\a b^\a_\b -\frac{1} {4}b^2)\left(-(P^\b_\a)^{(0)} -c^\b_\a +\frac{1} {2}b^\b_\g b^\g_\a -\frac{1} {4}b^\b_\a b\right) \right. \nonumber \\
&-& \left. b^\b_\a(-bc -6d +5b^\b_\a c^\a_\b -\frac{3}{2}b^\b_\a b^\a_\g b^\g_\b +\frac{1}{2}b^\b_\a b^\a_\b b -P^{(1)}) \right. \nonumber \\
&+& \left. a^{\b\g} \left[2\partial_\g \left(\partial_\a(-P^{(0)} -2c +\frac{3}{4}b^\b_\a b^\a_\b -\frac{1} {4}b^2)\right) \right. \right. \nonumber \\
&-& \left. \left. a^{\m\ep}A_{\a\b\ep}\partial_\m(-P^{(0)} -2c +\frac{3}{4}b^\b_\a b^\a_\b -\frac{1} {4}b^2)\right] \right. 
\nonumber \\
&+& \left. \d^\b_\a \left[ 4\big(-\frac{3}{2}bd -c^2 -12e +\frac{21}{2}b^\b_\a d^\a_\b +5c^\b_\a c^\a_\b 
-\frac{19}{2}b^\b_\a b^\a_\g c^\g_\b \right. \right. \nonumber \\
&+& \left. \left. \frac{9}{4}b^\b_\g b^\g_\a b^\a_\d b^\d_\b + \frac{3} {2}b^\b_\a c^\a_\b b+  b^\b_\a b^\a_\b c -\frac{1} {2}b^\g_\b b^\b_\a b^\a_\g b -\frac{1} {4}(b^\b_\a b^\a_\b)^2 -P^{(2)}) \right. \right. \nonumber \\
&-& \left. \left. \frac{1}{2}(-P^{(0)} -2c +\frac{3}{4}b^\b_\a b^\a_\b -\frac{1} {4}b^2)^2 \right. \right. \nonumber \\
&+& \left. \left. b(-bc -6d +5b^\b_\a c^\a_\b -\frac{3}{2}b^\b_\a b^\a_\g b^\g_\b +\frac{1}{2}b^\b_\a b^\a_\b b -P^{(1)}) \right. \right. \nonumber \\
&+& \left. \left. a^{\g\d} \left[a^{\m\ep}A_{\g\d\ep}\partial_\m(-P^{(0)} -2c +\frac{3}{4}b^\b_\a b^\a_\b 
-\frac{1} {4}b^2) \right. \right. \right. \nonumber \\
&-& \left. \left. \left. 2\partial_\g \left(\partial_\d(-P^{(0)} -2c +\frac{3}{4}b^\b_\a b^\a_\b -\frac{1} {4}b^2) \right) \right] \right] \right \}. 
\label{analzeroLbamixed}
\eq 


\subsection{Simplification of the terms}
\label{Simplification of the terms}

We can simplify the five relations (\ref{analzeroL00mixed}), (\ref{analoneL00mixed}), (\ref{analzeroL0amixed}), (\ref{analoneL0amixed}) and (\ref{analzeroLbamixed}) by using the trace of the higher-order field equations. Either working with the vacuum field equations (\ref{eq:FEs}) or with the field equations in case of radiation (\ref{eq:FEs_rad}), using Eq. (\ref{T general}) for $w=1/3$, the trace of the energy-momentum tensor reads,
\be
T = 0.
\label{Tequal0}
\ee
Then, the trace of the field equations in both cases gives the identity,
\be
R - 6\ep \Box_g R = 0.
\label{idBox}
\ee
Because of (\ref{BoxR}) and (\ref{eq:Rshort}) the last identity reads,
\be
R^{(0)} - 6\ep \left \{ 2R^{(2)} - a^{\a\b} \left[ \pa_\a \left( \pa_\b R^{(0)} \right) - \dfrac{1}{2}b_{\a\b}R^{(1)} - \dfrac{1}{2}a^{\m\ep}A_{\a\b\ep} \pa_\m R^{(0)} \right] \right \} = 0, 
\label{idBoxseries}
\ee 
and in terms of $a_{\a\b},b_{\a\b},c_{\a\b},d_{\a\b},e_{\a\b}$ we obtain,
\bq
&&6\ep \left \{ 2\left(-\frac{3}{2}bd -c^2 -12e +\frac{21}{2}b^\b_\a d^\a_\b +5c^\b_\a c^\a_\b 
- \frac{19}{2}b^\b_\a b^\a_\g c^\g_\b +\frac{9}{4}b^\b_\g b^\g_\a b^\a_\d b^\d_\b + \frac{3} {2}b^\b_\a c^\a_\b b \right. \right.    \nonumber  \\
&+& \left. \left. b^\b_\a b^\a_\b c -\frac{1} {2}b^\g_\b b^\b_\a b^\a_\g b - \frac{1} {4}(b^\b_\a b^\a_\b)^2-P^{(2)}\right) \right. \nonumber \\
&-& \left. a^{\a\b} \left[ \pa_\a \left( \pa_\b (-P^{(0)} -2c +\frac{3}{4}b^\d_\g b^\g_\d -\frac{1} {4}b^2) \right) \right. \right. \nonumber \\
&-& \left. \left. \dfrac{1}{2}b_{\a\b} \left( -bc -6d +5b^\d_\g c^\g_\d -\frac{3}{2}b^\ep_\d b^\d_\g b^\g_\ep + \frac{1}{2}b^\d_\g b^\g_\d b-P^{(1)} \right) \right. \right. \nonumber \\
&-& \left. \left. \dfrac{1}{2}a^{\m\ep}A_{\a\b\ep}\pa_\m \left( -P^{(0)} -2c +\frac{3}{4}b^\d_\g b^\g_\d -\frac{1} {4}b^2 \right) \right] \right \} \nonumber \\
&=&  -P^{(0)} -2c +\frac{3}{4}b^\b_\a b^\a_\b -\frac{1} {4}b^2.
\label{idBoxseriesanal}
\eq
Then, using the identity (\ref{idBoxseriesanal}), Eqns. (\ref{analzeroL00mixed}), (\ref{analoneL00mixed}), (\ref{analzeroL0amixed}), (\ref{analoneL0amixed}) and (\ref{analzeroLbamixed}) take the form, 
\bq
(L^0_0)^{(0)}&=&\frac{1}{6}P^{(0)} + \frac{1}{8}b^\b_\a b^\a_\b +\frac{1}{24}b^2 - \dfrac{2}{3}c 
+\ep \left [2(-P^{(0)} -2c +\frac{3}{4}b^\b_\a b^\a_\b -\frac{1} {4}b^2)(-c + \frac{1}{4}b^\d_\g b^\g_\d) \right. \nonumber \\ &-& \left. \frac{1}{2}(-P^{(0)} -2c +\frac{3}{4}b^\b_\a b^\a_\b -\frac{1} {4}b^2)^2 \right. \nonumber \\
&-& \left. 4\left(-\frac{3}{2}bd -c^2 -12e +\frac{21}{2}b^\b_\a d^\a_\b +5c^\b_\a c^\a_\b 
- \frac{19}{2}b^\b_\a b^\a_\g c^\g_\b +\frac{9}{4}b^\b_\g b^\g_\a b^\a_\d b^\d_\b 
+ \frac{3} {2}b^\b_\a c^\a_\b b \right. \right.     \nonumber  \\
&+& \left. \left. b^\b_\a b^\a_\b c -\frac{1} {2}b^\g_\b b^\b_\a b^\a_\g b 
- \frac{1} {4}(b^\b_\a b^\a_\b)^2-P^{(2)}\right) \right ],
\label{analzeroL00mixednew}
\eq
\bq
(L^0_0)^{(1)} &=& \frac{1}{2}P^{(1)}-\frac{1}{4}b^\b_\a b^\a_\b b+\frac{1}{4}b^\b_\a b^\a_\g b^\g_\b -\frac{1}{2}b^\b_\a c^\a_\b +\frac{1}{2}bc \nonumber \\
&+& \ep \left \{ 2(-P^{(0)} -2c +\frac{3}{4}b^\ep_\d b^\d_\ep -\frac{1} {4}b^2)(-3d + 2b_\a^\b c_\b^\a - \frac{1}{2}b_\a^\b b_\g^\a b_\b^\g) \right. \nonumber \\
&+& \left. 2(-bc -6d +5b^\b_\a c^\a_\b -\frac{3}{2}b^\b_\a b^\a_\g b^\g_\b +
\frac{1}{2}b^\b_\a b^\a_\b b-P^{(1)})(-c + \frac{1}{4}b^\ep_\d b^\d_\ep) \right. \nonumber \\
&-& \left. (-P^{(0)} -2c +\frac{3}{4}b^\ep_\d b^\d_\ep -\frac{1} {4}b^2)(-bc -6d +5b^\b_\a c^\a_\b -\frac{3}{2}b^\b_\a b^\a_\g b^\g_\b + \frac{1}{2}b^\b_\a b^\a_\b b-P^{(1)}) \right. \nonumber \\
&+& \left. 2c(-bc -6d +5b^\b_\a c^\a_\b -\frac{3}{2}b^\b_\a b^\a_\g b^\g_\b + \frac{1}{2}b^\b_\a b^\a_\b b-P^{(1)}) \right. \nonumber \\
&+& \left. 2b(-\frac{3}{2}bd -c^2 -12e +\frac{21}{2}b^\b_\a d^\a_\b +5c^\b_\a c^\a_\b - \frac{19}{2}b^\b_\a b^\a_\g c^\g_\b +\frac{9}{4}b^\b_\g b^\g_\a b^\a_\d b^\d_\b + \frac{3} {2}b^\b_\a c^\a_\b b \right.    \nonumber  \\
&+& \left. b^\b_\a b^\a_\b c -\frac{1} {2}b^\g_\b b^\b_\a b^\a_\g b - \frac{1} {4}(b^\b_\a b^\a_\b)^2-P^{(2)}) \right. \nonumber \\
&+& \left. a^{\a\b} \left[ -2\partial_\a \left( \partial_\b(-bc -6d +5b^\d_\g c^\g_\d -\frac{3}{2}b^\ep_\d b^\d_\g b^\g_\ep + \frac{1}{2}b^\d_\g b^\g_\d b-P^{(1)})\right) \right. \right. \nonumber \\
&+& \left. \left. 2E^\m_{\a\b} \pa_\m (-P^{(0)} -2c +\frac{3}{4}b^\d_\g b^\g_\d -\frac{1} {4}b^2) \right. \right. \nonumber \\
&+& \left. \left. a^{\m\ep}A_{\a\b\ep} \pa_\m (-bc -6d +5b^\d_\g c^\g_\d -\frac{3}{2}b^\d_\g b^\g_\z b^\z_\d +
\frac{1}{2}b^\d_\g b^\g_\d b-P^{(1)}) \right] \right. \nonumber \\
&+& \left. b^{\a\b} \left[ 2\partial_\a \left( \partial_\b(-P^{(0)} -2c +\frac{3}{4}b^\d_\g b^\g_\d -\frac{1} {4}b^2) \right) \right. \right. \nonumber \\
&-& \left. \left. b_{\a\b}(-bc -6d +5b^\d_\g c^\g_\d -\frac{3}{2}b^\ep_\d b^\d_\g b^\g_\ep + \frac{1}{2}b^\d_\g b^\g_\d b-P^{(1)}) \right. \right. \nonumber \\
&-& \left. \left. a^{\m\ep}A_{\a\b\ep} \pa_\m (-P^{(0)} -2c +\frac{3}{4}b^\d_\g b^\g_\d -\frac{1} {4}b^2) \right] \right \},
\label{analoneL00mixednew}
\eq
and,
\bq
(L^0_\a)^{(0)} &=& \frac{1}{2}(\nabla_\b b^\b_\a-\nabla_\a b) 
+\ep \left[(-P_0 -2c +\frac{3}{4}b^\b_\a b^\a_\b -\frac{1} {4}b^2)(\nabla_\b b^\b_\a-\nabla_\a b) \right. \nonumber \\
&-& \left. 2\partial_\a(-bc -6d +5b^\b_\a c^\a_\b -\frac{3}{2}b^\b_\a b^\a_\g b^\g_\b 
+\frac{1}{2}b^\b_\a b^\a_\b b -P_1) \right. \nonumber \\
&+& \left. b^\b_\a \partial_\b(-P_0 -2c +\frac{3}{4}b^\b_\a b^\a_\b -\frac{1} {4}b^2) \right],
\label{analzeroL0amixednew}
\eq 
\bq
(L^0_\a)^{(1)} &=& (\nabla_\b c^\b_\a - \nabla_\a c) -\frac{1} {2}\nabla_\b (b_\g^\b b_\a^\g) 
+ \frac{1} {2}\nabla_\a (b_\g^\b b_\b^\g)  \nonumber \\
&+& \ep \left \{ 2(-P^{(0)} -2c +\frac{3}{4}b^\ep_\d b^\d_\ep -\frac{1} {4}b^2)\left[(\nabla_\b c^\b_\a - \nabla_\a c) 
-\frac{1} {2}\nabla_\b (b_\g^\b b_\a^\g) + \frac{1} {2}\nabla_\a (b_\g^\b b_\b^\g)\right] \right. \nonumber \\
&+& \left. (-bc -6d +5b^\d_\g c^\g_\d -\frac{3}{2}b^\ep_\d b^\d_\g b^\g_\ep +
\frac{1}{2}b^\d_\g b^\g_\d b-P^{(1)})(\nabla_\b b^\b_\a - \nabla_\a b)  \right. \nonumber \\
&-& \left. 4\partial_\a (-\frac{3}{2}bd -c^2 -12e +\frac{21}{2}b^\g_\b d^\b_\g +5c^\g_\b c^\b_\g -
\frac{19}{2}b^\d_\g b^\g_\b c^\b_\d +\frac{9}{4}b^\ep_\d b^\d_\g b^\g_\b b^\b_\ep + \frac{3} {2}b^\g_\b c^\b_\g b \right. \nonumber  \\
&+& \left. b^\g_\b b^\b_\g c -\frac{1} {2}b^\d_\g b^\g_\b b^\b_\d b - \frac{1} {4}(b^\g_\b b^\b_\g)^2-P^{(2)}) \right. \nonumber \\
&+& \left. b^\b_\a \partial_\b(-bc -6d +5b^\d_\g c^\g_\d -\frac{3}{2}b^\ep_\d b^\d_\g b^\g_\ep + \frac{1}{2}b^\d_\g b^\g_\d b-P^{(1)}) \right. \nonumber \\
&+& \left. (2c^\b_\a -b^\b_\g b^\g_\a) \partial_\b (-P^{(0)} -2c +\frac{3}{4}b^\ep_\d b^\d_\ep -\frac{1} {4}b^2) \right \},
\label{analoneL0amixednew}
\eq
and,
\bq
(L^\b_\a)^{(0)}&=&-(P^\b_\a)^{(0)} -c^\b_\a +\frac{1} {2}b^\b_\g b^\g_\a -\frac{1} {4}b^\b_\a b 
+ \d^\b_\a( \dfrac{1}{6}P^{(0)}- \frac{1}{8}b^\d_\g b^\g_\d + \frac{1}{24}b^2 + \dfrac{1}{3}c ) \nonumber \\
&+&\ep \left \{2(-P^{(0)} -2c +\frac{3}{4}b^\b_\a b^\a_\b -\frac{1} {4}b^2)\left(-(P^\b_\a)^{(0)} -c^\b_\a +\frac{1} {2}b^\b_\g b^\g_\a -\frac{1} {4}b^\b_\a b\right) \right. \nonumber \\
&-& \left. b^\b_\a(-bc -6d +5b^\b_\a c^\a_\b -\frac{3}{2}b^\b_\a b^\a_\g b^\g_\b +\frac{1}{2}b^\b_\a b^\a_\b b -P^{(1)}) \right. \nonumber \\
&+& \left. a^{\b\g} \left[2\partial_\g \left(\partial_\a(-P^{(0)} -2c +\frac{3}{4}b^\b_\a b^\a_\b -\frac{1} {4}b^2)\right) \right. \right. \nonumber \\
&-& \left. \left. a^{\m\ep}A_{\a\b\ep}\partial_\m(-P^{(0)} -2c +\frac{3}{4}b^\b_\a b^\a_\b -\frac{1} {4}b^2)\right] \right. 
\nonumber \\
&-& \left. \dfrac{1}{2} \d^\b_\a \left(-P^{(0)} -2c +\frac{3}{4}b^\d_\g b^\g_\d -\frac{1} {4}b^2 \right)^2 \right \}. 
\label{analzeroLbamixednew}
\eq  
We now proceed to find the form of the solution of the higher-order gravitanional field equations in vacuum, in the vicinity of a point that is regular in the time. 


\section{Regular vacuum field equations}
\label{Regular vacuum field equations}

In this Section, we find the degrees of freedom of the vacuum field equations in $R + \ep R^2$ theory through regular formal series expansion of the form (\ref{eq:3dimmetric}).


\subsection{First approximation of the vacuum field equations}
\label{First approximation of the vacuum field equations}

Taking into account the vacuum field equations (\ref{eq:Loo}), (\ref{eq:Loa}) and (\ref{eq:Lab}), and using the notation of Subsection \ref{Series form of field equations}, we write,
\be
0 = \mathcal{C}_0 = (L^0_0)^{(0)} + t(L^0_0)^{(1)}, 
\label{C0vacuumreg}
\ee
\be
0 = \mathcal{C}_\a  = (L^0_\a)^{(0)} + t(L^0_\a)^{(1)},
\label{Cavacuumreg} 
\ee
and also,
\be
0 = L^\b_\a = (L^\b_\a)^{(0)}. 
\label{LABvacuumreg}
\ee
Apparently, from Eqns. (\ref{C0vacuumreg}), (\ref{Cavacuumreg}) and (\ref{LABvacuumreg}), we find that,
\be
(L^0_0)^{(0)} = 0,  
\label{zeroL000vr}
\ee
\be
(L^0_0)^{(1)} = 0, 
\label{zeroL001vr}
\ee
\be
(L^0_\a)^{(0)} = 0,  
\label{zeroL0a0vr}
\ee
\be
(L^0_\a)^{(1)} = 0,  
\label{zeroL0a1vr}
\ee
\be
(L^\b_\a)^{(0)} = 0.
\label{zeroLab0vr}
\ee
These last five equations give $14$ relations between the initial data $a_{\a\b},b_{\a\b},c_{\a\b},d_{\a\b}, \\
e_{\a\b}$. However, in the following analysis, we show that the first-order terms $(L^0_0)^{(1)},\\ (L^0_\a)^{(1)}$ do not eventually imply four relations between the initial data, due to the fact that they are going to be proved identically equal to zero. Thus, we conclude that only $10$ relations of them remain.


\subsection{First-order terms of the vacuum field equations}
\label{First-order terms of the vacuum field equations}

In order to prove that the first-order terms $(L^0_0)^{(1)},(L^0_\a)^{(1)}$ vanish identically, we use the identity (\ref{clL}), namely that, 
\be
\nabla_0 L^0_j + \nabla_\b L^\b_j=0,
\label{idLijstart}
\ee
for $j=0$ and $j=\a$. In particular, for $j=0$ we find,
\be
\partial_t L^0_0-\g^{\a\b}\nabla_\b L^0_\a=0,
\label{idLio}
\ee
that is using (\ref{eq:3diminvmetric}), (\ref{C0vacuumreg}) and (\ref{Cavacuumreg}), we write,
\be
(L^0_0)^{(1)}-(\g^{\a\b})^{(0)}\nabla_\b (L^0_\a)^{(0)}=0.
\label{eq:identity_Lio_zeroth order}
\ee
Taking into account the relation (\ref{zeroL0a0vr}), we directly conclude that the term $(L^0_0)^{(1)}$ vanishes identically. 

Working analogously, for $j=\a$ we have, 
\bq
\partial_t L^0_\a -\frac{1}{2}K^\b_\a L^0_\b +\nabla_\b L^\b_\a=0,
\label{idLia}
\eq
that is using (\ref{eq:Kmixed}), (\ref{Cavacuumreg}) and (\ref{LABvacuumreg}) we find,
\be
(L^0_\a)^{(1)}-\frac{1}{2}(K^\b_\a)^{(0)}(L^0_\b)^{(0)}+\nabla_\b (L^\b_\a)^{(0)}=0.
\label{identity_Lia_zeroth order}
\ee
Because of the relations (\ref{zeroL0a0vr}) and (\ref{zeroLab0vr}), we find that the term $(L^0_\a)^{(1)}$ also vanishes identically.


\subsection{Degrees of freedom of the vacuum field equations}
\label{Degrees of freedom of the vacuum field equations}

Taking into consideration that the expression (\ref{eq:3dimmetric}) contains $30$ degrees of freedom (six of each spatial matrix $\g^{(n)}_{\a\b},n=0,\cdots,4$), the problem we are faced with is, what the initial number of thirty free functions becomes, after the imposition of the higher-order evolution and constraint equations, that is how it finally compares with the $16$ degrees of freedom that any general solution in vacuum must posses as shown in Section \ref{Function-counting}.
 
According to the results of the previous subsection, namely that the first-order terms $(L^0_0)^{(1)},(L^0_\a)^{(1)}$ of the vacuum field equations (\ref{C0vacuumreg}) and (\ref{Cavacuumreg}) are identically equal to zero, the number of the relations between the initial data $a_{\a\b},b_{\a\b},c_{\a\b},d_{\a\b},\\e_{\a\b}$ that the five Eqns. (\ref{zeroL000vr})-(\ref{zeroLab0vr}) provide is finally equal to $10$. In particular, the $10$ relations are,
\bq
\frac{1}{6}P^{(0)} + \frac{1}{8}b^\b_\a b^\a_\b +\frac{1}{24}b^2 - \dfrac{2}{3}c &=& 
\ep \left [2(-P^{(0)} -2c +\frac{3}{4}b^\b_\a b^\a_\b -\frac{1} {4}b^2)(c - \frac{1}{4}b^\d_\g b^\g_\d) \right. \nonumber \\ &+& \left. \frac{1}{2}(-P^{(0)} -2c +\frac{3}{4}b^\b_\a b^\a_\b -\frac{1} {4}b^2)^2 \right. \nonumber \\
&+& \left. 4\left(-\frac{3}{2}bd -c^2 -12e +\frac{21}{2}b^\b_\a d^\a_\b +5c^\b_\a c^\a_\b 
- \frac{19}{2}b^\b_\a b^\a_\g c^\g_\b  \right. \right. \nonumber \\
&+& \left. \left. \frac{9}{4}b^\b_\g b^\g_\a b^\a_\d b^\d_\b 
+ \frac{3} {2}b^\b_\a c^\a_\b b     
+ b^\b_\a b^\a_\b c -\frac{1} {2}b^\g_\b b^\b_\a b^\a_\g b  \right. \right. \nonumber \\
&-& \left. \left. \frac{1} {4}(b^\b_\a b^\a_\b)^2-P^{(2)}\right) \right ],
\label{eq for analzeroL00mixednew}
\eq
\bq
\frac{1}{2}(\nabla_\b b^\b_\a-\nabla_\a b) &=& 
\ep \left[(-P_0 -2c +\frac{3}{4}b^\b_\a b^\a_\b -\frac{1} {4}b^2)(\nabla_\a b - \nabla_\b b^\b_\a) \right. \nonumber \\
&+& \left. 2\partial_\a(-bc -6d +5b^\b_\a c^\a_\b -\frac{3}{2}b^\b_\a b^\a_\g b^\g_\b 
+\frac{1}{2}b^\b_\a b^\a_\b b -P_1) \right. \nonumber \\
&-& \left. b^\b_\a \partial_\b(-P_0 -2c +\frac{3}{4}b^\b_\a b^\a_\b -\frac{1} {4}b^2) \right],
\label{eq for analzeroL0amixednew}
\eq 
and,
\bq
&& \ep \left \{2(-P^{(0)} -2c +\frac{3}{4}b^\b_\a b^\a_\b -\frac{1} {4}b^2)\left(-(P^\b_\a)^{(0)} -c^\b_\a +\frac{1} {2}b^\b_\g b^\g_\a -\frac{1} {4}b^\b_\a b\right) \right. \nonumber \\
&-& \left. b^\b_\a(-bc -6d +5b^\b_\a c^\a_\b -\frac{3}{2}b^\b_\a b^\a_\g b^\g_\b +\frac{1}{2}b^\b_\a b^\a_\b b -P^{(1)}) \right. \nonumber \\
&+& \left. a^{\b\g} \left[2\partial_\g \left(\partial_\a(-P^{(0)} -2c +\frac{3}{4}b^\b_\a b^\a_\b -\frac{1} {4}b^2)\right) \right. \right. \nonumber \\
&-& \left. \left. a^{\m\ep}A_{\a\b\ep}\partial_\m(-P^{(0)} -2c +\frac{3}{4}b^\b_\a b^\a_\b -\frac{1} {4}b^2)\right] \right. 
\nonumber \\
&-& \left. \dfrac{1}{2} \d^\b_\a \left(-P^{(0)} -2c +\frac{3}{4}b^\b_\a b^\a_\b -\frac{1} {4}b^2 \right)^2 \right \} =
(P^\b_\a)^{(0)} + c^\b_\a - \frac{1} {2}b^\b_\g b^\g_\a + \frac{1} {4}b^\b_\a b \nonumber \\
&-& \d^\b_\a( \dfrac{1}{6}P^{(0)}- \frac{1}{8}b^\b_\a b^\a_\b + \frac{1}{24}b^2 + \dfrac{1}{3}c ). 
\label{eq for analzeroLbamixednew}
\eq 

Hence, in total, we find that the imposition of the field equations leads to $1$0 relations between the $30$ functions of the perturbation metric (\ref{eq:3dimmetric}), that is, we are left with $20$ free functions. Taking into account the freedom we have in performing $4$ diffeomorphism changes, we finally conclude that there are in total $16$ free functions in the solution  (\ref{eq:3dimmetric}). This means that the regular solution (\ref{eq:3dimmetric}) corresponds to a general solution of the problem. To put it differently, regularity is a generic feature of the $R+\ep R^2$ theory in vacuum, assuming analyticity (see \cite{ctt12,ctt13}).


\section{Regular field equations in case of radiation}
\label{Regular field equations in case of radiation}

In this Section, we find the degrees of freedom of the gravitational field equations in $R + \ep R^2$ theory plus radiation through regular formal series expansion of the form (\ref{eq:3dimmetric}). 


\subsection{Field equations in case of radiation}
\label{Field equations in case of radiation}

According to the relations (\ref{T^0_0 general}), (\ref{T^0_a general}) and (\ref{T^b_a general}) and the equation of state $(p=\dfrac{\rho}{3})$ in case of radiation models, the components of the energy-momentum tensor $T^i_j$ become,
\be
T^0_0 = \dfrac{1}{3}\rho(4{u_0}^2 - 1),
\label{T^0_0 generalrad} 
\ee
\be
T^0_\a = \dfrac{4}{3}\rho u_\a u_0,
\label{T^0_a generalrad} 
\ee
and,
\be
T^\b_\a = \dfrac{1}{3}\rho(4u^\b u_\a - \d^\b_\a).
\label{T^b_a generalrad}
\ee
Also for the trace of the energy-momentum tensor we obtain,
\be
T=trT_{ij}=0.
\label{T generalrad}
\ee 
Taking into account Eqns. (\ref{T^0_0 generalrad}), (\ref{T^0_a generalrad}) and (\ref{T^b_a generalrad}), the gravitational field equations (\ref{eq:Loo_rad}), (\ref{eq:Loa_rad}) and (\ref{eq:Lab_rad}), and the notation of Subsection \ref{Series form of field equations} we obtain,
\be
\dfrac{8\pi G}{3}\rho(4{u_0}^2 - 1) = \mathcal{C}_0 = (L^0_0)^{(0)} + t(L^0_0)^{(1)}, 
\label{C0radreg} 
\ee
\be
\dfrac{32\pi G}{3}\rho u_\a u_0 = \mathcal{C}_\a  = (L^0_\a)^{(0)} + t(L^0_\a)^{(1)},
\label{Caradreg} 
\ee
and,
\be
\dfrac{8\pi G}{3}\rho(4u^\b u_\a - \d^\b_\a) = L^\b_\a = (L^\b_\a)^{(0)}, 
\label{LABradreg}
\ee
where the terms appearing in these last three relations are given by (\ref{analzeroL00mixednew})-(\ref{analzeroLbamixednew}).

Further, using the identity (\ref{eq:velocities identity}), we get,
\bq
1 = u_i u^i & \approx & {u_0}^2 - \left[a^{\a\b} - b^{\a\b}t + \left(b^{\a\g}\;b_\g^\b - c^{\a\b}\right)t^2  \right. 
\nonumber \\
&+& \left. \left(-d^{\a\b} + b^{\a\g}\;c_\g^\b - b^{\a\g}\;b_\g^\d\;b_\d^\b +
 c^{\a\g}\;b_\g^\b\right)t^3 \right.   \nonumber     \\
&+& \left. \left(-e^{\a\b} + b^{\a\g}d_\g^\b - b^{\a\g}b_\g^\d\;c_\d^\b + c^{\a\g}\;c_\g^\b + d^{\a\g}\;b_\g^\b - b^{\a\g}\; c_\g^\d b_\d^\b + b^{\a\g}b_\g^\d b_\d^\ep b_\ep^\b \right. \right. \nonumber \\
&-& \left. \left. c^{\a\g}\;b_\g^\d b_\d^\b\right)t^4 \right]u_\a u_\b.
\label{velocities identity regular}
\eq
Considering,
\be
u_0 \approx 1,
\label{uoapprox1radreg}
\ee
namely that,
\be
a^{\a\b}u_\a u_\b \rightarrow 0
\label{vel tend to zero}
\ee
when $t$ tends to zero, the system of equations (\ref{C0radreg}), (\ref{Caradreg}) and (\ref{LABradreg}) gives (to lowest order),
\be
8\pi G\rho = \mathcal{C}_0 = (L^0_0)^{(0)} + t(L^0_0)^{(1)}, 
\label{C0radregfinal}
\ee
\be 
\dfrac{32\pi G}{3}\rho u_\a = \mathcal{C}_\a  = (L^0_\a)^{(0)} + t(L^0_\a)^{(1)},
\label{Caradregfinal} 
\ee
and also,
\be
-\dfrac{8\pi G}{3}\rho \d^\b_\a = L^\b_\a = (L^\b_\a)^{(0)}. 
\label{LABradregfinal}
\ee
We are now ready to find the degrees of freedom of the gravitational field equations in case of radiation. 


\subsection{Degrees of freedom of the field equations in case of radiation}
\label{Degrees of freedom of the vacuum field equations in case of radiation}

Taking into consideration that the $36$ initial data consisting of the quantities $a_{\a\b},b_{\a\b},c_{\a\b},d_{\a\b},e_{\a\b}$ together with $p,\rho,u^i$ are related through the Eqns. (\ref{C0radregfinal}), (\ref{Caradregfinal}), (\ref{LABradregfinal}), the equation of state (\ref{radiation eq}) and the identity (\ref{velocities identity regular}), the problem we are faced with is, what the initial number of $36$ free functions becomes, after the imposition of the higher-order evolution and constraint equations, that is how it finally compares with the $20$ degrees of freedom that any general solution in vacuum must posses as shown in Subsection \ref{Function counting in the presence of a fluid}.

Obviously, Eq. (\ref{C0radregfinal}) gives one relation for the energy density $\rho$. Then, if we consider $(L^0_\a)^{(0)} \neq 0$, and substitute $\rho$ from Eq. (\ref{C0radregfinal}) to Eq. (\ref{Caradregfinal}), we find that,
\be
u_\a = \dfrac{3(L^0_\a)^{(0)}}{4(L^0_0)^{(0)}}.
\label{velradreg}
\ee 
However, this last expression for the velocities $u_\a$ does not satisfy (\ref{vel tend to zero}), and therefore is not acceptable. Thus, we find three relations between the initial data $a_{\a\b},b_{\a\b},c_{\a\b},d_{\a\b}$, namely,
\bq
0 &=& (L^0_\a)^{(0)} = \frac{1}{2}(\nabla_\b b^\b_\a-\nabla_\a b) 
+\ep \left[(-P_0 -2c +\frac{3}{4}b^\b_\a b^\a_\b -\frac{1} {4}b^2)(\nabla_\b b^\b_\a-\nabla_\a b) \right. \nonumber \\
&-& \left. 2\partial_\a(-bc -6d +5b^\b_\a c^\a_\b -\frac{3}{2}b^\b_\a b^\a_\g b^\g_\b 
+\frac{1}{2}b^\b_\a b^\a_\b b -P_1) \right. \nonumber \\
&+& \left. b^\b_\a \partial_\b(-P_0 -2c +\frac{3}{4}b^\b_\a b^\a_\b -\frac{1} {4}b^2) \right].
\label{analzeroL0amixednew equal to zero}
\eq  
Then, because of the identity (\ref{eq:identity_Lio_zeroth order}), the term $(L^0_0)^{(1)}$ is identically equal to zero. This means that,
\be
8\pi G \rho = (L^0_0)^{(0)} = \textrm{constant},
\label{constant energy density}
\ee
with respect to $t$, which is consistent with the result obtained from the generalized Friedmann equation (\ref{gen Friedmann eq density}) regarding the form (\ref{gses}). This last equation reads,
\bq
8\pi G \rho &=& \frac{1}{6}P^{(0)} + \frac{1}{8}b^\b_\a b^\a_\b +\frac{1}{24}b^2 - \dfrac{2}{3}c 
+\ep \left [2(-P^{(0)} -2c +\frac{3}{4}b^\b_\a b^\a_\b -\frac{1} {4}b^2)(-c + \frac{1}{4}b^\d_\g b^\g_\d) \right. \nonumber \\ &-& \left. \frac{1}{2}(-P^{(0)} -2c +\frac{3}{4}b^\b_\a b^\a_\b -\frac{1} {4}b^2)^2 \right. \nonumber \\
&-& \left. 4\left(-\frac{3}{2}bd -c^2 -12e +\frac{21}{2}b^\b_\a d^\a_\b +5c^\b_\a c^\a_\b 
- \frac{19}{2}b^\b_\a b^\a_\g c^\g_\b +\frac{9}{4}b^\b_\g b^\g_\a b^\a_\d b^\d_\b 
+ \frac{3} {2}b^\b_\a c^\a_\b b \right. \right.     \nonumber  \\
&+& \left. \left. b^\b_\a b^\a_\b c -\frac{1} {2}b^\g_\b b^\b_\a b^\a_\g b 
- \frac{1} {4}(b^\b_\a b^\a_\b)^2-P^{(2)}\right) \right ].
\label{analzeroL00mixednew with energy density}
\eq
Further, using Eq. (\ref{analzeroL0amixednew equal to zero}), we find three accepted relations concerning the velocities $u_\a$, namely,
\be
u_\a = \dfrac{3(L^0_\a)^{(1)}}{4(L^0_0)^{(0)}}t,
\label{fin vel tend zero}
\ee
We note that the quantity $(L^0_0)^{(0)}$ must not be equal to zero. Equation (\ref{fin vel tend zero}) reads,
\bq
u_\a &=& \dfrac{3t}{4} \left \{(\nabla_\b c^\b_\a - \nabla_\a c) -\frac{1} {2}\nabla_\b (b_\g^\b b_\a^\g) 
+ \frac{1} {2}\nabla_\a (b_\g^\b b_\b^\g) \right.  \nonumber \\
&+& \left. \ep \left [ 2(-P^{(0)} -2c +\frac{3}{4}b^\ep_\d b^\d_\ep -\frac{1} {4}b^2)[(\nabla_\b c^\b_\a - \nabla_\a c) 
-\frac{1} {2}\nabla_\b (b_\g^\b b_\a^\g) + \frac{1} {2}\nabla_\a (b_\g^\b b_\b^\g)] \right. \right. \nonumber \\
&+& \left. \left. (-bc -6d +5b^\d_\g c^\g_\d -\frac{3}{2}b^\ep_\d b^\d_\g b^\g_\ep +
\frac{1}{2}b^\d_\g b^\g_\d b-P^{(1)})(\nabla_\b b^\b_\a - \nabla_\a b)  \right. \right. \nonumber \\
&-& \left. \left. 4\partial_\a (-\frac{3}{2}bd -c^2 -12e +\frac{21}{2}b^\g_\b d^\b_\g +5c^\g_\b c^\b_\g -
\frac{19}{2}b^\d_\g b^\g_\b c^\b_\d +\frac{9}{4}b^\ep_\d b^\d_\g b^\g_\b b^\b_\ep + \frac{3} {2}b^\g_\b c^\b_\g b \right. \right. \nonumber  \\
&+& \left. \left. b^\g_\b b^\b_\g c -\frac{1} {2}b^\d_\g b^\g_\b b^\b_\d b - \frac{1} {4}(b^\g_\b b^\b_\g)^2-P^{(2)}) \right. \right. \nonumber \\
&+& \left. \left. b^\b_\a \partial_\b(-bc -6d +5b^\d_\g c^\g_\d -\frac{3}{2}b^\ep_\d b^\d_\g b^\g_\ep + \frac{1}{2}b^\d_\g b^\g_\d b-P^{(1)}) \right. \right. \nonumber \\
&+& \left. \left. (2c^\b_\a -b^\b_\g b^\g_\a) \partial_\b (-P^{(0)} -2c +\frac{3}{4}b^\ep_\d b^\d_\ep -\frac{1} {4}b^2) \right ] \right \} \nonumber \\
& \times & \left \{ \frac{1}{6}P^{(0)} + \frac{1}{8}b^\b_\a b^\a_\b +\frac{1}{24}b^2 - \dfrac{2}{3}c 
+\ep \left [2(-P^{(0)} -2c +\frac{3}{4}b^\b_\a b^\a_\b -\frac{1} {4}b^2)(-c + \frac{1}{4}b^\d_\g b^\g_\d) \right. \right. \nonumber \\ 
&-& \left. \left. \frac{1}{2}(-P^{(0)} -2c +\frac{3}{4}b^\b_\a b^\a_\b -\frac{1} {4}b^2)^2 \right. \right. \nonumber \\
&-& \left. \left. 4\left(-\frac{3}{2}bd -c^2 -12e +\frac{21}{2}b^\b_\a d^\a_\b +5c^\b_\a c^\a_\b 
- \frac{19}{2}b^\b_\a b^\a_\g c^\g_\b +\frac{9}{4}b^\b_\g b^\g_\a b^\a_\d b^\d_\b 
+ \frac{3} {2}b^\b_\a c^\a_\b b \right. \right. \right.    \nonumber  \\
&+& \left. \left. \left. b^\b_\a b^\a_\b c -\frac{1} {2}b^\g_\b b^\b_\a b^\a_\g b 
- \frac{1} {4}(b^\b_\a b^\a_\b)^2-P^{(2)}\right) \right ] \right \}^{-1}.
\label{analoneL0amixednew with ua}
\eq
Also, substituting $\rho$ from (\ref{C0radregfinal}) to (\ref{LABradregfinal}) we find that the zeroth-order terms give,
\be
(L^\b_\a)^{(0)} = -\dfrac{1}{3}\d^\b_\a (L^0_0)^{(0)},
\label{LbaLooradreg}
\ee
namely that,

\bq
&&-(P^\b_\a)^{(0)} -c^\b_\a +\frac{1} {2}b^\b_\g b^\g_\a -\frac{1} {4}b^\b_\a b 
+ \d^\b_\a( \dfrac{1}{6}P^{(0)}- \frac{1}{8}b^\d_\g b^\g_\d + \frac{1}{24}b^2 + \dfrac{1}{3}c ) \nonumber \\
&+&\ep \left \{2(-P^{(0)} -2c +\frac{3}{4}b^\b_\a b^\a_\b -\frac{1} {4}b^2)\left(-(P^\b_\a)^{(0)} -c^\b_\a +\frac{1} {2}b^\b_\g b^\g_\a -\frac{1} {4}b^\b_\a b\right) \right. \nonumber \\
&-& \left. b^\b_\a(-bc -6d +5b^\b_\a c^\a_\b -\frac{3}{2}b^\b_\a b^\a_\g b^\g_\b +\frac{1}{2}b^\b_\a b^\a_\b b -P^{(1)}) \right. \nonumber \\
&+& \left. a^{\b\g} \left[2\partial_\g \left(\partial_\a(-P^{(0)} -2c +\frac{3}{4}b^\b_\a b^\a_\b -\frac{1} {4}b^2)\right) \right. \right. \nonumber \\
&-& \left. \left. a^{\m\ep}A_{\a\b\ep}\partial_\m(-P^{(0)} -2c +\frac{3}{4}b^\b_\a b^\a_\b -\frac{1} {4}b^2)\right] \right. 
\nonumber \\
&-& \left. \dfrac{1}{2} \d^\b_\a \left(-P^{(0)} -2c +\frac{3}{4}b^\d_\g b^\g_\d -\frac{1} {4}b^2 \right)^2 \right \} 
\nonumber \\
&=& -\dfrac{1}{3}\d^\b_\a \left \{ \frac{1}{6}P^{(0)} + \frac{1}{8}b^\d_\g b^\g_\d +\frac{1}{24}b^2 - \dfrac{2}{3}c \right.  \nonumber \\
&+& \left.  \ep \left [2(-P^{(0)} -2c +\frac{3}{4}b^\d_\g b^\g_\d -\frac{1} {4}b^2)(-c + \frac{1}{4}b^\d_\g b^\g_\d)  
- \frac{1}{2}(-P^{(0)} -2c +\frac{3}{4}b^\d_\g b^\g_\d -\frac{1} {4}b^2)^2 \right. \right. \nonumber \\
&-& \left. \left. 4\left(-\frac{3}{2}bd -c^2 -12e +\frac{21}{2}b^\d_\g d^\g_\d +5c^\d_\g c^\g_\d 
- \frac{19}{2}b^\ep_\d b^\d_\g c^\g_\ep +\frac{9}{4}b^\z_\ep b^\ep_\d b^\d_\g b^\g_\z 
+ \frac{3} {2}b^\d_\g c^\g_\d b \right. \right. \right.    \nonumber  \\
&+& \left. \left. \left. b^\d_\g b^\g_\d c -\frac{1} {2}b^\ep_\d b^\d_\g b^\g_\ep b 
- \frac{1} {4}(b^\d_\g b^\g_\d)^2-P^{(2)}\right) \right ] \right \}. 
\label{analzeroL00Lbamixednew}
\eq 
which constitute six additional relations between the initial data. Notice that the trace of Eq. (\ref{LbaLooradreg}) yields the identity (\ref{idBoxseries}).

To sum up, from Eqns. (\ref{analzeroL0amixednew equal to zero}), (\ref{constant energy density}), (\ref{fin vel tend zero}) and (\ref{LbaLooradreg}), we found one relation for the energy density $\rho$, three relations for the velocities $u_\a$ and nine more relations regarding the initial data $a_{\a\b},b_{\a\b},c_{\a\b},d_{\a\b},e_{\a\b}$. In adition to this, we have to count two additional relations: one relation from the equation of state (\ref{radiation eq}) and one more from the identity (\ref{velocities identity regular}). Thus, the imposition of the field equations in case of radiation leads to $15$ relations between the $36$ functions of the perturbation metric (\ref{eq:3dimmetric}), that is, we are left with $21$ free functions. Subtracting $4$ diffeomorphism transformations, we finally conclude that there are in total $17$ free functions in the solution (\ref{eq:3dimmetric}) in case of radiation. This means that, the regular solution (\ref{eq:3dimmetric}) does not correspond to a general solution of the problem. Therefore, assuming analyticity, regularity is not a generic feature of the $R+\ep R^2$ theory in case of radiation.


\section{Degrees of freedom with $\g_{\a\b}=\sum{\g^{(n)}_{\a\b} t^{n}},n\geq 4$}
\label{Degrees of freedom general}

In this Section, we show that the number of the degrees of freedom of the gravitational field equations in $R + \ep R^2$ theory is the same, if instead of the regular expression (\ref{eq:3dimmetric}), we consider a regular expression of the form,
\be
\g_{\a\b}=\sum{\g^{(n)}_{\a\b} t^{n}},\quad n\geq 4.
\label{gengab}
\ee
In particular, we prove this statement in both cases (vacuum-radiation). 


\subsection{Field equations with $\g_{\a\b}=\sum{\g^{(n)}_{\a\b} t^{n}},n\geq 4$}
\label{Field equations general regular}

We now consider what would change or not in the result of free functions, if we did not use the series in (\ref{eq:3dimmetric}) up to the fourth power of t, but up to the nth power of t, where n is a natural number, with $n\geq 4$. Then, we would have the 30 initial data from $a_{\a\b}, b_{\a\b}, c_{\a\b}, d_{\a\b},e_{\a\b}$ and more $6\times(n-4)$ initial data from the aditional spatial matrixes $\g^{(n)}_{\a\b},n>4$, which means $6n+6$ initial data in total for the spatial matrixes appearing in (\ref{gengab}). From (\ref{eq:hamiltonian}) and (\ref{eq:momentum}) we find that $\mathcal{C}_0$ and $\mathcal{C}_\a$ are third order differential equations with respect to {t} and from (\ref{eq:pW}) snap equation is a fourth order differential equation with respect to $t$, therefore relations (\ref{eq:C0}), (\ref{eq:Ca}) and (\ref{eq:LAB}) become: 
\be
\mathcal{C}_0    =  (L^0_0)^{(0)} + t(L^0_0)^{(1)} + \cdots + t^{n-3}(L^0_0)^{(n-3)}, 
\label{eq:C0n}
\ee
\be
\mathcal{C}_\a   =  (L^0_\a)^{(0)} + t(L^0_\a)^{(1)} +\cdots + t^{n-3}(L^0_\a)^{(n-3)}, 
\label{eq:Can}
\ee
and,
\be
L^\b_\a  =  (L^\b_\a)^{(0)}  + t(L^\b_\a)^{(1)}+ \cdots + t^{n-4}(L^\b_\a)^{(n-4)}. 
\label{eq:LABn}
\ee


\subsection{Degrees of freedom with $\g_{\a\b}=\sum{\g^{(n)}_{\a\b} t^{n}},n\geq 4$ in vacuum}
\label{Degrees of freedom general-regular in vacuum}

In order to have the same conclusion as in the case of $n=4$, which is 16 arbitrary functions for the regular solution of the higher-order equations in vacuum, we must show that the derived system of equations in vacuum, namely,
\be
0=\mathcal{C}_0    =  (L^0_0)^{(0)} + t(L^0_0)^{(1)} + \cdots + t^{n-3}(L^0_0)^{(n-3)}, 
\label{eq:C0nregvac}
\ee
\be
0=\mathcal{C}_\a   =  (L^0_\a)^{(0)} + t(L^0_\a)^{(1)} +\cdots + t^{n-3}(L^0_\a)^{(n-3)}, 
\label{eq:Canregvac}
\ee
and,
\be
0=L^\b_\a  =  (L^\b_\a)^{(0)}  + t(L^\b_\a)^{(1)}+ \cdots + t^{n-4}(L^\b_\a)^{(n-4)}, 
\label{eq:LABnregvac}
\ee
provide $6n-14$ relations between the initial data $\g^{(m)}_{\a\b}$, $m=0,1,...,n$. To make it clearer, if we subtract from the initial $6n+6$ data the number $6n-14$, we find $20$ data. Further, taking into consideration four diffeomorphism transformations, we end up with the number $16$.

Notice that the relations that we get from $(L^0_0)^{(0)},(L^0_\a)^{(0)} $ and from $(L^\b_\a)^{(i)}$, for $i=0,1,...,n-4$ are $6n-14$, that is the number we would like to have. Therefore we must prove the statement that the terms $(L^0_0)^{(j)}$ and $(L^0_\a)^{(j)}$ vanish identically for any $j=1,2,...,n-3$ and consequently they do not change the expected number of the initial data.

Following the method of induction, in step 1, for $n=4$, we have already shown that this holds, due to the fact that the terms $(L^0_0)^{(1)}$ and $(L^0_\a)^{(1)}$ vanish identically by using the identity (\ref{idLijstart}).

In step 2, we suppose that our statement holds when $n=k$, where k is a natural number, with $k\geq 4$, namely, that the terms $(L^0_0)^{(j)}$ and $(L^0_\a)^{(j)}$ vanish identically for any $j=1,2,...,k-3$. 

We now prove that our statement stands for $n=k+1$. Indeed, using the identity (\ref{idLio}), we find,
\bq
&&(L^0_0)^{(1)}-(\g^{\a\b})^{(0)}\nabla_\b (L^0_\a)^{(0)} \nonumber \\ &+&\sum^{n-4}_{l=1}\{[(l+1)(L^0_0)^{(l+1)}-\sum^{l}_{m=0}(\g^{\a\b})^{(m)}\nabla_\b (L^0_\a)^{(l-m)}]t^l\}=0,
\label{idLioseries}
\eq
and then, for the $k-3$ order term ($l=k-3$), we obtain,
\bq
(k-2)(L^0_0)^{(k-2)} - \sum^{k-3}_{m=0}(\g^{\a\b})^{(m)}\nabla_\b (L^0_\a)^{(k-3-m)}=0.
\label{idL00k-2}
\eq 
From step 2, due to the fact that $k-3-m\leq k-3$, we get that the term $(L^0_\a)^{(k-3-m)}$ is identically zero and thus the term ${L^0_0}^{(k-2)}$ vanishes identically.
 
Also, using the other identity (\ref{idLia}), we have,
\bq
&&(L^0_\a)^{(1)}-\frac{1}{2}(K^\b_\a)^{(0)}(L^0_\b)^{(0)}+\nabla_\b (L^\b_\a)^{(0)} \nonumber \\ &+&\sum^{n-4}_{l=1}\{[(l+1)(L^0_\a)^{(l+1)}-\frac{1}{2}\sum^{l}_{m=0}(K^\b_\a)^{(m)}(L^0_\b)^{(l-m)}
+\nabla_\b (L^\b_\a)^{(l)}]t^l\}=0, \nonumber \\
&&
\label{idLiaseries}
\eq
and also for the $k-3$ order term, we get,
\bq
(k-2)(L^0_\a)^{(k-2)} - \frac{1}{2}\sum^{k-3}_{m=0}(K^\b_\a)^{(m)}(L^0_\b)^{(k-3-m)} + \nabla_\b (L^\b_\a)^{(k-3)}=0.
\label{idLoak-2}
\eq
According to step 2, due to the fact that $k-3-m\leq k-3$, we also get that the term $(L^0_\b)^{(k-3-m)}$ is identically zero. In addition to this, we have already counted the six relations given by the term $(L^\b_\a)^{(k-3)}$ ($(L^\b_\a)^{(i)}=0$, for $i=0,1,...,n-4$, but not identically), and therefore the remaining term $(L^0_\a)^{(k-2)}$ vanishes identically.

Hence, without loss of generality, we can study the problem of finding the number of the arbitrary functions in the regular case in vacuum, by using (\ref{eq:3dimmetric}).


\subsection{Degrees of freedom with $\g_{\a\b}=\sum{\g^{(n)}_{\a\b} t^{n}},n\geq 4$ in the case of radiation}
\label{Degrees of freedom general-regular in the case of radiation}

In the case of radiation the system of the gravitational field equations becomes,
\be
\dfrac{8\pi G}{3}\rho(4{u_0}^2 - 1)=\mathcal{C}_0    =  (L^0_0)^{(0)} + t(L^0_0)^{(1)} + \cdots + t^{n-3}(L^0_0)^{(n-3)}, 
\label{eq:C0nregrad}
\ee
\be
\dfrac{32\pi G}{3}\rho u_\a u_0=\mathcal{C}_\a   =  (L^0_\a)^{(0)} + t(L^0_\a)^{(1)} +\cdots + t^{n-3}(L^0_\a)^{(n-3)}, 
\label{eq:Canregrad}
\ee
and,
\be
\dfrac{8\pi G}{3}\rho(4u^\b u_\a - \d^\b_\a)=L^\b_\a  =  (L^\b_\a)^{(0)}  + t(L^\b_\a)^{(1)}+ \cdots + t^{n-4}(L^\b_\a)^{(n-4)}. 
\label{eq:LABnregrad}
\ee
Taking into account identity (\ref{velocities identity regular}) and relation (\ref{uoapprox1radreg}), the above system is modified in its final form as follows:
\be
8\pi G\rho=\mathcal{C}_0    =  (L^0_0)^{(0)} + t(L^0_0)^{(1)} + \cdots + t^{n-3}(L^0_0)^{(n-3)}, 
\label{eq:C0nregrad2}
\ee
\be
\dfrac{32\pi G}{3}\rho u_\a = \mathcal{C}_\a   =  (L^0_\a)^{(0)} + t(L^0_\a)^{(1)} +\cdots + t^{n-3}(L^0_\a)^{(n-3)}, 
\label{eq:Canregrad2}
\ee
and,
\be
-\dfrac{8\pi G}{3}\rho \d^\b_\a=L^\b_\a  =  (L^\b_\a)^{(0)}  + t(L^\b_\a)^{(1)}+ \cdots + t^{n-4}(L^\b_\a)^{(n-4)}. 
\label{eq:LABnregrad2}
\ee
Here we have $6n+12$ initial data in total ($6n+6$ by the spatial matrixes $\g^{(m)}_{\a\b}$, $m=1,2,...,n$ and $6$ more by $p,\rho,u^i$). Because of the equation of state (\ref{radiation eq}), the identity (\ref{velocities identity regular}) and the four diffeomorphism changes, we also need to prove that the derived relations by the system (\ref{eq:C0nregrad2})-(\ref{eq:LABnregrad2}) are equal to $6n-11$.

From Eq. (\ref{eq:C0nregrad2}) we find one relation for the energy density $\rho$. Because of (\ref{vel tend to zero}), we must have that,
\be
(L^0_\a)^{(0)} = 0,
\label{zero Loa gen rad reg}
\ee
namely that,
\bq
0 &=& (L^0_\a)^{(0)} = \frac{1}{2}(\nabla_\b b^\b_\a-\nabla_\a b) 
+\ep \left[(-P_0 -2c +\frac{3}{4}b^\b_\a b^\a_\b -\frac{1} {4}b^2)(\nabla_\b b^\b_\a-\nabla_\a b) \right. \nonumber \\
&-& \left. 2\partial_\a(-bc -6d +5b^\b_\a c^\a_\b -\frac{3}{2}b^\b_\a b^\a_\g b^\g_\b 
+\frac{1}{2}b^\b_\a b^\a_\b b -P_1) \right. \nonumber \\
&+& \left. b^\b_\a \partial_\b(-P_0 -2c +\frac{3}{4}b^\b_\a b^\a_\b -\frac{1} {4}b^2) \right],
\label{analzeroL0amixednew equal to zero new}
\eq
and due to the identity (\ref{eq:identity_Lio_zeroth order}), the term $(L^0_0)^{(1)}$ vanishes identically.

Then, substituting $\rho$ from Eq. (\ref{eq:C0nregrad2}) in (\ref{eq:Canregrad2}), we obtain three relations for the velocities, namely,
\be
u_\a = \dfrac{3(L^0_\a)^{(1)}}{4(L^0_0)^{(0)}}t,
\label{fin vel tend zero general}
\ee
or,

\bq
u_\a &=& \dfrac{3t}{4} \left \{(\nabla_\b c^\b_\a - \nabla_\a c) -\frac{1} {2}\nabla_\b (b_\g^\b b_\a^\g) 
+ \frac{1} {2}\nabla_\a (b_\g^\b b_\b^\g) \right.  \nonumber \\
&+& \left. \ep \left [ 2(-P^{(0)} -2c +\frac{3}{4}b^\ep_\d b^\d_\ep -\frac{1} {4}b^2)[(\nabla_\b c^\b_\a - \nabla_\a c) 
-\frac{1} {2}\nabla_\b (b_\g^\b b_\a^\g) + \frac{1} {2}\nabla_\a (b_\g^\b b_\b^\g)] \right. \right. \nonumber \\
&+& \left. \left. (-bc -6d +5b^\d_\g c^\g_\d -\frac{3}{2}b^\ep_\d b^\d_\g b^\g_\ep +
\frac{1}{2}b^\d_\g b^\g_\d b-P^{(1)})(\nabla_\b b^\b_\a - \nabla_\a b)  \right. \right. \nonumber \\
&-& \left. \left. 4\partial_\a (-\frac{3}{2}bd -c^2 -12e +\frac{21}{2}b^\g_\b d^\b_\g +5c^\g_\b c^\b_\g -
\frac{19}{2}b^\d_\g b^\g_\b c^\b_\d +\frac{9}{4}b^\ep_\d b^\d_\g b^\g_\b b^\b_\ep + \frac{3} {2}b^\g_\b c^\b_\g b \right. \right. \nonumber  \\
&+& \left. \left. b^\g_\b b^\b_\g c -\frac{1} {2}b^\d_\g b^\g_\b b^\b_\d b - \frac{1} {4}(b^\g_\b b^\b_\g)^2-P^{(2)}) \right. \right. \nonumber \\
&+& \left. \left. b^\b_\a \partial_\b(-bc -6d +5b^\d_\g c^\g_\d -\frac{3}{2}b^\ep_\d b^\d_\g b^\g_\ep + \frac{1}{2}b^\d_\g b^\g_\d b-P^{(1)}) \right. \right. \nonumber \\
&+& \left. \left. (2c^\b_\a -b^\b_\g b^\g_\a) \partial_\b (-P^{(0)} -2c +\frac{3}{4}b^\ep_\d b^\d_\ep -\frac{1} {4}b^2) \right ] \right \} \nonumber \\
& \times & \left \{ \frac{1}{6}P^{(0)} + \frac{1}{8}b^\b_\a b^\a_\b +\frac{1}{24}b^2 - \dfrac{2}{3}c 
+\ep \left [2(-P^{(0)} -2c +\frac{3}{4}b^\b_\a b^\a_\b -\frac{1} {4}b^2)(-c + \frac{1}{4}b^\d_\g b^\g_\d) \right. \right. \nonumber \\ 
&-& \left. \left. \frac{1}{2}(-P^{(0)} -2c +\frac{3}{4}b^\b_\a b^\a_\b -\frac{1} {4}b^2)^2 \right. \right. \nonumber \\
&-& \left. \left. 4\left(-\frac{3}{2}bd -c^2 -12e +\frac{21}{2}b^\b_\a d^\a_\b +5c^\b_\a c^\a_\b 
- \frac{19}{2}b^\b_\a b^\a_\g c^\g_\b +\frac{9}{4}b^\b_\g b^\g_\a b^\a_\d b^\d_\b 
+ \frac{3} {2}b^\b_\a c^\a_\b b \right. \right. \right.    \nonumber  \\
&+& \left. \left. \left. b^\b_\a b^\a_\b c -\frac{1} {2}b^\g_\b b^\b_\a b^\a_\g b 
- \frac{1} {4}(b^\b_\a b^\a_\b)^2-P^{(2)}\right) \right ] \right \}^{-1}.
\label{analoneL0amixednew with ua new}
\eq
Finally, substituting $\rho$ from (\ref{eq:C0nregrad2}) to (\ref{eq:LABnregrad2}), we find that,
\be
(L^\b_\a)^{(k)} = -\dfrac{1}{3}\d^\b_\a (L^0_0)^{(k)},
\label{eq for Lab rad reg}
\ee
where $k=0,1,...,n-4$. Thus, Eqns. (\ref{eq for Lab rad reg}) give $6n-18$ relations between the initial data $a_{\a\b}, b_{\a\b}, c_{\a\b}, d_{\a\b},e_{\a\b}$. Then, Eqns. (\ref{eq:C0nregrad2}), (\ref{zero Loa gen rad reg}), (\ref{fin vel tend zero general}) and (\ref{eq for Lab rad reg}) give $6n-11$ relations in total.

Therefore, similarly to the case of vacuum, without loss of generality, we can study the problem of finding the number of the arbitrary functions in the regular case in radiation models, using (\ref{eq:3dimmetric}). 

To show the connection between the results in general relativity and higher-order gravity, in Appendix \ref{AppendixA} we give details concerning the regular case of vacuum and radiation models in general relativity.


\section{Choice of the initial data}
\label{Choice of the initial data}

The question emerging now is: In case of vacuum, out of the 30 different functions $a_{\a\b},b_{\a\b},c_{\a\b},d_{\a\b},e_{\a\b}$, and in case of radiation out of the 36 different functions $a_{\a\b},b_{\a\b},c_{\a\b},d_{\a\b},e_{\a\b},p,\rho,u^i$ which sixteen-twenty of those should be chosen as our initial data?


\subsection{Initial data in vacuum}
\label{Initial data in vacuum}

We have shown in Subsection \ref{Function-counting} that the vacuum higher-order gravity equations (\ref{eq:pg}), (\ref{eq:pK}), (\ref{eq:pD}) and (\ref{eq:pW}) together with the constraints  (\ref{eq:hamiltonian}) and (\ref{eq:momentum}), admit a regular formal series expansion of the form  (\ref{eq:3dimmetric}) as a general solution requiring $16$ smooth initial data. If we prescribe the thirty data
 \be a_{\a\b}, \quad  b_{\a\b},\quad  c_{\a\b},\quad  d_{\a\b},\quad  e_{\a\b},
\ee
initially, we still have the freedom to fix $14$ of them. We choose to leave the six components of the metric $a_{\a\b}$ free, and we choose the four symmetric space tensors $b_{\a\b},c_{\a\b},d_{\a\b}$ and $e_{\a\b}$ to be traceless with respect to $a_{\a\b}$. Then, we proceed to count the number of free functions in several steps, starting from these $6+4\times 5=26$ functions. Firstly,  (\ref{zeroL000vr}) fixes one of the components of $b_{\a\b}$ and Eq. (\ref{zeroL0a0vr}) fixes $3$ more components of $ b_{\a\b}$, thus leaving  $ b_{\a\b}$ with one component. Further, we use  the $6$ relations in (\ref{zeroLab0vr}) to completely fix the remaining $5$ components of $c_{\a\b}$ and the last of $b_{\a\b}$. Summing up the free functions we have found, we end up with,
\be
\underbrace{6}_{\textrm{from}\ a_{\a\b}}+\underbrace{0}_{\textrm{from}\ b_{\a\b}}+\underbrace{0}_{\textrm{from}\ c_{\a\b}}+\underbrace{10}_{\textrm{from}{\ d_{\a\b}}\ \textrm{and}\ e_{\a\b}}=16
\label{eq:notation_initial_data}
 \ee
  suitable free data as required for the solution to be a general one.
We thus arrive at the following result which summarizes what we have shown in this Section, and generalizes a theorem of Rendall \cite{ren1} for higher order gravity theories that derive from the lagrangian $R+\ep R^2$.
\begin{theorem}
Let $a_{\a\b}$ be a smooth Riemannian metric , $b_{\a\b},c_{\a\b},d_{\a\b}$ and $e_{\a\b}$  be symmetric smooth tensor fields which are traceless  with respect to the metric $a_{\a\b}$, i.e., they satisfy $b=c=d=e=0$.  Then there exists a formal power series expansion solution of the vacuum higher order gravity equations of the form (\ref{eq:3dimmetric}) such that:
\begin{enumerate}
\item It is unique
\item The coefficients $\g^{(n)\,\a\b}$ are all smooth
\item It holds that $\g^{(0)}_{\a\b}=a_{\a\b}$ and $\g^{(1)}_{\a\b}=b_{\a\b},$ $\g^{(2)}_{\a\b}=c_{\a\b}$, $\g^{(3)}_{\a\b}=d_{\a\b}$ and  $\g^{(4)}_{\a\b}=e_{\a\b}$.
\end{enumerate}
\end{theorem}
In the course of the proof of this result, uniqueness followed because all coefficients were found recursively, while smoothness follows because in no step of the proof did we find it necessary to lower the $\mathcal{C}^\infty$ assumption. We also note that $b_{\a\b}$ and $c_{\a\b}$ are necessarily transverse with respect to $a_{\a\b}$.

In addition to this, taking into account that the {3}-metric $\g_{\a\b}$ is considered equal to the spatial matrix $a_{\a\b}$ when $t\rightarrow 0$, we obtain the following additional conditions for the six free components of the metric $a_{\a\b}$,
\be
a_{11}>0,\quad 
\det{\left(\begin{array}{cc}
a_{11} & a_{12} \\ 
a_{21} & a_{22}
\end{array}\right)}>0,\quad 
\det{(a_{\a\b})}>0,
\label{eq:determinant_a_ab}
\ee
which are directly coming through the three inequalities (\ref{eq:determinant_g_ab}). These conditions inform us that the three diagonal components of the $a_{\a\b}$ satisfy,
\be
a_{11}>0,\quad a_{22}>0,\quad a_{33}>0,
\label{diagonal_a_ab}
\ee
which, of course, allow us to determine more properly  the unknown matrix $a_{\a\b}$, but they don' t give us the opportunity to lower the number of the arbitrary functions in our analysis (such a subtraction is being acceptable only for equalities).
    
All in all, in order to clarify what we mentioned in Section \ref{Analytic solution of the local Cauchy problem} regarding the difference in analysis which we have to follow between the Jordan and the conformal frame, interpreting the notation of the initial data (\ref{eq:notation_initial_data}), we can easily understand why there is a difference between the number of the arbitrary functions in Jordan ($16$ d.o.f.) and its conformal frame ($6$ d.o.f.). We realize, that the ten more initial data, which the Jordan frame requires, emerge from the spatial matrixes $D_{\a\b}$ and $W_{\a\b}$, quantities that do not appear in the pertubative analysis of the conformal frame, because of the second order equations that describe it.


\subsection{Initial data in the case of radiation}
\label{Initial data in the case of radiation}

In Subsection \ref{Function counting in the presence of a fluid} we showed that the higher-order gravity equations (\ref{eq:pg}), (\ref{eq:pK}), (\ref{eq:pD}) and (\ref{eq:pW with T}) together with the constraints  (\ref{eq:hamiltonian_rad}) and (\ref{eq:momentum_rad}), admit a regular formal series expansion of the form (\ref{eq:3dimmetric}) as a general solution requiring $17$ smooth initial data. If we prescribe the $36$ data
 \be a_{\a\b}, \quad  b_{\a\b},\quad  c_{\a\b},\quad  d_{\a\b},\quad  e_{\a\b},\quad p,\quad \rho,\quad u^i
\ee
initially, we have the freedom to fix $19$ of them. We also choose to leave the six components of the metric $a_{\a\b}$ free. Then we proceed to count the number of free functions in several steps. 

First, (\ref{constant energy density}) fixes the function $\rho$ and Eq. (\ref{fin vel tend zero}) fixes $3$ more components of $u^i$. Further, we use the 6 relations in (\ref{LbaLooradreg}) to completely fix the $6$ components of $c_{\a\b}$ and the three relations in (\ref{analzeroL0amixednew equal to zero}) to fix three components of $b_{\a\b}$. Taking into account the identity (\ref{velocities identity regular}), the equation of state (\ref{radiation eq}) and four diffeomorphism transformations that we use to fix two components for each one of $d_{\a\b},e_{\a\b}$, we end up with,
\be
\underbrace{6}_{\textrm{from}\ a_{\a\b}}+\underbrace{3}_{\textrm{from}\ b_{\a\b}}+
\underbrace{0}_{\textrm{from}\ c_{\a\b}}+\underbrace{4}_{\textrm{from}\ d_{\a\b}}+\underbrace{4}_{\textrm{from}\ e_{\a\b}}+
\underbrace{0}_{\textrm{from}{\ p}\ \textrm{and}\ \rho}+\underbrace{0}_{\textrm{from}\ u^i}=17
\label{eq:notation_initial_data}
 \ee
suitable free data as required for the solution.

Obviously, in this case too, the three inequalities (\ref{eq:determinant_a_ab}) are still valid as well as the inequalities (\ref{diagonal_a_ab}), giving a better approach for the components of the tensor $a_{\a\b}$. \\
However, the choice of the initial data presented in both cases above (vacuum and radiation)is not unique, but there are many ways to choose them.

For instance, in the case of radiation, if we use the four diffeomorphism transformations not to fix two components for each one of $d_{\a\b},e_{\a\b}$, but to fix four of the components of $e_{\a\b}$, then we would resulted with,
\be
\underbrace{6}_{\textrm{from}\ a_{\a\b}}+\underbrace{3}_{\textrm{from}\ b_{\a\b}}+
\underbrace{0}_{\textrm{from}\ c_{\a\b}}+\underbrace{6}_{\textrm{from}\ d_{\a\b}}+\underbrace{2}_{\textrm{from}\ e_{\a\b}}+
\underbrace{0}_{\textrm{from}{\ p}\ \textrm{and}\ \rho}+\underbrace{0}_{\textrm{from}\ u^i}=17.
\label{eq:notation_initial_data}
 \ee
suitable free data as required for the solution.

An other option is to use three of the four diffeomorphism transformations in order to bring $a_{\a\b}$ to diagonal form and the last one to fix one more component of the matrix $b_{\a\b}$. Then, we end up with,
\be
\underbrace{3}_{\textrm{from}\ a_{\a\b}}+\underbrace{2}_{\textrm{from}\ b_{\a\b}}+
\underbrace{0}_{\textrm{from}\ c_{\a\b}}+\underbrace{6}_{\textrm{from}\ d_{\a\b}}+\underbrace{6}_{\textrm{from}\ e_{\a\b}}+
\underbrace{0}_{\textrm{from}{\ p}\ \textrm{and}\ \rho}+\underbrace{0}_{\textrm{from}\ u^i}=17
\label{eq:notation_initial_data}
 \ee
suitable free data as required for the solution.


\chapter{Singular universes in higher-order gravity} 

\label{Chapter6} 

\lhead{Chapter 6. \emph{Singular universes in higher-order gravity}} 

In this Chapter, we study the form and the type of the solution of the gravitational field equations of higher-order gravity in vacuum as well as in the case of radiation, in the vicinity of a point that is not regular, but singular in the time. In Section \ref{Analytic field equations through singular formal series expansion}, we present the higher-order field equations by using the $3$-metric (\ref{spatial metric rad}), and we calculate the various quantities of these equations in terms of the initial data $a_{\a\b},b_{\a\b},c_{\a\b},d_{\a\b},p ,\rho,u^i$. In Section \ref{Regular vacuum field equations}, we show that the form (\ref{spatial metric rad}) can not provide any kind of solution of the higher-order gravity equations in vacuum, and also, in Section \ref{Regular field equations in case of radiation}, we use the same {3}-metric to search for the form of the solution in the case of radiation. In Section \ref{Degrees of freedom general}, we show that the degrees of freedom in case of radiation remain the same, if the order of the singular {3}-metric becomes greater than $4$, and finally, in Section \ref{Choice of the initial data}, we present the choice of the initial data between the initial functions, taking into account the corresponding function-counting problem of Chapter \ref{Chapter2}.


\section{Analytic field equations through singular formal series expansion}
\label{Analytic field equations through singular formal series expansion}

In this Section we present the gravitational field equations of higher-order gravity theory in vacuum derived from the analytic lagrangian $f(R)=R+\ep R^2$, that is the Eqns. (\ref{eq:Loo}), (\ref{eq:Loa}) and (\ref{eq:Lab}), through singular formal series expansion (\ref{spatial metric rad}) and find the form of their solution. We also present the corresponding field equations in the case of radiation by using the same series.


\subsection{Series form of field equations}
\label{Series form of field equations2}

Similarly to the previous Chapter, in order to simplify our further analysis, we take into account the order of the higher-order differential equations in vacuum. In particular, Eqns. (\ref{eq:Loo}), (\ref{eq:Loa}) are third-order differential equations with respect to the proper time $t$ and Eq. (\ref{eq:Lab}) is fourth-order differential equation with respect to $t$. 

Therefore, we can use the following notation:
\be
\mathcal{C}_0 = (L^0_0)^{(-3)}\frac{1}{t^3} + (L^0_0)^{(-2)}\frac{1}{t^2} + (L^0_0)^{(-1)}\frac{1}{t},
\label{eq:C0_s}
\ee
\be
\mathcal{C}_\a = (L^0_\a)^{(-2)}\frac{1}{t^2} + (L^0_\a)^{(-1)}\frac{1}{t} + (L^0_\a)^{(0)},
\label{eq:Ca_s}
\ee
and,
\be
L^\b_\a = (L^\b_\a)^{(-3)}\frac{1}{t^3} + (L^\b_\a)^{(-2)}\frac{1}{t^2} + (L^\b_\a)^{(-1)}\frac{1}{t}.
\label{eq:LAB_s}
\ee
We note that this notation pertains to both cases of vacuum and radiation, because the order of the differential equations (\ref{eq:Loo_rad}), (\ref{eq:Loa_rad}) and (\ref{eq:Lab_rad}) is correspondingly the same with the order of vacuum differential equations. Apparently, the five terms $(L^0_0)^{(0)},(L^0_0)^{(1)},(L^0_\a)^{(0)},(L^0_\a)^{(1)},(L^\b_\a)^{(0)}$ appearing in (\ref{eq:C0}), (\ref{eq:Ca}) and (\ref{eq:LAB}) are not the same in both cases, but they have the same dependence on the matrices $a_{\a\b},b_{\a\b},c_{\a\b},d_{\a\b}$. Thus, as in the case of regular series expansion, we also present this dependence both for vacuum and radiation models once.


\subsection{Calculation of the series terms}
\label{Calculation of the series terms2}

In terms of $a_{\a\b},b_{\a\b},c_{\a\b},d_{\a\b}$, using relations (\ref{eq:R00mixed_rad}), (\ref{eq:R0amixed_rad}), (\ref{eq:Rabmixed_rad}) and (\ref{Rscalar_rad}), the quantities appearing in Eq. (\ref{eq:C0_s}) become,
\bq
(L^0_0)^{(-3)} &=& \ep \big[2R^{(-1)}(R^0_0)^{(-2)} - K^{(-1)}R^{(-1)}\big] \nonumber \\
               &=& \dfrac{3}{2}\ep(\widetilde{P} + 2b),
\label{L^0_0(-3)_rad}
\eq
\bq
(L^0_0)^{(-2)} &=& (R^0_0)^{(-2)} + \ep \left[2\left(R^{(-1)}(R^0_0)^{(-1)} + R^{(0)}(R^0_0)^{(-2)}\right) - \dfrac{1}{2}(R^{(-1)})^2  \right. \nonumber \\
&-& \left.  2(\g^{\a\b})^{(-1)}\pa_\a(\pa_\b R^{(-1)}) +2(\g^{\a\b})^{(-1)}(\G^\m_{\a\b})^{(0)}\pa_\m R^{(-1)} - K^{(0)}R^{(-1)}\right] \nonumber \\
&=& \dfrac{3}{4} + \ep \left[-\dfrac{1}{2}(\widetilde{P}^2 - 4b^2)
+ \dfrac{3}{2}(-6c - \dfrac{1}{4}b^2 + \dfrac{11}{4}b^\b_\a b^\a_\b - \k)
+ 2a^{\a\b}\pa_\a \big(\pa_\b (\widetilde{P} + 2b)\big) \right. \nonumber \\
&-& \left. 2a^{\a\b}\widetilde{\G}^\m_{\a\b}\pa_\m(\widetilde{P} + 2b) \right],
\label{L^0_0(-2)_rad}
\eq
\bq
(L^0_0)^{(-1)} &=& (R^0_0)^{(-1)} - \dfrac{1}{2}R^{(-1)} + \ep \left \{ 2\left[R^{(-1)}(R^0_0)^{(0)} + R^{(0)}(R^0_0)^{(-1)}
+ R^{(1)}(R^0_0)^{(-2)}\right]  \right. \nonumber \\
&-& \left. R^{(-1)}R^{(0)} - 2\left[(\g^{\a\b})^{(-1)}\pa_\a(\pa_\b R^{(0)}) + (\g^{\a\b})^{(0)}\pa_\a(\pa_\b R^{(-1)})\right] \right. \nonumber \\
&+& \left. 2\left[(\g^{\a\b})^{(-1)}(\G^\m_{\a\b})^{(0)}\pa_\m R^{(0)} + (\g^{\a\b})^{(0)}(\G^\m_{\a\b})^{(0)}\pa_\m R^{(-1)}
 \right. \right. \nonumber \\
&+&\left. \left.  (\g^{\a\b})^{(-1)}(\G^\m_{\a\b})^{(1)}\pa_\m R^{(-1)}\right] +\left[K^{(-1)}R^{(1)} - K^{(1)}R^{(-1)}\right] \right \}  \nonumber \\
&=& \dfrac{1}{2}(\widetilde{P}+b) + \ep \left[ (\widetilde{P} + 2b)(\dfrac{1}{4}b^\b_\a b^\a_\b - \dfrac{1}{4}b^2 - \k)
- b(-6c -\frac{1}{4}b^2 + \frac{11}{4}b^\b_\a b^\a_\b - \k) \right. \nonumber \\
&+& \left. \dfrac{9}{2}(-12d -bc + 11b^\b_\a c^\a_\b - \frac{7}{2}b^\b_\g b^\g_\a b^\a_\b + \frac{1}{2}bb^\b_\a b^\a_\b - \l)
- 2b^{\a\b}\pa_\a \left(\pa_\b(\widetilde{P}+2b)\right) \right. \nonumber \\
&-& \left. 2a^{\a\b}\pa_\a \left(\pa_\b (-6c -\frac{1}{4}b^2 + \frac{11}{4}b^\d_\g b^\g_\d - \k)\right)
+ 2b^{\a\b}\widetilde{\G}^\m_{\a\b}\pa_\m(\widetilde{P} + 2b) \right. \nonumber \\
&+& \left. 2a^{\a\b}\widetilde{\G}^\m_{\a\b}\pa_\m (-6c -\frac{1}{4}b^2 + \frac{11}{4}b^\d_\g b^\g_\d - \k)
- 2a^{\a\b}E^\m_{\a\b}\pa_\m (\widetilde{P} + 2b) \right].
\label{L^0_0(-1)_rad}
\eq
Further, the quantities appearing in Eq. (\ref{eq:Ca_s}) become,
\bq
(L^0_\a)^{(-2)} &=& \ep \big[2\pa_\a R^{(-1)} + (K^\b_\a)^{(-1)}\pa_\b R^{(-1)} \big] \nonumber \\
&=& -3\ep \pa_\a (\widetilde{P} + 2b),
\label{L^0_a(-2)_rad}
\eq
\bq
(L^0_\a)^{(-1)} &=& \ep \left[2R^{(-1)}(R^0_\a)^{(0)} + (K^\b_\a)^{(-1)}\pa_\b R^{(0)} + (K^\b_\a)^{(0)}\pa_\b R^{(-1)} \right] \nonumber \\
&=& \ep \left[-(\widetilde{P} + 2b)(\nabla_\b b^\b_\a - \nabla_\a b)
+ \pa_\a (-6c -\frac{1}{4}b^2 + \frac{11}{4}b^\d_\g b^\g_\d - \k) \right. \nonumber \\
&-& \left. b^\b_\a \pa_\b (\widetilde{P} + 2b) \right],
\label{L^0_a(-1)_rad}
\eq
\bq
(L^0_\a)^{(0)} &=& (R^0_\a)^{(0)} + \ep \left[2\left(R^{(-1)}(R^0_\a)^{(1)} + R^{(0)}(R^0_\a)^{(0)}\right) - 2\pa_\a R^{(1)} \right. \nonumber \\
&+& \left. \left((K^\b_\a)^{(-1)}\pa_\b R^{(1)} + (K^\b_\a)^{(0)}\pa_\b R^{(0)} + (K^\b_\a)^{(1)}\pa_\b R^{(-1)}\right) \right] \nonumber \\
&=& \frac{1}{2}(\nabla_\b b^\b_\a -\nabla_\a b) + \ep \left[-2(\widetilde{P} + 2b)\big[(\nabla_\b c^\b_\a -\nabla_\a c) 
-\frac{1}{2}\nabla_\b (b^\b_\g b^\g_\a) +\frac{1}{2}\nabla_\a (b^\b_\g b^\g_\b)\big] \right. \nonumber \\
&+& \left. (-6c -\frac{1}{4}b^2 + \frac{11}{4}b^\d_\g b^\g_\d - \k)(\nabla_\b b^\b_\a -\nabla_\a b) \right. \nonumber \\
&-& \left. \pa_\a (-12d -bc + 11b^\d_\g c^\g_\d - \frac{7}{2}b^\b_\g b^\g_\d b^\d_\b + \frac{1}{2}bb^\d_\g b^\g_\d - \l)
\right. \nonumber \\
&+& \left. b^\b_\a \pa_\b (-6c -\frac{1}{4}b^2 + \frac{11}{4}b^\d_\g b^\g_\d - \k)
- (2c^\b_\a - b^\b_\g b^\g_\a)\pa_\b (\widetilde{P} + 2b) \right],
\label{L^0_a(0)_rad}
\eq
and finally for the quantities of Eq. (\ref{eq:LAB_s}) we have,
\bq
(L^\b_\a)^{(-3)} &=& \ep \big[2R^{(-1)}(R^\b_\a)^{(-2)} + (K^\b_\a)^{(-1)}R^{(-1)} + 4\d^\b_\a R^{(-1)}
- \d^\b_\a K^{(-1)}R^{(-1)} \big] \nonumber \\
&=& -\dfrac{3}{2}\ep \d^\b_\a (\widetilde{P} + 2b),
\label{L^b_a(-3)_rad}
\eq
\bq
(L^\b_\a)^{(-2)} &=& (R^\b_\a)^{(-2)} + \ep \left\{2\left[R^{(-1)}(R^\b_\a)^{(-1)} + R^{(0)}(R^\b_\a)^{(-2)} \right]
-\dfrac{1}{2}\d^\b_\a (R^{(-1)})^2 \right. \nonumber \\
&+& \left. 2(\g^{\b\g})^{(-1)}\pa_\g (\pa_\a R^{(-1)}) - 2(\g^{\b\g})^{(-1)}(\G^\m_{\g\a})^{(0)}\pa_\m R^{(-1)}
+ (K^\b_\a)^{(0)}R^{(-1)} \right. \nonumber \\
&-& \left. 2\d^\b_\a(\g^{\g\d})^{(-1)}\pa_\g (\pa_\d R^{(-1)})
+ 2\d^\b_\a(\g^{\g\d})^{(-1)}(\G^\m_{\g\d})^{(0)} (\pa_\m R^{(-1)}) - \d^\b_\a K^{(0)}R^{(-1)} \right\} \nonumber \\
&=& -\dfrac{1}{4}\d^\b_\a + \ep \left[(\widetilde{P} + 2b)(2\widetilde{P}^\b_\a + \dfrac{1}{2}b^\b_\a + \dfrac{1}{2}b\d^\b_\a
- \dfrac{1}{2}\widetilde{P}\d^\b_\a) \right. \nonumber \\
&-& \left. \dfrac{1}{2}\d^\b_\a (-6c -\frac{1}{4}b^2 + \frac{11}{4}b^\d_\g b^\g_\d - \k) \right. \nonumber \\
&-& \left. 2a^{\b\g}\pa_\g \left(\pa_\a(\widetilde{P} + 2b)\right) + 2a^{\b\g}\widetilde{\G}^\m_{\g\a}\pa_\m(\widetilde{P} + 2b)
+2\d^\b_\a a^{\g\d}\pa_\g \left(\pa_\d(\widetilde{P} + 2b)\right) \right. \nonumber \\
&-& \left. 2\d^\b_\a a^{\g\d}\widetilde{\G}^\m_{\g\d}\pa_\m(\widetilde{P} + 2b) \right],
\label{L^b_a(-2)_rad}
\eq
\bq
(L^\b_\a)^{(-1)} &=& (R^\b_\a)^{(-1)} - \dfrac{1}{2}\d^\b_\a R^{(-1)} + \ep \left\{2\left[R^{(-1)}(R^\b_\a)^{(0)} + R^{(0)}(R^\b_\a)^{(-1)} + R^{(1)}(R^\b_\a)^{(-2)}\right] \right. \nonumber \\
&-& \left. R^{(-1)}R^{(0)}\d^\b_\a + 2\left[(\g^{\b\g})^{(-1)}\pa_\g(\pa_\a R^{(0)}) + (\g^{\b\g})^{(0)}\pa_\g(\pa_\a R^{(-1)}) \right] \right. \nonumber \\
&-& \left. 2\left[(\g^{\b\g})^{(-1)}(\G^\m_{\g\a})^{(0)}\pa_\m R^{(0)} + (\g^{\b\g})^{(0)}(\G^\m_{\g\a})^{(0)}\pa_\m R^{(-1)}
\right. \right. \nonumber \\
&+& \left. \left. (\g^{\b\g})^{(-1)}(\G^\m_{\g\a})^{(1)}\pa_\m R^{(-1)} \right] - (K^\b_\a)^{(-1)}R^{(1)} + (K^\b_\a)^{(1)}R^{(-1)} \right.  \nonumber \\
&-& \left. 2\d^\b_\a \left[(\g^{\g\d})^{(-1)}\pa_\g(\pa_\d R^{(0)})
+ (\g^{\g\d})^{(0)}\pa_\g(\pa_\d R^{(-1)}) \right] \right. \nonumber \\
&+& \left. 2\d^\b_\a \left[(\g^{\g\d})^{(-1)}(\G^\m_{\g\d})^{(0)}\pa_\m R^{(0)}
+ (\g^{\g\d})^{(0)}(\G^\m_{\g\d})^{(0)}\pa_\m R^{(-1)} \right.  \right. \nonumber \\
&+& \left. \left. (\g^{\g\d})^{(-1)}(\G^\m_{\g\d})^{(1)}\pa_\m R^{(-1)} \right] + \d^\b_\a \left(K^{(-1)}R^{(1)} - K^{(1)}R^{(-1)} \right) \right\} \nonumber \\
&=& -\widetilde{P}^\b_\a + \dfrac{1}{2}\widetilde{P}\d^\b_\a - \dfrac{3}{4}b^\b_\a + \dfrac{3}{4}b\d^\b_\a \nonumber \\
&+& \ep \left[\dfrac{3}{2}\d^\b_\a(-12d - bc + 11b^\d_\g c^\g_\d - \dfrac{7}{2}b^\d_\g b^\g_\ep b^\ep_\d
+ \dfrac{1}{2}bb^\d_\g b^\g_\d - \l) \right. \nonumber \\
&+& \left. (\widetilde{P} + 2b)(3c^\b_\a - 3c\d^\b_\a + \dfrac{5}{4}b^\d_\g b^\g_\d \d^\b_\a - \dfrac{1}{4}b^2 \d^\b_\a
- \dfrac{3}{2}b^\b_\g b^\g_\a + \dfrac{1}{2}bb^\b_\a - \d^\b_\a \k + 2\k^\b_\a) \right. \nonumber \\
&-& \left. \dfrac{1}{2}(4\widetilde{P}^\b_\a + 3b^\b_\a + b\d^\b_\a)(-6c - \dfrac{1}{4}b^2 + \dfrac{11}{4}b^\d_\g b^\g_\d - \k)
- 2\d^\b_\a a^{\g\d}E^\m_{\g\d}\pa_\m (\widetilde{P} + 2b) \right. \nonumber \\
&+& \left. 2b^{\b\g}\pa_\g \left(\pa_\a(\widetilde{P} + 2b) \right)
- 2a^{\b\g}\widetilde{\G}^\m_{\g\a}\pa_\m (-6c - \dfrac{1}{4}b^2 + \dfrac{11}{4}b^\d_\ep b^\ep_\d - \k) \right. \nonumber \\
&-& \left. 2b^{\b\g}\widetilde{\G}^\m_{\g\a}\pa_\m (\widetilde{P} + 2b)
+ 2a^{\b\g}E^\m_{\g\a}\pa_\m (\widetilde{P} + 2b) \right. \nonumber \\
&-& \left. 2\d^\b_\a a^{\g\d}\pa_\g \left(\pa_\d (-6c - \dfrac{1}{4}b^2 + \dfrac{11}{4}b^\z_\ep b^\ep_\z - \k) \right) \right. 
\nonumber \\
&-& \left. 2\d^\b_\a b^{\g\d}\pa_\g \left(\pa_\d (\widetilde{P} + 2b) \right)
+ 2\d^\b_\a a^{\g\d}\widetilde{\G}^\m_{\g\d}\pa_\m (-6c - \dfrac{1}{4}b^2 + \dfrac{11}{4}b^\z_\ep b^\ep_\z - \k) \right. \nonumber \\
&+& \left. 2\d^\b_\a b^{\g\d}\widetilde{\G}^\m_{\g\d}\pa_\m (\widetilde{P} + 2b)
+ 2a^{\b\g}\pa_\g \left(\pa_\a(-6c - \dfrac{1}{4}b^2 + \dfrac{11}{4}b^\d_\ep b^\ep_\d - \k) \right) \right]. 
\label{L^b_a(-1)_rad}
\eq
We now proceed to find the form of the solution of the higher-order gravitanional field equations in vacuum, in the vicinity of a point that is singular in the time.


\subsection{Simplification of the terms}
\label{Simplification of the terms2} 

We can simplify these nine relations (\ref{L^0_0(-3)_rad})-(\ref{L^b_a(-1)_rad}) by using the trace of the higher-order field equations. In both cases (vacuum and radiation) the trace of the fields equation gives the identity,
\be
R - 6\ep \Box_g R = 0.
\label{idBox2}
\ee
Then, for the $t^{-3}$ order terms we obtain the identity,
\be
\widetilde{P} + 2b = 0,
\label{eq for P}
\ee 
for the $t^{-2}$ order terms we obtain,
\be
-12d -bc + 11b^\b_\a c^\a_\b - \frac{7}{2}b^\b_\g b^\g_\a b^\a_\b + \frac{1}{2}bb^\b_\a b^\a_\b - \l = 0,
\label{eq for R1}
\ee
and finally for the $t^{-1}$ order terms we have,
\be
\dfrac{1}{2} a^{\a\b} a^{\m\ep}A_{\a\b\ep}\pa_\m \left( -6c -\frac{1}{4}b^2 + \frac{11}{4}b^\b_\a b^\a_\b - \k \right) = a^{\a\b}\pa_\a \left( \pa_\b (-6c -\frac{1}{4}b^2 + \frac{11}{4}b^\d_\g b^\g_\d - \k)\right).
\label{eq for paR0}
\ee

Then, using identities (\ref{eq for P}), (\ref{eq for R1}) and (\ref{eq for paR0}), the quantities $(L^0_0)^{(-3)},(L^0_\a)^{(-2)}$ and $(L^\b_\a)^{(-3)}$ vanish identically. The other six quantities appearing in (\ref{L^0_0(-3)_rad})-(\ref{L^b_a(-1)_rad}) become,
\be
(L^0_0)^{(-2)} = \dfrac{3}{4} + \dfrac{3}{2}\ep( -6c -\frac{1}{4}b^2 + \frac{11}{4}b^\b_\a b^\a_\b - \k ),
\label{L^0_0(-2)_radnew}
\ee
\be
(L^0_0)^{(-1)} = -\dfrac{1}{2}b -\ep b( -6c -\frac{1}{4}b^2 + \frac{11}{4}b^\b_\a b^\a_\b - \k ),
\label{L^0_0(-1)_radnew}
\ee
and,
\be
(L^0_\a)^{(-1)} = \ep \pa_\a ( -6c -\frac{1}{4}b^2 + \frac{11}{4}b^\d_\g b^\g_\d - \k ),
\label{L^0_a(-1)_radnew}
\ee
\bq
(L^0_\a)^{(0)} &=& \frac{1}{2}(\nabla_\b b^\b_\a -\nabla_\a b) 
+ \ep \left[ (-6c -\frac{1}{4}b^2 + \frac{11}{4}b^\d_\g b^\g_\d - \k)(\nabla_\b b^\b_\a -\nabla_\a b) \right. \nonumber \\ 
&+& \left. b^\b_\a \pa_\b (-6c -\frac{1}{4}b^2 + \frac{11}{4}b^\d_\g b^\g_\d - \k) \right],
\label{L^0_a(0)_radnew}
\eq
and also,
\be
(L^\b_\a)^{(-2)} = -\dfrac{1}{4}\d^\b_\a - \dfrac{1}{2}\ep \d^\b_\a ( -6c -\frac{1}{4}b^2 + \frac{11}{4}b^\d_\g b^\g_\d - \k ),
\label{L^b_a(-2)_radnew}
\ee
\bq
(L^\b_\a)^{(-1)} &=& -\widetilde{P}^\b_\a - \dfrac{3}{4}b^\b_\a - \dfrac{1}{4}b\d^\b_\a \nonumber \\
&+& \ep \left[ \dfrac{1}{2}(4\widetilde{P}^\b_\a + 3b^\b_\a + b\d^\b_\a)(-6c - \dfrac{1}{4}b^2 + \dfrac{11}{4}b^\d_\g b^\g_\d - \k) \right. \nonumber \\
&-& \left. a^{\b\g}a^{\m\ep}A_{\g\a\ep}\pa_\m (-6c - \dfrac{1}{4}b^2 + \dfrac{11}{4}b^\d_\ep b^\ep_\d - \k) \right. \nonumber \\
&+& \left. 2a^{\b\g}\pa_\g \left(\pa_\a(-6c - \dfrac{1}{4}b^2 + \dfrac{11}{4}b^\d_\ep b^\ep_\d - \k)\right) \right]
\label{L^b_a(-1)_radnew}
\eq


\subsection{Further simplification of the terms}
\label{Further simplification of the terms}

Using the identity (\ref{clL}), we obtain, 
\be
\nabla_0 L^0_j + \nabla_\b L^\b_j=0,
\label{idLij}
\ee
and for $j=\a$ we find,
\be
\pa_t L^0_\a - \dfrac{1}{2}K^\b_\a L^0_\b + \pa_\b L^\b_\a + KL^0_\a - \G^\g_{\a\b}L^\b_\g + \G^\b_{\b\g}L^\g_\a = 0.
\label{identity of L^i_a}
\ee
We substitute the series of the various terms appearing in (\ref{identity of L^i_a}) and then, for the $t^{-2}$ order we find that the coefficient $(L^0_\a)^{(-1)}$ vanishes identically, namely,
\be
\pa_\a ( -6c -\frac{1}{4}b^2 + \frac{11}{4}b^\d_\g b^\g_\d - \k ) = 0.
\label{id zero paR0}
\ee
Consequently, the terms of (\ref{L^0_a(0)_radnew}) and (\ref{L^b_a(-1)_radnew}) become,
\be
(L^0_\a)^{(0)} = \frac{1}{2}(\nabla_\b b^\b_\a -\nabla_\a b) 
+ \ep  (-6c -\frac{1}{4}b^2 + \frac{11}{4}b^\d_\g b^\g_\d - \k)(\nabla_\b b^\b_\a -\nabla_\a b),
\label{L^0_a(0)_radnewfinal}
\ee
and,
\bq
(L^\b_\a)^{(-1)} &=& -\widetilde{P}^\b_\a - \dfrac{3}{4}b^\b_\a - \dfrac{1}{4}b\d^\b_\a \nonumber \\
&-& \dfrac{1}{2} \ep (4\widetilde{P}^\b_\a + 3b^\b_\a + b\d^\b_\a)(-6c - \dfrac{1}{4}b^2 + \dfrac{11}{4}b^\d_\g b^\g_\d - \k). 
\label{L^b_a(-1)_radnewfinal}
\eq
We can now proceed to find the form of the solution of the field equations in vacuum, in the vicinity of a point that is singular in the time.


\section{Singular vacuum field equations}
\label{Singular vacuum field equations}

In this Section, we find the degrees of freedom of the vacuum field equations in $R + \ep R^2$ theory through singular formal series expansion of the form (\ref{spatial metric rad}).


\subsection{First approximation of the vacuum field equations}
\label{First approximation of the vacuum field equations2}

From the vacuum field equations (\ref{eq:Loo}), (\ref{eq:Loa}) and (\ref{eq:Lab}), and the notation of the previous Section, we obtain,
\be
0 = \mathcal{C}_0 = (L^0_0)^{(-3)}\frac{1}{t^3} + (L^0_0)^{(-2)}\frac{1}{t^2} + (L^0_0)^{(-1)}\frac{1}{t},
\label{eq:C0_svac}
\ee
\be
0 = \mathcal{C}_\a = (L^0_\a)^{(-2)}\frac{1}{t^2} + (L^0_\a)^{(-1)}\frac{1}{t} + (L^0_\a)^{(0)},
\label{eq:Ca_svac}
\ee
and,
\be
0 = L^\b_\a = (L^\b_\a)^{(-3)}\frac{1}{t^3} + (L^\b_\a)^{(-2)}\frac{1}{t^2} + (L^\b_\a)^{(-1)}\frac{1}{t}.
\label{eq:LAB_svac}
\ee
Then, from Eqns. (\ref{eq:C0_svac}), (\ref{eq:Ca_svac}) and (\ref{eq:LAB_svac}), we find that,
\be
(L^0_0)^{(-3)} = 0,  
\label{zeroL00-3vs}
\ee
\be
(L^0_0)^{(-2)} = 0,  
\label{zeroL00-2vs}
\ee
\be
(L^0_0)^{(-1)} = 0,  
\label{zeroL00-1vs}
\ee
and,
\be
(L^0_\a)^{(-2)} = 0,  
\label{zeroL0a-2vs}
\ee
\be
(L^0_\a)^{(-1)} = 0,  
\label{zeroL0a-1vs}
\ee
\be
(L^0_\a)^{(0)} = 0,  
\label{zeroL0a0vs}
\ee
while,
\be
(L^\b_\a)^{(-3)} = 0,  
\label{zeroLba-3vs}
\ee
\be
(L^\b_\a)^{(-2)} = 0,  
\label{zeroLba-2vs}
\ee
\be
(L^\b_\a)^{(-1)} = 0.  
\label{zeroLba-1vs}
\ee
However, in the previous Section, we proved that the quantities $(L^0_0)^{(-3)},(L^0_\a)^{(-2)},\\ (L^0_\a)^{(-1)}$ and $(L^\b_\a)^{(-3)}$ vanish identically. Therefore, relations (\ref{zeroL00-3vs}), (\ref{zeroL0a-2vs}), (\ref{zeroL0a-1vs}) and (\ref{zeroLba-3vs}) do not provide any information regarding the initial data. In the following analysis we prove that only one of the previous equations give relation between the initial data $a_{\a\b},b_{\a\b},c_{\a\b},d_{\a\b}$.


\subsection{Zeroth-components of the vacuum field equations}
\label{Zeroth-components of the vacuum field equations}

Using (\ref{zeroL00-2vs}), we get one relation between the initial data, namely,
\be
-6c -\frac{1}{4}b^2 + \frac{11}{4}b^\b_\a b^\a_\b - \k = -\dfrac{1}{2\ep}. 
\label{R0 constant}
\ee
Then, from (\ref{zeroL00-1vs}) the term $(L^0_0)^{(-1)}$ becomes identically equal to zero. Also, using (\ref{R0 constant}), we find that the remaining three components $(L^0_\a)^{(0)},(L^\b_\a)^{(-2)},(L^\b_\a)^{(-1)}$ vanish identically. However, below we prove that the previous results can not be acceptable.


\subsection{Non existence of the singular vacuum field equations}
\label{Non existence of the singular vacuum field equations}

Our starting point is the following \emph{conformal transformation theorem} given in \cite{cphd}.
\begin{theorem}
 Consider the gravitational field equations derived from the gravitational lagrangian density $L_g=f(R)(-g)^{1/2}$ in vacuum on a $4$-dimensional spacetime $(\mathcal{V},g_{ij})$. Under a conformal transformation $\tilde{g}_{ij}={\Omega}^2 g_{ij}$, where $\Omega=f'(R)$, the transformed field equations on the $4$-dimensional spacetime $(\mathcal{V},\tilde{g}_{ij})$ become Einstein equations with a minimally coupled scalar field matter source derived from a lagrangian density $\tilde{L}_{\tilde{g}} = \left(\tilde{R} - \dfrac{3}{2}\nabla_i \phi \nabla^i \phi - V(\phi)\right)(-g)^{1/2}$. The scalar field potential is calculated to be \\
$V(\phi)=\dfrac{1}{2}\left(Rf'(R)-f(R)\right)[f'(R)]^{-2}$.  
  \end{theorem}
Then, taking into account the identities (\ref{eq for P}), (\ref{eq for R1}) and Eq. (\ref{R0 constant}), we note that the scalar curvature given by (\ref{Rscalar_rad}) satisfies the following equation,
\be
1 + 2\ep R = 0.
\label{con eq for R}
\ee
This last equation reads,
\be
f'(R)=0.
\label{f prime R zero}
\ee  
Due to the above conformal transformation theorem, this final result is not acceptable.

Hence, we conclude that in the vacuum gravitational field theory derived from the lagrangian $R + \ep R^2$, the formal series expansion (\ref{spatial metric rad}) lead to a system of equations in the form (\ref{eq:C0_svac})-(\ref{eq:LAB_svac}), which does not provide any kind of solution for the initial data $a_{\a\b},b_{\a\b},c_{\a\b},d_{\a\b}$.

It is note worthy that in case of vacuum in general relativity $(\ep=0)$ we get the same conclusion. In Appendix \ref{AppendixB}, we give more details for the solution of the field equations in case of general relativity in vacuum with a singular formal series expansion.


\section{Singular field equations in case of radiation}
\label{Singular field equations in case of radiation}

In this Section, we find the degrees of freedom of the gravitational field equations in $R + \ep R^2$ theory plus radiation through singular formal series expansion of the form (\ref{spatial metric rad}). 


\subsection{Field equations in case of radiation}
\label{Field equations in case of radiation2}  
 
In case of radiation models, the components of the energy-momentum tensor $T^i_j$ and its trace are given by (\ref{T^0_0 generalrad})-(\ref{T generalrad}). Then, using the notation (\ref{eq:C0_s})-(\ref{eq:LAB_s}), we obtain,
\be
\dfrac{8\pi G}{3}\rho(4{u_0}^2 - 1) =\mathcal{C}_0 = (L^0_0)^{(-3)}\frac{1}{t^3} + (L^0_0)^{(-2)}\frac{1}{t^2} + (L^0_0)^{(-1)}\frac{1}{t},
\label{eq:C0_srad}
\ee
\be
\dfrac{32\pi G}{3}\rho u_\a u_0 =\mathcal{C}_\a = (L^0_\a)^{(-2)}\frac{1}{t^2} + (L^0_\a)^{(-1)}\frac{1}{t} + (L^0_\a)^{(0)},
\label{eq:Ca_srad}
\ee
and,
\be
\dfrac{8\pi G}{3}\rho(4u^\b u_\a - \d^\b_\a) = L^\b_\a = (L^\b_\a)^{(-3)}\frac{1}{t^3} + (L^\b_\a)^{(-2)}\frac{1}{t^2}
+ (L^\b_\a)^{(-1)}\frac{1}{t},
\label{eq:LAB_srad}
\ee 
where the nine terms appearing in these last three relations are given by (\ref{L^0_0(-3)_rad})-(\ref{L^b_a(-1)_rad}).

In view of the identity (\ref{eq:velocities identity}) we get,
\be
1 = u_i u^i \approx {u_0}^2 - \dfrac{1}{t}a^{\a\b} u_\a u_\b .
\label{velocities identity singular}
\ee
Then, the system of equations (\ref{eq:C0_srad}), (\ref{eq:Ca_srad}) and (\ref{eq:LAB_srad}) becomes,
\be
8\pi G\rho =\mathcal{C}_0 = (L^0_0)^{(-3)}\frac{1}{t^3} + (L^0_0)^{(-2)}\frac{1}{t^2} + (L^0_0)^{(-1)}\frac{1}{t},
\label{eq:C0_sradf}
\ee
\be
\dfrac{32\pi G}{3}\rho u_\a =\mathcal{C}_\a = (L^0_\a)^{(-2)}\frac{1}{t^2} + (L^0_\a)^{(-1)}\frac{1}{t} + (L^0_\a)^{(0)},
\label{eq:Ca_sradf}
\ee
and,
\be
-\dfrac{8\pi G}{3}\rho \d^\b_\a = L^\b_\a = (L^\b_\a)^{(-3)}\frac{1}{t^3} + (L^\b_\a)^{(-2)}\frac{1}{t^2}
+ (L^\b_\a)^{(-1)}\frac{1}{t}.
\label{eq:LAB_sradf}
\ee
From Eq. (\ref{eq:C0_sradf}), we note that the degree of the energy density $\rho$ seems to be equal to $-3$, which is not in accordance with the Friedmann solution (\ref{a prop t^1/2}) and Eq. (\ref{eq ra^4}). However, in the following analysis, we prove that the term $(L^0_0)^{(-3)}$ is equal to zero and in the same time the term $(L^0_0)^{(-2)}$ does not vanish, and so, as we expected, the degree of the energy density becomes equal to $-2$.

We are now ready to find the relations which the terms appearing in the system of equations (\ref{eq:C0_sradf})-(\ref{eq:LAB_sradf}) satisfy and then the degrees of freedom of the gravitational field equations in case of radiation.


\subsection{Degrees of freedom of the field equations in case of radiation}
\label{Degrees of freedom of the vacuum field equations in case of radiation2}

Our starting point are the three identities (\ref{eq for P}), (\ref{eq for R1}) and (\ref{eq for paR0}) provided by (\ref{idBox2}), as well as the identity (\ref{id zero paR0}). Using these, the terms $(L^0_0)^{(-3)},(L^0_\a)^{(-2)},(L^\b_\a)^{(-3)}$ and $(L^0_\a)^{(-1)}$ vanish identically.

Then, the system of equations (\ref{eq:C0_sradf})-(\ref{eq:LAB_sradf}), which describes the dependence between the initial data, is modified in its final form as follows:
\be
8\pi G\rho = (L^0_0)^{(-2)}\frac{1}{t^2} + (L^0_0)^{(-1)}\frac{1}{t},
\label{eq:C0_sradf2}
\ee
\be
\dfrac{32\pi G}{3}\rho u_\a = (L^0_\a)^{(0)},
\label{eq:Ca_sradf2}
\ee
\be
-\dfrac{8\pi G}{3}\rho \d^\b_\a = (L^\b_\a)^{(-2)}\frac{1}{t^2} + (L^\b_\a)^{(-1)}\frac{1}{t},
\label{eq:LAB_sradf2}
\ee
where for the remaining five terms we have,
\be
(L^0_0)^{(-2)} = \dfrac{3}{4} + \dfrac{3}{2}\ep( -6c -\frac{1}{4}b^2 + \frac{11}{4}b^\b_\a b^\a_\b - \k ),
\label{L^0_0(-2)_radnew2}
\ee
\be
(L^0_0)^{(-1)} =-\dfrac{1}{2}b -\ep b( -6c -\frac{1}{4}b^2 + \frac{11}{4}b^\b_\a b^\a_\b - \k ),
\label{L^0_0(-1)_radnew2}
\ee
and,
\be
(L^0_\a)^{(0)} = \frac{1}{2}(\nabla_\b b^\b_\a -\nabla_\a b) 
+ \ep  (-6c -\frac{1}{4}b^2 + \frac{11}{4}b^\d_\g b^\g_\d - \k)(\nabla_\b b^\b_\a -\nabla_\a b),
\label{L^0_a(0)_radnewfinal2}
\ee
and also,
\be
(L^\b_\a)^{(-2)} = -\dfrac{1}{4}\d^\b_\a - \dfrac{1}{2}\ep \d^\b_\a ( -6c -\frac{1}{4}b^2 + \frac{11}{4}b^\d_\g b^\g_\d - \k ),
\label{L^b_a(-2)_radnew2}
\ee
\bq
(L^\b_\a)^{(-1)} &=& -\widetilde{P}^\b_\a - \dfrac{3}{4}b^\b_\a - \dfrac{1}{4}b\d^\b_\a \nonumber \\
&-& \dfrac{1}{2} \ep (4\widetilde{P}^\b_\a + 3b^\b_\a + b\d^\b_\a)(-6c - \dfrac{1}{4}b^2 + \dfrac{11}{4}b^\d_\g b^\g_\d - \k). 
\label{L^b_a(-1)_radnewfinal2}
\eq
Then, from Eq. (\ref{eq:C0_sradf2}) we find one relation for the energy density $\rho$, namely,
\bq
8\pi G\rho &=&  \left [ \dfrac{3}{4} + \dfrac{3}{2}\ep( -6c -\frac{1}{4}b^2 + \frac{11}{4}b^\b_\a b^\a_\b - \k ) \right ] \frac{1}{t^2} \nonumber \\
&-& \left [ \dfrac{1}{2}b +\ep b( -6c -\frac{1}{4}b^2 + \frac{11}{4}b^\b_\a b^\a_\b - \k ) \right ] \frac{1}{t}.
\label{eq for energy density rad with L00(-2) and L00(-1)}
\eq
Substituting $\rho$ from Eq. (\ref{eq:C0_sradf2}) to Eq. (\ref{eq:Ca_sradf2}), we find three more relations for the velocities $u_\a$ and the spatial tensors $a_{\a\b},b_{\a\b},c_{\a\b},d_{\a\b}$, namely,
\be
u_\a = \dfrac{3(L^0_\a)^{(0)}}{4(L^0_0)^{(-2)}}t^2.
\label{final eq of u_a}
\ee
These last relations concerning the velocities $u_\a$ and the spatial tensors $a_{\a\b},b_{\a\b},c_{\a\b},d_{\a\b}$ read,
\bq
u_\a &=& \dfrac{3t^2}{4} \left [ \frac{1}{2}(\nabla_\b b^\b_\a -\nabla_\a b) 
+ \ep  (-6c -\frac{1}{4}b^2 + \frac{11}{4}b^\d_\g b^\g_\d - \k)(\nabla_\b b^\b_\a -\nabla_\a b) \right ] \nonumber \\
& \times & \left [ \dfrac{3}{4} + \dfrac{3}{2}\ep( -6c -\frac{1}{4}b^2 + \frac{11}{4}b^\b_\a b^\a_\b - \k ) \right ]^{-1}.
\label{eq for velocities rad with L0a(0) and L00(-2)}
\eq 
Continuing, we substitute $\rho$ from Eq. (\ref{eq:C0_sradf2}) to Eq. (\ref{eq:LAB_sradf2}) and we obtain,
\be
(L^\b_\a)^{(-2)}\frac{1}{t^2} + (L^\b_\a)^{(-1)}\frac{1}{t} = -\dfrac{1}{3}\d^\b_\a \left[ (L^0_0)^{(-2)}\frac{1}{t^2} + (L^0_0)^{(-1)}\frac{1}{t} \right],
\label{eq:L00_LAB}
\ee   
Using the last equation and Eqns. (\ref{L^0_0(-2)_radnew2})-(\ref{L^b_a(-1)_radnewfinal2}), we find that the terms of order $t^{-2}$ cancel, while the terms of order $t^{-1}$ give,
\be
-3(\widetilde{P}^\b_\a + \dfrac{3}{4}b^\b_\a + \dfrac{5}{12}b\d^\b_\a)
-\dfrac{1}{2} \ep(12\widetilde{P}^\b_\a + 9b^\b_\a + 5b\d^\b_\a)(-6c - \dfrac{1}{4}b^2 + \dfrac{11}{4}b^\d_\g b^\g_\d
- \k) = 0.
\label{eq:L00_LAB(-1)}
\ee
Notice that the trace of Eq. (\ref{eq:L00_LAB(-1)}) gives the identity (\ref{eq for P}).

Eventually, to sum up, from equations (\ref{eq:C0_sradf2}), (\ref{final eq of u_a}) and (\ref{eq:L00_LAB(-1)}) we have already found $1$ relation for the energy density $\rho$, $3$ for the velocities $u^\a$ and additional $6$ concerning the initial data $a_{\a\b},b_{\a\b},c_{\a\b},d_{\a\b}$. Also, we should not make the mistake to count as extra connections between the initial data the identities (\ref{eq for P}), (\ref{eq for R1}) and (\ref{eq for paR0}), because the trace of the gravitational field equations yields directly to the same relations. 

Nevertheless, we have to take into consideration the fact that the choice of the time $t$ in the metric (\ref{spatial metric rad}) is completely determined by the condition $t=0$ at the singularity, while the space coordinates still permit arbitrary transformations that do not involve the time. These arbitrary transformations can be used, for example, to bring tensor $a_{\a\b}$ to diagonal form.

Thus, from the initial $28$ functions $a_{\a\b},b_{\a\b},c_{\a\b},d_{\a\b},\rho,u^\a$, we have to subtract $13$ that are not arbitrary, and we finally find that the solution contains all together $15$ physically different arbitrary functions. In comparison with the number $20$ which corresponds to a general solution of the problem, we see that after the imposition of the higher order evolution and constraint equations, tensor $\g_{\a\b}$ of the form (\ref{spatial metric rad}) can not be an appropriate form for the general solution.

This is something that we expected since the begining of this analysis, because of the form of the energy-momentum tensor $T^i_j$. In particular, $T^i_j$ is a diagonal matrix of the form,
\be 
T^i_j = \textrm{diag}(\rho,-p,-p,-p),
\label{diagT^i_j}
\ee
while tensor $L^i_j$ consists of terms that include the matrix $\g_{\a\b}$ and its derivatives and so it is not diagonal in general. Therefore, from the first equation (\ref{eq:FEs_rad}) it is clear that these two members can not finally provide the general solution. Furthermore, we note that in this solution the spatial metric is inhomogeneous and anisotropic, but the energy density tends to become homogeneous as $t\rightarrow 0$. 

Finally, it is worth noting that in case we set $\ep=0$, we immediately obtain from the basic equations (\ref{eq:C0_sradf2}), (\ref{final eq of u_a}) and (\ref{eq:L00_LAB(-1)}) the exact same results with \cite{ll} in general relativity. 
 
We continue quering if the order of $\g_{\a\b}$ in (\ref{spatial metric rad}) affects or not the number of the arbitrary functions.


\section{Degrees of freedom with $\g_{\a\b}=\sum{\g^{(n)}_{\a\b} t^{n}},n\geq 4$}
\label{Degrees of freedom general2}

In this Section, we show that the number of the degrees of freedom of the gravitational field equations in $R + \ep R^2$ theory plus radiation is the same, if instead of the singular form (\ref{spatial metric rad}), we consider a singular expression of the form,
\be
\g_{\a\b}=\sum{\g^{(n)}_{\a\b} t^{n}},\quad n\geq 4.
\label{gengab2}
\ee


\subsection{Field equations with $\g_{\a\b}=\sum{\g^{(n)}_{\a\b} t^{n}},n\geq 4$}
\label{Field equations general singular}

We assume a formal series representation of the spatial metric of the form:
\be
\g_{\a\b} = (\g_{\a\b})^{(1)}t + (\g_{\a\b})^{(2)}t^2 + (\g_{\a\b})^{(3)}t^3 + (\g_{\a\b})^{(4)}t^4 + \cdots
+ (\g_{\a\b})^{(n)}t^n,
\label{general spatial metric rad}
\ee
where $n$ is a natural number, with $n>4$. Obviously, we have, 
\be
(\g_{\a\b})^{(1)}=a_{\a\b}, \quad (\g_{\a\b})^{(2)}=b_{\a\b}, \quad (\g_{\a\b})^{(3)}=c_{\a\b}, \quad (\g_{\a\b})^{(4)}=d_{\a\b}.
\label{terms of general spatial metric rad}
\ee
We further note that the expression (\ref{general spatial metric rad}) contains $6n$ degrees of freedom ($6$ of each one of the $n$ spatial matrixes). Adding $4$ additional degrees of freedom regarding the energy density $\rho$ and the velocities $u^\a$, the main question is how many of these $6n+4$ data are independent when (\ref{general spatial metric rad}) is taken to be a possible solution of the evolution equations (\ref{eq:pg}), (\ref{eq:pK}), (\ref{eq:pD}) and (\ref{eq:pW with T}) together with the constraint equations (\ref{eq:hamiltonian_rad}) and (\ref{eq:momentum_rad}) and the relations (\ref{radiation eq}), (\ref{eq:velocities identity}) on each slice $\mathcal{M}_{t}$?
\\
To answer this question, we consider our basic system of equations (\ref{eq:C0_sradf2}), (\ref{eq:Ca_sradf2}) and (\ref{eq:LAB_sradf2}), which now, due to (\ref{general spatial metric rad}), takes the form,
\be
8\pi G\rho = (L^0_0)^{(-2)}\frac{1}{t^2} + (L^0_0)^{(-1)}\frac{1}{t} + \cdots + (L^0_0)^{(n-5)}t^{n-5},
\label{eq:C0_rad_new2_gen2}
\ee
\be
\dfrac{32\pi G}{3}\rho u_\a = (L^0_\a)^{(0)} + \cdots + (L^0_\a)^{(n-4)}t^{n-4},
\label{eq:Ca_rad_new2_gen2}
\ee
\be
-\dfrac{8\pi G}{3}\rho \d^\b_\a = (L^\b_\a)^{(-2)}\frac{1}{t^2} + (L^\b_\a)^{(-1)}\frac{1}{t} + \cdots
+ (L^\b_\a)^{(n-5)}t^{n-5}.
\label{eq:LAB_rad_new2_gen2}
\ee


\subsection{Degrees of freedom with $\g_{\a\b}=\sum{\g^{(n)}_{\a\b} t^{n}},n\geq 4$ in radiation models}
\label{Degrees of freedom singular in radiation}

We notice that (\ref{eq:C0_rad_new2_gen2}) give us $1$ relation for the energy density $\rho$, which we substitute in (\ref{eq:Ca_rad_new2_gen2}) and we find the $3$ relations for the velocities,
\be
u_\a = \dfrac{3(L^0_\a)^{(0)}}{4(L^0_0)^{(-2)}}t^2.
\label{final eq of u_a_gen}
\ee
The final thing we have to do, is to substitute $\rho$ from (\ref{eq:C0_rad_new2_gen2}) to (\ref{eq:LAB_rad_new2_gen2}). Then, we obtain,
\bq
(L^\b_\a)^{(-2)}\frac{1}{t^2} + (L^\b_\a)^{(-1)}\frac{1}{t} + \cdots
+ (L^\b_\a)^{(n-5)}t^{n-5} &=& -\dfrac{1}{3}\d^\b_\a (L^0_0)^{(-2)}\frac{1}{t^2}
-\dfrac{1}{3}\d^\b_\a (L^0_0)^{(-1)}\frac{1}{t} - \cdots \nonumber \\
&-& \dfrac{1}{3}\d^\b_\a(L^0_0)^{(n-5)}t^{n-5}.
\label{eq:L00_LAB_gen}
\eq 
As we have already seen at the previous section, the terms of order $t^{-2}$ in (\ref{eq:L00_LAB_gen}) cancel and the $t^{-1}$ order terms give $6$ relations between the initial data, which is the equation (\ref{eq:L00_LAB(-1)}). For the $t^k$ order terms of (\ref{eq:L00_LAB_gen}), where $k=0,1,...,n-5$, we obtain the corresponding following equations,
\be
(L^\b_\a)^{(k)} = -\dfrac{1}{3}\d^\b_\a (L^0_0)^{(k)}.
\label{eq:L00_LAB_gen_0,1,...,n-5}
\ee
Apparently, each one of the previous $n-4$ equations gives $6$ relations that contain the initial data. Therefore, from (\ref{eq:L00_LAB_gen_0,1,...,n-5}) we find $6\times(n-4)=6n-24$ relations between the data $(\g_{\a\b})^{(0)},...,(\g_{\a\b})^{(n)}$ and so, from (\ref{eq:L00_LAB_gen}) we have $0+6+(6n-24)=6n-18$ relations in total.

Consequently, taking further into account one relation from (\ref{eq:C0_rad_new2_gen2}), three relations from (\ref{final eq of u_a_gen}) and $(6n-24)$ relations from (\ref{eq:L00_LAB_gen}), the counting give us $1+3+(6n-18)=6n-14$ relations in total between the initial data. Subtracting from the total $6n+4$ data from which we started, we obtain that only $18$ functions can be arbitrary. However the $3$ diffeormorphisms give us the final number of the free functions in the problem, which is obviously the number $15$. Conclusively, without loss of generality we can stop the series in (\ref{general spatial metric rad}) at the order $4$.

In Appendix \ref{AppendixB}, we give details for the solution of the field equations in general relativity plus radiation with a singular formal series expansion.


\section{Choice of the initial data}
\label{Choice of the initial data with radiation}

We now ask: Out of the $28$ different functions $a_{\a\b},b_{\a\b},c_{\a\b},d_{\a\b},\rho,u^\a$, which $15$ of those should be chosen as our initial data? We have shown in this Section that in case of radiation the higher order gravity equations
(\ref{eq:pg}), (\ref{eq:pK}), (\ref{eq:pD}) and (\ref{eq:pW with T}) together with the constraint equations (\ref{eq:hamiltonian_rad}) and (\ref{eq:momentum_rad}), admit a singular formal series expansion of the form  (\ref{spatial metric rad}) and the solution, which is not general, requires 15 smooth initial data. 

If we prescribe the $28$ data
\be 
a_{\a\b}, \quad  b_{\a\b},\quad  c_{\a\b},\quad  d_{\a\b},\quad  \rho,\quad  u^\a
\ee
initially, we still have the freedom to fix $13$ of them. We choose to leave the twelve components of the metrics $c_{\a\b}$ and $d_{\a\b}$ free, and we choose the symmetric space tensor $a_{\a\b}$ to be diagonal. Then we proceed to count the number of free functions in three steps, starting from these $3+0+2\times 6 +1+3=19$ functions. First, (\ref{eq:C0_sradf2}) fixes the function $\rho$ and secondly Eq. (\ref{final eq of u_a}) fixes the $3$ more components of $ u^\a$. Lastly, we use  the 6 relations in (\ref{eq:L00_LAB(-1)}) to completely fix the $6$ components of $b_{\a\b}$. Summing up the free functions we have found, we end up with,
\be
\underbrace{3}_{\textrm{from}\ a_{\a\b}} + \underbrace{0}_{\textrm{from}\ b_{\a\b}}
+\underbrace{12}_{\textrm{from}{\ c_{\a\b}}\ \textrm{and}\ d_{\a\b}} + \underbrace{0}_{\textrm{from}\ \rho}
+ \underbrace{0}_{\textrm{from}\ u^\a} = 15,
\label{eq:notation_initial_data_rad}
\ee
which are suitable free data as required for the specific particular solution.
Obviously this is not the only way to choose the initial data. For instance, we can choose the space tensor $a_{\a\b}$ not to be diagonalnot and fix three components of $c_{\a\b}$. Then, we end up with,
\be
\underbrace{6}_{\textrm{from}\ a_{\a\b}} + \underbrace{0}_{\textrm{from}\ b_{\a\b}}
+\underbrace{3}_{\textrm{from}\ c_{\a\b}} + \underbrace{6}_{\textrm{from}\ d_{\a\b}} + \underbrace{0}_{\textrm{from}\ \rho}
+ \underbrace{0}_{\textrm{from}\ u^\a} = 15,
\label{eq:notation_initial_data_rad}
\ee 
suitable free data.
 

\chapter{Conclusions and future directions} 

\label{Chapter7} 

\lhead{Chapter 7. \emph{Conclusions and future directions}} 

\section{Conclusions}
\label{Conclusions}

Throughout this Thesis we extensively studied properties of formal power series solutions of higher-order $f(R) = R + \ep R^2$ gravity theory as well of the standard Einstein theory in vacuum and in the presence of radiation. In particular, we used the Landau-Lifschitz method presented in \cite{ll} in order to explore the form and the kind of solution in each case separately. We applied this method both to regular and singular cases. Therefore, our approach provides information about eight different cases:

\begin{center}
\begin{tabular}{|c|c|c|}\hline
    General Relativity   & Regular case       & Vacuum                       \\\hline
    General Relativity   & Singular case      & Vacuum                       \\\hline
    General Relativity   & Regular case       & Radiation                     \\\hline
    General Relativity   & Singular case      & Radiation                     \\\hline
    Higher-order gravity & Regular case       & Vacuum                        \\\hline 
    Higher-order gravity & Singular case      & Vacuum                        \\\hline  
    Higher-order gravity & Regular case       & Radiation                     \\\hline
    Higher-order gravity & Singular case      & Radiation                     \\\hline
\end{tabular}      
\end{center}

In Chapter \ref{Chapter2}, we introduced the ADM (Arnowitt-Deser-Misner) formalism in order to study the evolution problem of cosmological asymptotics in higher-order gravity theories in vacuum as well as in the presence of a fluid. We used normal coordinates and the synchronous reference system in which the time coordinate is the proper time along the fluid world lines. Then, we splitted the higher-order field equations (\ref{eq:FEs}) in vacuum derived from the analytic lagrangian $f(R)=R+\ep R^2$ into constraints and evolution equations using the (3+1)-splitting formulation. Because of the order of these differential equations, to describe them as a dynamical system, apart from the first variational equation,
\be
\partial_t \g_{\a\b} = K_{\a\b},
\label{first variational equation conclusion}
\ee
which is used in general relativity, we needed the second and third variational equations, which were respectively found to be:
\be
\partial_t K_{\a\b} = D_{\a\b}, \qquad \partial_t D_{\a\b} = W_{\a\b}. 
\label{eq:pK and eq:pD conclusion}
\ee 
Next, we dealt with the function-counting problem of the quadratic theory in vacuum. We found that the exact number of the degrees of freedom according to the vacuum theory is equal to $16$. Finally, we concluded that in the presence of a fluid the exact number of degrees of freedom is $20$.

In Chapter \ref{Chapter3}, we studied the Cauchy-Kovalevskaya formulation in higher-order gravity theory in vacuum as well as in the presence of a fluid. Prior to anything else, we dealt with the mathematical content of higher-order gravity equations, the initial value problem of Cauchy as in the case of general relativity. In order to study the Cauchy problem of the higher-order gravity theory, we followed the standard approach and we exploited harmonic coordinates. Then, we proved that the dynamical system consisting of the four evolution equations (\ref{eq:pg}), (\ref{eq:pK}), (\ref{eq:pD}) and (\ref{eq:pW}) together with the constraints equations (\ref{eq:hamiltonian}) and (\ref{eq:momentum}), which describes the higher-order gravity equations in vacuum, has the Cauchy-Kovalevskaya property, namely that none of the terms appearing in the right hand side of this system includes time derivatives. We also conducted the same analysis in the case of a fluid and found that the corresponding system of the four evolution equations (\ref{eq:pg}), (\ref{eq:pK}), (\ref{eq:pD}) and (\ref{eq:pW with T}) together with the constraints equations (\ref{eq:hamiltonian_rad}) and (\ref{eq:momentum_rad}) is of the Cauchy-Kovalevskaya type.

Next, we showed that in a synchronous reference system if we prescribe initial data which satisfy the constraint equations in both cases (vacuum and fluid) on some initial slice, then there is a neighborhood of this slice such that the corresponding evolution equations of each case have an analytic solution in this neighborhood consistent with the data. We also discussed the equivalence between general relativity plus a scalar field and higher-order gravity theory in vacuum from this viewpoint. Specifically, through the conformal transformation theorem, we discovered the reason for which these two theories do not have the same degrees of freedom.

In Chapter \ref{Chapter4}, we introduced the perturbative formulation in higher-order gravity assuming a formal series representation of the spatial metric $\g_{\a\b}=-g_{\a\b}$ following the Landau-Lifschitz method. In doing so, we started with the Robertson-Walker metric (\ref{rwmetric}) to find the connection between the various components of the Ricci tensor $R^i_j$ and the scalar curvature $R$ as well with the scale factor $a=a(t)$ and its derivatives. Despite the fact that we are interested in higher-order cosmologies that are not homogeneous and isotropic, we used (\ref{rwmetric}) as a motivation in what followed in the next two Chapters. We then presented the solutions of the Friedmann equations in general relativity in cases of vacuum and radiation and we found the generalized Friedmann equations of higher-order $f(R)=R+\ep R^2$ gravity. We showed that the solutions of the Friedmann equations in general relativity in cases of vacuum and radiation are particular solutions for the generalized Friedmann equations as well.

This fact led us to use the same formal power series expansion for the $3$-metric $\g_{\a\b}$ in our analysis for the field equations of higher-order gravity in vacuum as in the case of general relativity. Hence, in case of vacuum we started by considering a formal series expansion of the form,
\be
\g_{\a\b}= \g^{(0)}_{\a\b} +\g^{(1)}_{\a\b}\;t + \g^{(2)}_{\a\b}\;t^2 + \g^{(3)}_{\a\b}\;t^3 + \g^{(4)}_{\a\b}\;t^4 + \cdots,
\label{eq:3dimmetric conclusion}
\ee
where the $ \g^{(0)}_{\a\b} , \g^{(1)}_{\a\b} , \g^{(2)}_{\a\b} , \g^{(3)}_{\a\b} , \g^{(4)}_{\a\b},\cdots$ are functions of the space coordinates, or else we used the Landau-Lifschitz pertubative method in the vicinity of a point that is regular in the time. Using (\ref{eq:3dimmetric conclusion}) up to the fourth power of $t$, we made all the calculations of the quantities $K_{\a\b},D_{\a\b},W_{\a\b}$ and, so, we found the series of the $(3+1)$-splitting various components of the Ricci tensor $R^0_0,R^0_\a,R^\b_\a$ and the scalar curvature $R$ as well. Consequently, we ended up with the series of the evolution equations (\ref{eq:Kab}), (\ref{eq:Dab}), (\ref{eq:trL}) and the constraints equations (\ref{eq:L00mixed}) and (\ref{eq:L0amixed}) too. 

Next, for the case of radiation, we started by considering power series expansion that corresponds to the Landau-Lifschitz pertubative method in the vicinity of a point that is singular in the time, namely,
\be
\g_{\a\b} = \g^{(1)}_{\a\b}t + \g^{(2)}_{\a\b}t^2 + \g^{(3)}_{\a\b}t^3 + \g^{(4)}_{\a\b}t^4 + \cdots
\label{spatial metric rad conclusion}
\ee 
where the $\g^{(1)}_{\a\b} , \g^{(2)}_{\a\b} , \g^{(3)}_{\a\b} , \g^{(4)}_{\a\b},\cdots$ are functions of the space coordinates. We also used the form (\ref{spatial metric rad conclusion}) up to the fourth power of $t$ and found the form of the quantities $K_{\a\b},D_{\a\b},W_{\a\b}$ and the form of the various components of the Ricci tensor $R^i_j$ and the scalar curvature $R$. Finally, we presented the series of the evolution equations (\ref{eq:Kab_rad}), (\ref{eq:Dab_rad}), (\ref{snap rad all}) and the constraints equations (\ref{C0rad all}) and (\ref{Ca rad all}) as well.

In Chapter \ref{Chapter5}, we studied the form of the solution of the gravitational field equations of higher-order gravity in vacuum, as well as in the case of radiation, in the vicinity of a point that is regular in the time by using the $3$-metric (\ref{eq:3dimmetric conclusion}). First of all, because of the order of the differential equations in higher-order gravity in vacuum (Eqns. (\ref{eq:Loo}), (\ref{eq:Loa}) are third-order differential equations with respect to the proper time $t$, while Eq. (\ref{eq:Lab}) is a fourth-order differential equation with respect to $t$), we introduced the following notation,
\be
\mathcal{C}_0    =  (L^0_0)^{(0)} + t(L^0_0)^{(1)},  
\label{eq:C0 conclusion}
\ee
\be
\mathcal{C}_\a   =  (L^0_\a)^{(0)} + t(L^0_\a)^{(1)}, 
\label{eq:Ca conclusion}
\ee
and,
\be
L^\b_\a  =  (L^\b_\a)^{(0)}, 
\label{eq:LAB conclusion}
\ee
that includes only the terms that suffice to describe the solution of the higher-order field equations in cases of vacuum and radiation.

After calculating the series of these five terms $(L^0_0)^{(0)},(L^0_0)^{(1)},(L^0_\a)^{(0)},(L^0_\a)^{(1)},\\ (L^\b_\a)^{(0)}$, we were led to a significant simplification of them by using a well-known identity that follows from the higher-order field equations themselves, namely,
\be
R - 6\ep \Box_g R = 0.
\label{idBox conclusion}
\ee
Then, we dealt with the higher-order gravity equations in vacuum. In the begining of this analysis we had five equations, namely,
\be
(L^0_0)^{(0)} = (L^0_0)^{(1)} = (L^0_\a)^{(0)} = (L^0_\a)^{(1)} = (L^\b_\a)^{(0)} = 0,
\label{all equal to zero conclusion}
\ee
which correspond to $14$ relations between the $30$ initial data $ \g^{(0)}_{\a\b} , \g^{(1)}_{\a\b} , \g^{(2)}_{\a\b} , \g^{(3)}_{\a\b} , \g^{(4)}_{\a\b} $. However, through the basic identity,
\be
\nabla_i L^i_j = 0,
\label{basic identity conclusion}
\ee 
we proved that the quantities $(L^0_0)^{(1)}$ and $(L^0_\a)^{(1)}$ are identically equal to zero, so they do not provide any additional information concerning the $30$ initial data. Thus, we were left with $10$ relations between the initial data. Subtracting $10$ from the initial $30$ data and taking into account $4$ diffeomorphism changes, we finally found that the $3$-metric (\ref{eq:3dimmetric conclusion}) lead to a solution of the higher-order field equations in vacuum which corresponds to the general solution ($16$ arbitrary functions). 

Further, we dealt with the higher-order gravity equations in case of radiation. Using the identity,
\be
1 = u_i u^i = {u_0}^2 - \g^{\a\b}u_\a u_\b,
\label{second basic identity conclusion}
\ee
we found that the term $(L^0_\a)^{(0)}$ is necessarily equal to zero, while the term $(L^0_0)^{(1)}$ appearing in the above notation vanishes identically. Additionally, we obtained one relation concerning the energy density $\rho$, three relations for the velocities $u_\a$ and six relations between the initial data $\g^{(0)}_{\a\b} , \g^{(1)}_{\a\b} , \g^{(2)}_{\a\b} , \g^{(3)}_{\a\b} , \g^{(4)}_{\a\b}$. Therefore, taking into account four diffeomorphism transformations, from the initial $34$ data $\g^{(0)}_{\a\b} , \g^{(1)}_{\a\b} , \g^{(2)}_{\a\b} , \g^{(3)}_{\a\b} , \g^{(4)}_{\a\b},\rho,u^\a$ we were left with $17$ free functions. Hence, the form (\ref{eq:3dimmetric conclusion}) leads to a solution of the higher-order field equations in case of radiation which does not correspond to the general solution ($20$ arbitrary functions). 

Subsequently, we considered the same two problems (vacuum and radiation) by using the form (\ref{eq:3dimmetric conclusion}) up to the nth power of $t$. What we showed is that in both cases the results are quantitatively the same as the corresponding previous results. Specifically, in case of vacuum we used the method of induction. Finally, we presented the choice of the initial data between the initial functions of the two cases. In particular, in case of vacuum, because of the general behaviour of the solution, we generalized a theorem of Rendall \cite{ren1}.

In Chapter \ref{Chapter6}, we studied the form and the type of the solution of the gravitational field equations of higher-order gravity in vacuum as well as in the case of radiation, in the vicinity of a point that is not regular, but singular in the time by using the $3$-metric (\ref{spatial metric rad conclusion}). Bearing in mind the order of the differential equations in higher-order gravity in case of radiation (\ref{eq:Loo_rad}), (\ref{eq:Loa_rad}) are third-order differential equations with respect to the proper time $t$ and Eq. (\ref{eq:Lab_rad}) is fourth-order differential equation with respect to $t$), we were led to the following notation,
\be
\mathcal{C}_0 = (L^0_0)^{(-3)}\frac{1}{t^3} + (L^0_0)^{(-2)}\frac{1}{t^2} + (L^0_0)^{(-1)}\frac{1}{t},
\label{eq:C0_s conclusion}
\ee
\be
\mathcal{C}_\a = (L^0_\a)^{(-2)}\frac{1}{t^2} + (L^0_\a)^{(-1)}\frac{1}{t} + (L^0_\a)^{(0)},
\label{eq:Ca_s conclusion}
\ee
and,
\be
L^\b_\a = (L^\b_\a)^{(-3)}\frac{1}{t^3} + (L^\b_\a)^{(-2)}\frac{1}{t^2} + (L^\b_\a)^{(-1)}\frac{1}{t}.
\label{eq:LAB_s conclusion}
\ee 
We calculated the series of the nine terms $(L^0_0)^{(-3)},(L^0_0)^{(-2)},(L^0_0)^{(-1)}$ and $(L^0_\a)^{(-2)}, \\ (L^0_\a)^{(-1)}, (L^0_\a)^{(0)}$ and $(L^\b_\a)^{(-3)},(L^\b_\a)^{(-2)},(L^\b_\a)^{(-1)}$ appearing in the notation, we did the simplification of them by using the identity (\ref{idBox conclusion}). We futher simplified these terms by using the identities (\ref{basic identity conclusion}). Then, we dealt with the higher-order gravity equations in vacuum. Initially, we had nine equations, namely,
\be
(L^0_0)^{(-3)} = (L^0_0)^{(-2)} = (L^0_0)^{(-1)} = 0,
\label{all 00 equal to zero conclusion}
\ee
\be
(L^0_\a)^{(-2)} = (L^0_\a)^{(-1)} = (L^0_\a)^{(0)} = 0,
\label{all 0a equal to zero conclusion}
\ee
and,
\be
(L^\b_\a)^{(-3)} = (L^\b_\a)^{(-2)} = (L^\b_\a)^{(-1)} = 0,
\label{all ba equal to zero conclusion}
\ee
which correspond to $30$ relations between the $24$ initial data $ \g^{(1)}_{\a\b} , \g^{(2)}_{\a\b} , \g^{(3)}_{\a\b} , \g^{(4)}_{\a\b} $. However, we showed that all the quantities, apart from $(L^0_0)^{(-2)}$, vanish identically, so they do not provide information regarding the $24$ initial data. But the equation,
\be
(L^0_0)^{(-2)} = 0,
\label{L00(-2) equal to zero conclusion}
\ee
directly led us to the relation,
\be
1 + 2\ep R = 0,
\label{con eq for R conclusion}
\ee
namely that,
\be
f'(R)=0.
\label{f prime R zero conclusion}
\ee
This last relation cannot be acceptable due to the conformal transformation theorem (see subsection \ref{Non existence of the singular vacuum field equations}). Therefore, we concluded that in the vacuum gravitational field theory derived from the lagrangian $R + \ep R^2$, the formal series expansion (\ref{spatial metric rad conclusion}) leads to a system of equations which does not provide any kind of solution for the initial data $ \g^{(1)}_{\a\b} , \g^{(2)}_{\a\b} , \g^{(3)}_{\a\b} , \g^{(4)}_{\a\b}$. 

Further, we dealt with the higher-order gravity equations in case of radiation. Using the identity (\ref{idBox conclusion}), we found that the terms $(L^0_0)^{(-3)},(L^0_\a)^{(-2)},(L^\b_\a)^{(-3)}$ and $(L^0_\a)^{(-1)}$ vanish identically. So, we obtained one relation concerning the energy density $\rho$, three relations for the velocities $u_\a$ and six additional relations between the initial data $\g^{(1)}_{\a\b} , \g^{(2)}_{\a\b} , \g^{(3)}_{\a\b}$. Hence, taking into account three diffeomorphism transformations (the choice of the time in the $3$-metric (\ref{spatial metric rad conclusion}) is completely determined by the condition $t=0$ at the singularity) from the initial $28$ data $\g^{(1)}_{\a\b} , \g^{(2)}_{\a\b} , \g^{(3)}_{\a\b} , \g^{(4)}_{\a\b},\rho,u^\a$, we were left with $15$ free functions. Therefore, the form (\ref{spatial metric rad conclusion}) leads to a solution of the higher-order field equations in case of radiation which does not correspond to the general solution ($20$ arbitrary functions). Then, we considered the same problem in case of radiation by using the form (\ref{spatial metric rad conclusion}) up to the nth power of the proper time $t$. In this approach, we found exactly the same result about the number of the arbitrary functions of the theory with the corresponding previous result. Finally, we presented the choice of the initial data between the initial functions in case of radiation.

In Appendices \ref{AppendixA} and \ref{AppendixB}, we gave details about the solution of the Einstein equations in cases of vacuum and radiation by using the regular form (\ref{eq:3dimmetric conclusion}) and the singular form (\ref{spatial metric rad conclusion}) respectively. In particular, in Appendix \ref{AppendixA}, we showed that the form (\ref{eq:3dimmetric conclusion}) corresponds to the genaral solution of the problem in case of vacuum ($4$ arbitrary functions). In case of radiation, we ended up with $5$ free functions which correspond to a particular solution ($8$ free functions for general solution).
In Appendix \ref{AppendixB}, we showed that the singular form (\ref{spatial metric rad conclusion}) of the $3$-metric does not give any kind of solution in case of vacuum, but give a particular solution in case of radiation (we left with $3$ free functions). 

What is interesting from the discussion above is that in case of vacuum both general relativity and higher-order $f(R)=R+\ep R^2$ gravity give exactly the same results concerning the kind of the solution (general solution with the regular formal series expansion (\ref{eq:3dimmetric conclusion}) and no solution for the singular form (\ref{spatial metric rad conclusion})). In addition to this, both theories also give the same results in case of radiation (particular solution with the regular power series expansion (\ref{eq:3dimmetric conclusion}) with $3$ less arbitrary functions from those that a general solution requires and particular solution with the singular form (\ref{spatial metric rad conclusion}) with $5$ less free functions). Specifically, in case of radiation-dominated models, we conclude that the regular solution is "closer" to being a generic feature of the field equations (both in general relativity and in higher-order gravity in the analytic case) than the singular solution. 

In fact, this is a result that we did not expect due to the fact that the standard model which describes the evolution of the universe informs us about the fact that in the early universe (proper time $t\rightarrow 0$) radiation got hotter as the scale factor $a=a(t)$ decreased, and should have been the dominant contribution to the energy density of the universe. Hence, one would have expected that in the early universe the radiation-filled model should be a very good approximation for the dynamics of the universe. 

However, owing to the results of this Thesis, this picture does not seem to be generic at least in the analytic case. 

Besides, according to R. M. Wald (\cite{wald}, chap. 5), `\textit{It would also be self-consistent to assume that no radiation was present in the very early universe and that only cold matter, such as baryons, was present. However, a "cold big bang" model would have the major tasks of accounting for the present existence of the cosmic microwave background and the correct helium abundance of the universe, both of which are naturally explained by the standard "hot big bang" model described here}'. \\
We present the results of this Thesis in the following matrix:

\begin{center}
\begin{tabular}{|c|c|c|c|c|}\hline
Theory               & Case               & Kind       & Solution         & Arbitrary functions      \\\hline
General Relativity   & Regular case       & Vacuum     & general          & $4$                      \\\hline
General Relativity   & Singular case      & Vacuum     & no solution      & -                        \\\hline
General Relativity   & Regular case       & Radiation  & particular       & $5$ ($8$ for gen. sol.)  \\\hline
General Relativity   & Singular case      & Radiation  & particular       & $3$ ($8$ for gen. sol.)  \\\hline
Higher-order gravity & Regular case       & Vacuum     & general          & $16$                     \\\hline 
Higher-order gravity & Singular case      & Vacuum     & no solution      & -                        \\\hline  
Higher-order gravity & Regular case       & Radiation  & particular       & $17$ ($20$ for gen. sol.) \\\hline
Higher-order gravity & Singular case      & Radiation  & particular       & $15$ ($20$ for gen. sol.) \\\hline
\end{tabular}      
\end{center}


\section{Future directions}
\label{Future directions}

Up to now, we analysed the structure of higher-order gravity theory in vacuum and with radiation content and study them through formal series expansions. It would be interesting to apply the Landau-Lifschitz method in higher-order gravity for the case of dust $p=0$, where the corresponding expansions would be singular, and start with a term in the form $a_{\a\b}t^{4/3}$.

Another future problem would be to apply the same method for the higher-order gravity field equations in order to generalize the results presented in \cite{bct10} and deal with the general sudden cosmological singularity. In this case the formal series expansion is similar to the form (\ref{eq:3dimmetric conclusion}).

Finally, our results currently would nicely complement those obtained by the application of the method of asymptotic splittings \cite{cb07} in the study of the behaviour of dynamical systems concerning singularities of varying light speed cosmologies \cite{mcots07}, but also various systems with symmetry as in \cite{lcf00}. This method is completely different from the Landau-Lifschitz method developed here, due to the fact that it doesn't deal with series that include tensor terms, and it is applied to many kinds of dynamical systems that correspond to interesting cosmological cases (see \cite{ckt,ckt2013,ckkt16,ckit12,ckit13,ckit13ar,akk10,akk13}) which would lead to further interesting results.


\addtocontents{toc}{\vspace{2em}} 

\appendix 



\chapter{Regular solution in general relativity} 

\label{AppendixA} 

\lhead{Appendix A. \emph{Regular solution in general relativity}} 

In this Appendix, we present details of the regural solution of the field equations in general relativity in cases of vacuum and radiation. Here we look for $\g_{\a\b}$ in the form,
\be
\g_{\a\b} = a_{\a\b} + t b_{\a\b} + t^2 c_{\a\b} + \cdots
\label{gr spatial metric regular}
\ee 
where $a_{\a\b},b_{\a\b},c_{\a\b}$ are functions of the space coordinates. The reciprocal tensor is given by,
\be
\g^{\a\b} = a^{\a\b} - t b^{\a\b} + t^2\left(b^{\a\g}\;b_\g^\b - c^{\a\b}\right).
\label{gr spatial metric rec regular}
\ee
Then the extrinsic curvature becomes,
\be
K_{\a\b}=\partial_t\g_{\a\b}=b_{\a\b} + 2c_{\a\b}t,
\label{gr ext cur}
\ee
while,
\be
K_\b^\a=\g^{\a\g}K_{\g\b}=b_\b^\a + t(2c_\b^\a - b^{\a\g}b_{\g\b}).
\label{gr ext cur mixed}
\ee

In case of vacuum, the Einstein equations,
\be
R^0_0 = -\frac{1}{2}\partial_t K -\frac{1} {4}K_\a^\b K_\b^\a = 8\pi G(T^0_0 -\frac{1}{2}T),
\label{eq:RooT2}
\ee
\be
R^0_\a  = \frac{1}{2} (\nabla_\b K^\b_\a - \nabla_\a K) = 8\pi G T^0_\a,
\label{eq:RoaT2}
\ee
\be
R^\b_\a = -P^\b_\a - \frac{1}{2\sqrt{\g}} \partial_t(\sqrt{\g}K^\b_\a) = 8\pi G (T^\b_\a -\frac{1}{2}\d^\b_\a T),
\label{eq:RbaT2}
\ee
lead to the following relations,
\be
R^0_0 = -c + \frac{1}{4}b_\a^\b b_\b^\a = 0,
\label{eq:RooTvacreg}
\ee
\be
R^0_\a  = \frac{1}{2}\left(\nabla_\b b^\b_\a - \nabla_\a b\right) +
t\left(\nabla_\b c^\b_\a - \nabla_\a c -\frac{1} {2}\nabla_\b (b_\g^\b b_\a^\g) +
\frac{1}{2}\nabla_\a (b_\g^\b b_\b^\g)\right) = 0,
\label{eq:RoaTvacreg}
\ee
and also,
\be
R^\b_\a = -P^\b_\a - c^\b_\a + \frac{1} {2}b^\b_\g b^\g_\a -\frac{1} {4}b^\b_\a b = 0,
\label{eq:RbaTvacreg}
\ee
where tensor $P_{\a\b}$ is defined with respect to $a_{\a\b}$.

From (\ref{eq:RbaTvacreg}), the components of the spatial tensor $c_{\a\b}$ are completely determined in terms of the components of $a_{\a\b}$ and $b_{\a\b}$. Further, from the zeroth-order terms in (\ref{eq:RoaTvacreg}) we find three relations between the components of the tensor $b_{\a\b}$, namely,
\be
\nabla_\b b^\b_\a = \nabla_\a b.
\label{bab eq vacreg}
\ee
Then, using (\ref{eq:RooTvacreg}), (\ref{eq:RbaTvacreg}), (\ref{bab eq vacreg}) and the identity,
\be
\nabla_i G^i_j = 0,
\label{id gr}
\ee
where,
\be
G^i_j = R^i_j - \dfrac{1}{2}\d^i_j R,
\label{Eistein tensor}
\ee
the terms of the first-order in (\ref{eq:RoaTvacreg}) vanish identically.\\
Thus, Eqns. (\ref{eq:RooTvacreg})-(\ref{eq:RbaTvacreg}) give $10$ relations between the data $a_{\a\b},b_{\a\b},c_{\a\b}$, so that from the initially $18$ data there remain $8$ arbitrary spatial functions. Because of the four diffeomorphism transformations, we are left with the correct number of four arbitrary functions. Therefore, the form (\ref{gr spatial metric regular}) leads to the general solution of the Einstein equations in vacuum.

On the other hand, in case of radiation, using relations (\ref{T^0_0 generalrad})-(\ref{T^b_a generalrad}), the Einstein equations give,
\be
R^0_0 = -c + \frac{1}{4}b_\a^\b b_\b^\a = \dfrac{8\pi G}{3}\rho(4{u_0}^2 - 1),
\label{eq:RooTradreg}
\ee
\be
R^0_\a  = \frac{1}{2}\left(\nabla_\b b^\b_\a - \nabla_\a b\right) +
t\left(\nabla_\b c^\b_\a - \nabla_\a c -\frac{1} {2}\nabla_\b (b_\g^\b b_\a^\g) +
\frac{1}{2}\nabla_\a (b_\g^\b b_\b^\g)\right) = \dfrac{32\pi G}{3}\rho u_\a u_0,
\label{eq:RoaTradreg}
\ee
and,
\be
R^\b_\a = -P^\b_\a - c^\b_\a + \frac{1} {2}b^\b_\g b^\g_\a -\frac{1} {4}b^\b_\a b = \dfrac{8\pi G}{3}\rho(4u^\b u_\a - \d^\b_\a).
\label{eq:RbaTradreg}
\ee
Using the identity (\ref{eq:velocities identity}), we get,
\be
1 = u_i u^i  \approx  {u_0}^2 - \left[a^{\a\b} - b^{\a\b}t + \left(b^{\a\g}\;b_\g^\b - c^{\a\b}\right)t^2 \right] u_\a u_\b.
\label{gr velocities identity regular}
\ee
Considering,
\be
u_0 \approx 1,
\label{uoapprox1radreg2}
\ee
namely that,
\be
a^{\a\b}u_\a u_\b \rightarrow 0
\label{vel tend to zero2}
\ee
when $t$ tends to zero, the system of Einstein equations (\ref{eq:RooTradreg})-(\ref{eq:RbaTradreg}) become,
\be
R^0_0 = -c + \frac{1}{4}b_\a^\b b_\b^\a = 8\pi G\rho,
\label{eq:RooTradreg2}
\ee
\be
R^0_\a  = \frac{1}{2}\left(\nabla_\b b^\b_\a - \nabla_\a b\right) +
t\left(\nabla_\b c^\b_\a - \nabla_\a c -\frac{1} {2}\nabla_\b (b_\g^\b b_\a^\g) +
\frac{1}{2}\nabla_\a (b_\g^\b b_\b^\g)\right) = \dfrac{32\pi G}{3}\rho u_\a,
\label{eq:RoaTradreg2}
\ee
and,
\be
R^\b_\a = -P^\b_\a - c^\b_\a + \frac{1} {2}b^\b_\g b^\g_\a -\frac{1} {4}b^\b_\a b = -\dfrac{8\pi G}{3}\rho \d^\b_\a.
\label{eq:RbaTradreg2}
\ee
Then, from Eq. (\ref{eq:RooTradreg2}) we find one relation for the energy density $\rho$. Further, substituting $\rho$ to Eq. (\ref{eq:RoaTradreg2}), we obtain three relations between the components of $b_{\a\b}$, namely,
\be
\nabla_\b b^\b_\a = \nabla_\a b
\label{gr com bab rad reg}
\ee
and three more relations for the velocities $u_\a$ and the spatial tensors $a_{\a\b},b_{\a\b},c_{\a\b}$, which are,
\be
u_\a = \dfrac{3\left(\nabla_\b c^\b_\a - \nabla_\a c -\frac{1} {2}\nabla_\b (b_\g^\b b_\a^\g) +
\frac{1}{2}\nabla_\a (b_\g^\b b_\b^\g)\right)}{4(-c + \frac{1}{4}b_\a^\b b_\b^\a)}t.
\label{gr velradreg}
\ee
In addition to that, substituting $\rho$ from (\ref{eq:RooTradreg2}) to (\ref{eq:RbaTradreg2}), we find that,
\be
3(-P^\b_\a - c^\b_\a + \frac{1} {2}b^\b_\g b^\g_\a -\frac{1} {4}b^\b_\a b) + \d^\b_\a (-c + \frac{1}{4}b^\d_g b^g_\d) = 0,
\label{RbaRooradreg}
\ee
which constitute six additional relations between the initial data.\\
Thus, there are $13$ relations in total between the $22$ initial data $a_{\a\b},b_{\a\b},c_{\a\b},\rho,u_a$, and so we are left with $9$ arbitrary functions. Taking into account $4$ diffeomorphism changes there remain $5$ arbitrary functions. This is not the correct number for the general solution of the Einstein equations in case of radiation. Specifically, the correct number for the general solution is the number $8$. We note that in case of general reletivity plus radiation, as well as in case of higher-order gravity plus radiation, the correct number for the general solution of the theories ($8$ in general relativity and $20$ in higher-order gravity) is increased by three. 

Hence, using (\ref{gr spatial metric regular}), in case of vacuum, we are left with the correct number of arbitrary functions concerning the general solution of the Einstein equations, while this is not true in case of radiation . Therefore, only in the case of vacuum, regularity is a generic feature of general relativity as well as the $R + \ep R^2$ theory, assuming analyticity. 


\chapter{Singular solutions in general relativity} 

\label{AppendixB} 

\lhead{Appendix B. \emph{Singular solutions in general relativity}} 

In this Appendix, we give details about the singular solution of the field equations of general relativity in cases of vacuum and radiation. In particular, we look for a solution near the singularity $(t=0)$ in the form,
\be
\g_{\a\b} = t a_{\a\b} + t^2 b_{\a\b} + \cdots
\label{gr spatial metric singular}
\ee
where $a_{\a\b},b_{\a\b}$ are functions of the space coordinates. For the reciprocal tensor we find,
\be
\g^{\a\b} = \dfrac{1}{t}a^{\a\b} -  b^{\a\b},
\label{gr spatial metric rec singular}
\ee
while for the extrinsic curvature we have explicitly,
\be
K_{\a\b} = \partial_t \g_{\a\b}= a_{\a\b} + 2tb_{\a\b},
\label{eq:Kab_rad gr}
\ee
and,
\be
K_\b^\a = \dfrac{1}{t}\d^\b_\a + b^\b_\a.
\label{eq:Kmixed_rad}
\ee
In case of vacuum, the Einstein equations (\ref{eq:RooT2})-(\ref{eq:RbaT2}) give the following relations:
\be
R^0_0 = \dfrac{3}{4t^2} - \dfrac{b}{2t} = 0,
\label{eq:RooTvacsin}
\ee
\be
R^0_\a  = \frac{1}{2}\left(\nabla_\b b^\b_\a - \nabla_\a b\right) = 0
\label{eq:RoaTvacsin}
\ee
and also,
\be
R^\b_\a = -\frac{1}{4t^2}\d^\b_\a - \frac{1}{4t}(4P^\b_\a + 3b^\b_\a +b\d^\b_\a) = 0,
\label{eq:RbaTvacsin}
\ee
where tensor $P_{\a\b}$ is defined with respect to $a_{\a\b}$. But from the $(00)$-component of the Einstein equations, we see that the system is unsolvable. Therefore, the form (\ref{gr spatial metric singular}) can not give any kind of solution of the Einstein equations in vacuum.

Nevertheless, in case of radiation, using relations (\ref{T^0_0 generalrad})-(\ref{T^b_a generalrad}), the Einstein equations give,
\be
R^0_0 = \dfrac{3}{4t^2} - \dfrac{b}{2t} = \dfrac{8\pi G}{3}\rho(4{u_0}^2 - 1),
\label{eq:RooTradsin}
\ee
\be
R^0_\a  = \frac{1}{2}\left(\nabla_\b b^\b_\a - \nabla_\a b\right) = \dfrac{32\pi G}{3}\rho u_\a u_0,
\label{eq:RoaTradsin}
\ee
and,
\be
R^\b_\a = -\frac{1}{4t^2}\d^\b_\a - \frac{1}{4t}(4P^\b_\a + 3b^\b_\a +b\d^\b_\a) = \dfrac{8\pi G}{3}\rho(4u^\b u_\a - \d^\b_\a).
\label{eq:RbaTradsin}
\ee 
In view of the identity,
\be
1 = u_i u^i  \approx  {u_0}^2 - \dfrac{1}{t}a^{\a\b}u_\a u_\b,
\label{uoapprox1radsin2}
\ee
the system of the Einstein equations (\ref{eq:RooTradsin})-(\ref{eq:RbaTradsin}) becomes,
\be
R^0_0 = \dfrac{3}{4t^2} - \dfrac{b}{2t} = 8\pi G \rho,
\label{eq:RooTradsin2}
\ee
\be
R^0_\a  = \frac{1}{2}\left(\nabla_\b b^\b_\a - \nabla_\a b\right) = \dfrac{32\pi G}{3}\rho u_\a,
\label{eq:RoaTradsin2}
\ee
and,
\be
R^\b_\a = -\frac{1}{4t^2}\d^\b_\a - \frac{1}{4t}(4P^\b_\a + 3b^\b_\a +b\d^\b_\a) = -\dfrac{8\pi G}{3}\rho \d^\b_\a.
\label{eq:RbaTradsin2}
\ee 
Then, the energy density $\rho$ is found by (\ref{eq:RooTradsin2}), whereas for the velocities we have,
\be
u_\a = \dfrac{t^2}{2}\left(\nabla_\b b^\b_\a - \nabla_\a b\right).
\label{gr vel rad sin}
\ee
Further, substituting $\rho$ from Eq. (\ref{eq:RooTradsin2}) to Eq. (\ref{eq:RbaTradsin2}), we find that the terms of order $t^{-2}$ cancel, while the terms of order $t^{-1}$ give,
\be
P^\b_\a + \dfrac{3}{4}b^\b_\a + \dfrac{5}{12}\d^\b_\a b = 0.
\label{eq for Pab gr rad sin}
\ee
Taking into account the trace of (\ref{eq for Pab gr rad sin}) we find,
\be
P + 2b = 0, 
\label{trPab gr rad sin}
\ee
and then, using (\ref{eq for Pab gr rad sin}) and (\ref{trPab gr rad sin}) we obtain,
\be
b^\b_\a = -\dfrac{4}{3}P^\b_\a + \dfrac{5}{18} \d^\b_\a P.
\label{eq for bab gr rad sin}
\ee
Due to the identity (\ref{id gr}), we find that,
\be
\nabla_\b b^\b_\a = \dfrac{7}{9}\nabla_\a b.
\label{rel between bab gr rad sin}
\ee
Using Eqns. (\ref{gr vel rad sin}) and (\ref{rel between bab gr rad sin}), we obtain the final form for the velocities, which is,
\be
u_\a = -\dfrac{t^2}{9}\nabla_\a b.
\label{final gr vel rad sin}
\ee
Thus, all six functions $a_{\a\b}$ remain arbitrary, while the coefficients $b_{\a\b}$ are determined in terms of them. Additionally, the diffeomorphism transformations are not four, but three, due to the fact that the choice of the time in the $3$-metric (\ref{gr spatial metric singular}) is completely determined by the condition $t=0$ at the singularity.

Hence, Eqns. (\ref{eq:RooTradsin2}), (\ref{eq for bab gr rad sin}) and (\ref{final gr vel rad sin}) give $10$ relations in total between the $16$ initial data $a_{\a\b},b_{\a\b},\rho,u_\a$. Taking into account the $3$ diffeomorphism changes, we are left with the number $3$, which of course does not correspond to the general solution of the theory (the general solution includes $8$ arbitrary functions).

\addtocontents{toc}{\vspace{2em}} 

\backmatter


\label{Bibliography}



\end{document}